\documentclass{aa}

\usepackage{twoopt,natbib}
\usepackage{graphicx}
\usepackage{xcolor}
\usepackage{txfonts}
\usepackage[breaklinks=true,colorlinks=true,linkcolor=blue,citecolor=blue,filecolor=blue,urlcolor=blue]{hyperref}

\usepackage{amssymb}	
\usepackage{textcomp}
\usepackage{amsmath}
\usepackage[normalem]{ulem}
\usepackage{physics}
\usepackage{subcaption}

\newcommand{\arepo}{\texttt{AREPO}}
\newcommand{\mesa}{\texttt{MESA}}

\newcommand{\Msun}{\,\mathrm{M}_\odot}
\newcommand{\Rsun}{\,\mathrm{R}_\odot}
\newcommand{\au}{\,\textsc{au}}
\newcommand{\hr}{\,\mathrm{hr}}
\newcommand{\yr}{\,\mathrm{yr}}

\newcommand{\Gyr}{\,\mathrm{Gyr}}
\newcommand{\yrinv}{\,\mathrm{yr}^{-1}}
\newcommand{\MWGalinv}{\,\mathrm{MWGal}^{-1}}
\newcommand{\pcinvcb}{\,\mathrm{pc}^{-3}}
\newcommand{\kmsinv}{\,\mathrm{km}\,\mathrm{s}^{-1}}
\newcommand{\msinv}{\,\mathrm{m}\,\mathrm{s}^{-1}}
\newcommand{\gcmsqsinv}{\,\mathrm{g}\,\mathrm{cm}^{2}\,\mathrm{s}^{-1}}

\newcommand{\mbh}{m_{\mathrm{BH}}}
\newcommand{\ms}{m_{\mathrm{\star}}}
\newcommand{\rs}{r_{\mathrm{\star}}}
\newcommand{\Zs}{Z_{\mathrm{\star}}}
\newcommand{\ts}{t_{\mathrm{\star}}}
\newcommand{\Ls}{L_{\mathrm{\star}}}
\newcommand{\tdyn}{t_{\mathrm{dyn},\star}}
\newcommand{\rt}{r_{\mathrm{t}}}
\newcommand{\rp}{r_{\mathrm{p}}}

\newcommand{\rg}{r_{\mathrm{g}}}
\newcommand{\msr}{m_{\mathrm{\diamond}}}
\newcommand{\Dmsr}{\Delta m_{\mathrm{\diamond}}}
\newcommand{\Lsr}{L_{\mathrm{\diamond}}}
\newcommand{\Hc}{\mathrm{H}_{\mathrm{c}}}
\newcommand{\Eorb}{\varepsilon_{\mathrm{orb}}}
\newcommand{\Etide}{\varepsilon_{\mathrm{tide}}}
\newcommand{\Ebind}{\varepsilon_{\mathrm{bind}}}
\newcommand{\Ekick}{\varepsilon_{\mathrm{kick}}}
\newcommand{\horb}{h_{\mathrm{orb}}}
\newcommand{\rhoconc}{\rho_{\mathrm{conc}}}
\newcommand{\rhoconcexp}{(\rho_{\mathrm{c}}/\overline{\rho})^{1/3}}

\newcommand\Tstrut{\rule{0pt}{2.9ex}}         
\newcommand\Bstrut{\rule[-1.2ex]{0pt}{0pt}}   
\newcommand\TBstrut{\Tstrut\Bstrut}           

\begin{document}

   \title{Simulating the tidal disruption of stars by stellar-mass black holes using moving-mesh hydrodynamics}
   \titlerunning{Dynamics of $\mu$TDEs}

   \author{Pavan Vynatheya\inst{1}\fnmsep\thanks{corresponding author},
            Taeho Ryu\inst{1,2},
            Rüdiger Pakmor\inst{1},
            Selma E. de Mink\inst{1,3}
            \and
            Hagai B. Perets\inst{4}
          }
    \authorrunning{Vynatheya et al.}

   \institute{Max-Planck-Institut für Astrophysik, Karl-Schwarzschild-Straße 1, 85748 Garching bei München, Germany
         \and
            Physics and Astronomy Department, Johns Hopkins University, Baltimore, MD 21218, USA
        \and
            Anton Pannekoek Institute for Astronomy, University of Amsterdam, Science Park 904, 1098 XH Amsterdam, Netherlands
        \and
            Physics Department, Technion - Israel Institute of Technology, Haifa 3200003, Israel\\
             }

   \date{Received XXX; accepted YYY}

 
  \abstract
    {In the centers of dense star clusters, close encounters between stars and compact objects are likely to occur. We study tidal disruption events of main-sequence (MS) stars by stellar-mass black holes (termed $\mu$TDEs), which can shed light on the processes occurring in these clusters, including being an avenue in the mass growth of stellar-mass BHs. Using the moving-mesh hydrodynamics code \texttt{AREPO}, we perform a suite of  58 hydrodynamics simulations of partial $\mu$TDEs of realistic, \texttt{MESA}-generated MS stars by varying the initial mass of the star ($0.5\,{\rm M}_{\rm \odot}$ and $1\,{\rm M}_{\rm \odot}$), the age of the star (zero-age, middle-age and terminal-age), the mass of the black hole ($10\,{\rm M}_{\rm \odot}$ and $40\,{\rm M}_{\rm \odot}$) and the impact parameter (yielding almost no mass loss to full disruption). We then examine the dependence of the masses, spins, and orbital parameters of the partially disrupted remnant on the initial encounter parameters. We find that the mass lost from a star decreases  roughly exponentially with increasing distance of approach and that a $1\,{\rm M}_{\rm \odot}$ star loses lesser mass than a $0.5\,{\rm M}_{\rm \odot}$. Moreover, a more evolved star is less susceptible to mass loss. Tidal torques at the closest approach spin up the remnant very close to break-up velocity when the impact parameter is low. The remnant star can be bound (eccentric) or unbound (hyperbolic) to the black hole: hyperbolic orbits occur when the star's central density concentration is relatively low and the black hole-star mass ratio is high, which is the case for the disruption of a $0.5\,{\rm M}_{\rm \odot}$ star. Finally, we provide best-fit analytical formulae for  the aforementioned range of parameters that can be incorporated into cluster codes to model star-black hole interaction more accurately.}


   \maketitle
%

\section{Introduction} \label{sec:introduction}

Tidal disruption events (TDEs) of stars by supermassive black holes (SMBHs) have been a subject of significant interest in the past decade (see \citealp{2019GReGr..51...30S,2021ARA&A..59...21G} for reviews). TDEs by SMBHs are observed as transients in multiple wavelengths \citep{2021ARA&A..59...21G}, with $\sim 100$ such events having been detected by observatories in optical (ASAS-SN: \citealp{2017PASP..129j4502K}, ZTF: \citealp{2019PASP..131a8002B}, PTF: \citealp{2009PASP..121.1395L}, PS: \citealp{2016arXiv161205560C}, ATLAS: \citealp{2018PASP..130f4505T}, SDSS: \citealp{2019BAAS...51g.274K}, OGLE: \citealp{2015AcA....65....1U}), X-ray (ROSAT: \citealp{1999A&A...349..389V}, Swift: \citealp{2004ApJ...611.1005G}, XMM: \citealp{2001A&A...365L...1J}, Chandra: \citealp{2000SPIE.4012....2W}) and UV (GALEX: \citealp{2005ApJ...619L...1M}).

In this work, we examine TDEs of stars by stellar-mass black holes or SBHs (termed $\mu$TDEs). These are less studied and have only garnered interest recently. Although $\mu$TDEs have lower observable rates \citep{2016ApJ...823..113P}, they can shed light on the processes occurring in the centers of globular and nuclear star clusters where they are most likely to occur. In particular, they are an important avenue in the mass growth of SBHs to form intermediate-mass black holes (IMBHs, e.g., \citealp{2017MNRAS.467.4180S,2023MNRAS.521.2930R}). However, these studies make simplistic assumptions about the mass accreted onto a black hole after a TDE, which highlights the need for detailed simulations to model this interaction more accurately for the next generation of globular cluster simulations. Partial tidal disruptions (PTDEs), which are more likely than full tidal disruptions (FTDEs), can be responsible for the tidal capture, and subsequent encounters, of the remnant star by the BH if the remnant ends up in a bound orbit. The remnant stars of such interactions tend to be spun up from the torque due to the BH and can have peculiar internal structures (e.g., \citealp{2001ApJ...549..948A}).

$\mu$TDEs are also of interest in the context of low-mass X-ray binaries (LMXBs) and the newly-discovered Gaia BHs \citep{2023MNRAS.521.4323E,2023MNRAS.518.1057E,2023AJ....166....6C}. Isolated binary evolution cannot satisfactorily explain the observed rates of binaries with BHs and low-mass stellar companions (e.g., \citealp{2003MNRAS.341..385P,2006MNRAS.369.1152K}). Hence, the dynamical formation of such binaries in clusters, through PTDEs and subsequent tidal captures, may very well be a crucial process to explain them.

Accurate modeling of post-disruption orbits and the internal structure of remnant stars requires detailed hydrodynamics simulations. The first detailed study on $\mu$TDEs was carried out by \citet{2016ApJ...823..113P}. They estimated that these events could occur in Milky Way globular clusters at a rate of $10^{-6}\yrinv\MWGalinv$, and might observationally resemble ultra-long GRB events. Similar rates were obtained for scenarios where a supernova natal kick to a newly formed BH results in a chance encounter with its binary companion. \citet{2021MNRAS.503.6005W} used smoothed-particle hydrodynamics (SPH) to find the fallback rates onto the BH in the case of partial $\mu$TDEs of polytropic stars. \citet{2022ApJ...933..203K} performed a large suite of SPH simulations of partial- and full-disruptions, with varying BH and stellar masses, stellar polytropic indices, and impact parameters. Among other things, they observed that stellar structure plays a crucial role in the boundedness of the remnant stars after partial disruption, with the possibility of less centrally `concentrated' stars ending up in hyperbolic orbits. \citet{2024ApJ...961..149X} looked at low-eccentricity `tidal-peeling' events of realistic MS stars, with the star being slowly stripped away in multiple orbits.

Hydrodynamics studies on three-body encounters involving SBHs in globular clusters have also garnered interest, motivated by the significantly larger cross-sections of binary-single tidal interactions. \citet{2019ApJ...877...56L} studied the interaction of binary BHs and polytropic stars and showed that tidal disruption can alter the spin of one of the BHs. An extensive series of studies was carried out by \citet{2022MNRAS.516.2204R,2023MNRAS.519.5787R,2023MNRAS.525.5752R,2024MNRAS.527.2734R}, using both SPH and moving-mesh codes, on the different combinations of close encounters between realistic MS stars and SBHs.

In our study, we use the moving-mesh code \arepo{} \citep{2010MNRAS.401..791S,2016MNRAS.455.1134P,2020ApJS..248...32W} to simulate $\mu$TDEs of low-mass MS stars with SBHs. It should be noted that \citet{2023MNRAS.519.5787R,2023MNRAS.525.5752R,2024MNRAS.527.2734R} also employed \arepo{} to model TDEs with remarkable success. We generate realistic stellar models from detailed 1D \mesa{} \citep{2011ApJS..192....3P} profiles. TDEs of \mesa{}-generated stars by SMBHs and IMBHs have been investigated in the past (e.g., \citealp{2018ApJ...857..109G,2019ApJ...882L..26G,2019ApJ...882L..25L,2020ApJ...905..141L,2020ApJ...904...98R,2020ApJ...904...99R,2020ApJ...904..100R,2020ApJ...904..101R,2023ApJ...948...89K}), but $\mu$TDE studies have typically probed polytropic stars. In this paper, we focus on the dependence of post-disruption mass, spin, and orbital parameters on the initial conditions and provide best-fit functions for the same. These fits can be incorporated into  $N$-body or other cluster codes for better treatment of star-BH encounters.

The paper is organized as follows. In Section \ref{sec:methods}, we detail our simulation methodology, including our grid of initial conditions. We briefly describe the analysis of certain quantities in Section \ref{sec:analysis} and present our results and best-fit functions in Section \ref{sec:results}. Section \ref{sec:implications} and Section \ref{sec:summary} discuss the implications of our work and conclude, respectively.

\section{Methods} \label{sec:methods}

\begin{figure*}
    \begin{subfigure}{\columnwidth}
        \includegraphics[width=\columnwidth]{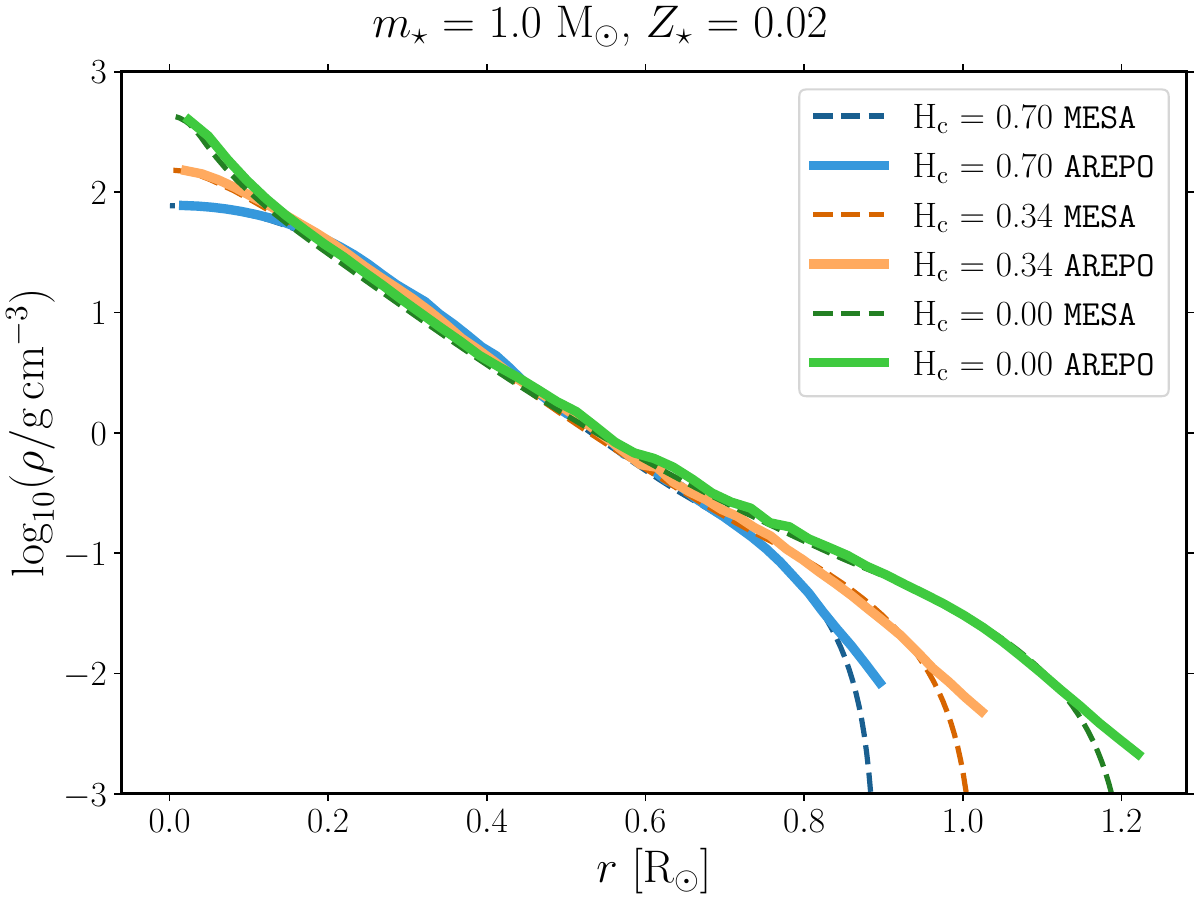}
        \caption{}
        \label{fig:den_profile_1M}
    \end{subfigure}
    \hfill
    \begin{subfigure}{\columnwidth}
        \includegraphics[width=\columnwidth]{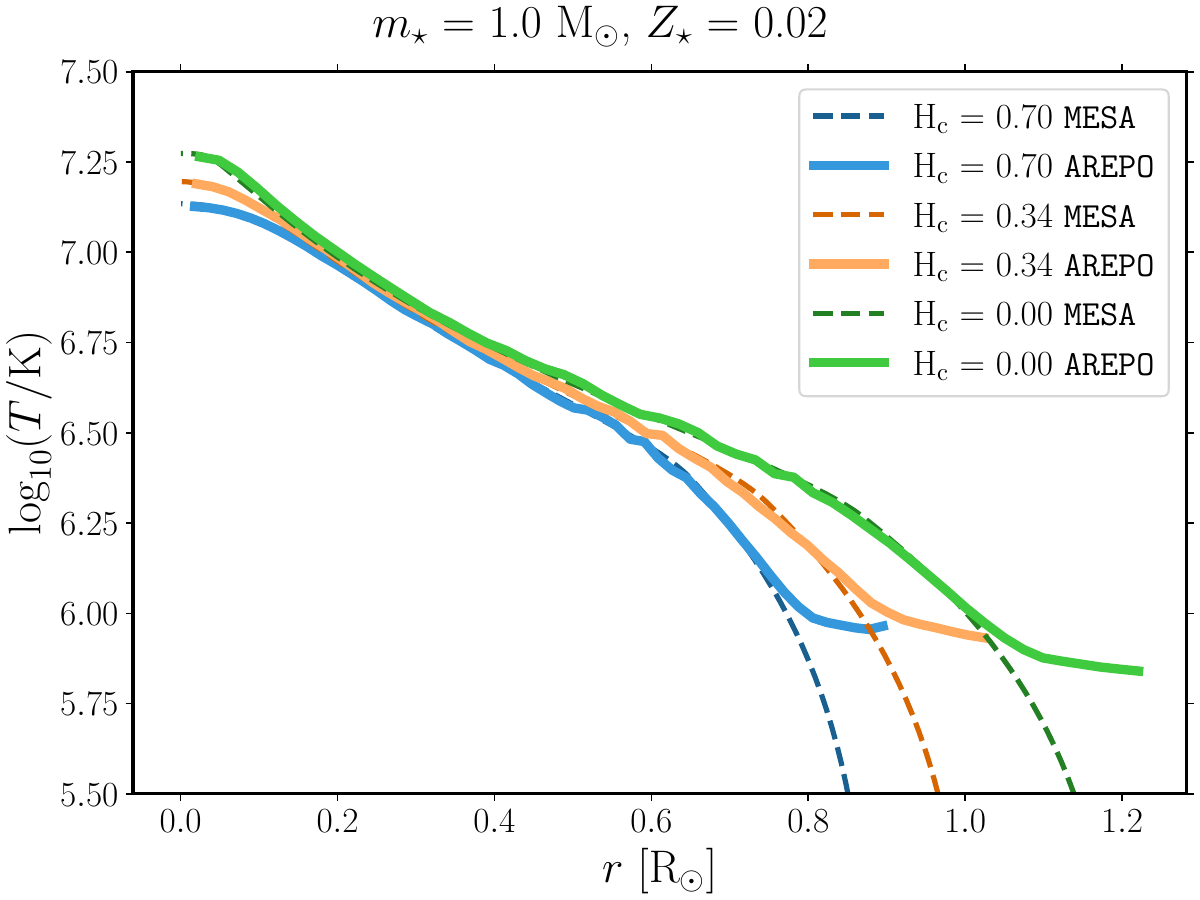}
        \caption{}
        \label{fig:temp_profile_1M}
    \end{subfigure}
    \centering
    \caption{Density (left) and temperature (right) profiles of a $1 \Msun$ MS star of different models (see Table \ref{tab:star_details}). The dashed and solid lines represent the 1D profiles generated by \mesa{} and initialized (and subsequently relaxed for five dynamical timescales) in 3D in \arepo{} respectively. They agree very well over most of the range except close to the stellar surface because of smoothing during relaxation in \arepo{}. An older MS star (with lower $\Hc$) has a denser and hotter core and a puffier outer layer than a younger MS star (with higher $\Hc$).}
    \label{fig:den_temp_profiles_1M}
\end{figure*}

\begin{figure*}
    \begin{subfigure}{\columnwidth}
        \includegraphics[width=\columnwidth]{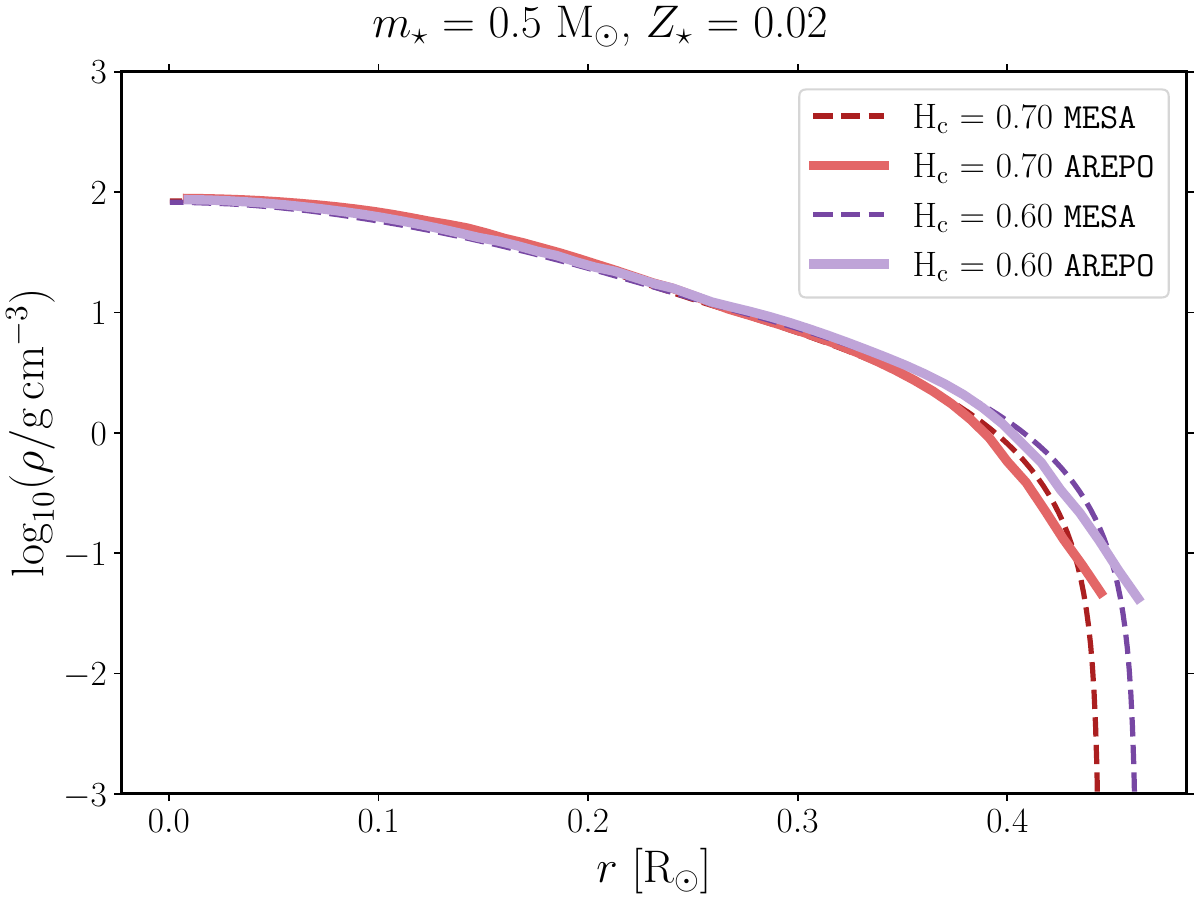}
        \caption{}
        \label{fig:den_profile_0.5M}
    \end{subfigure}
    \hfill
    \begin{subfigure}{\columnwidth}
        \includegraphics[width=\columnwidth]{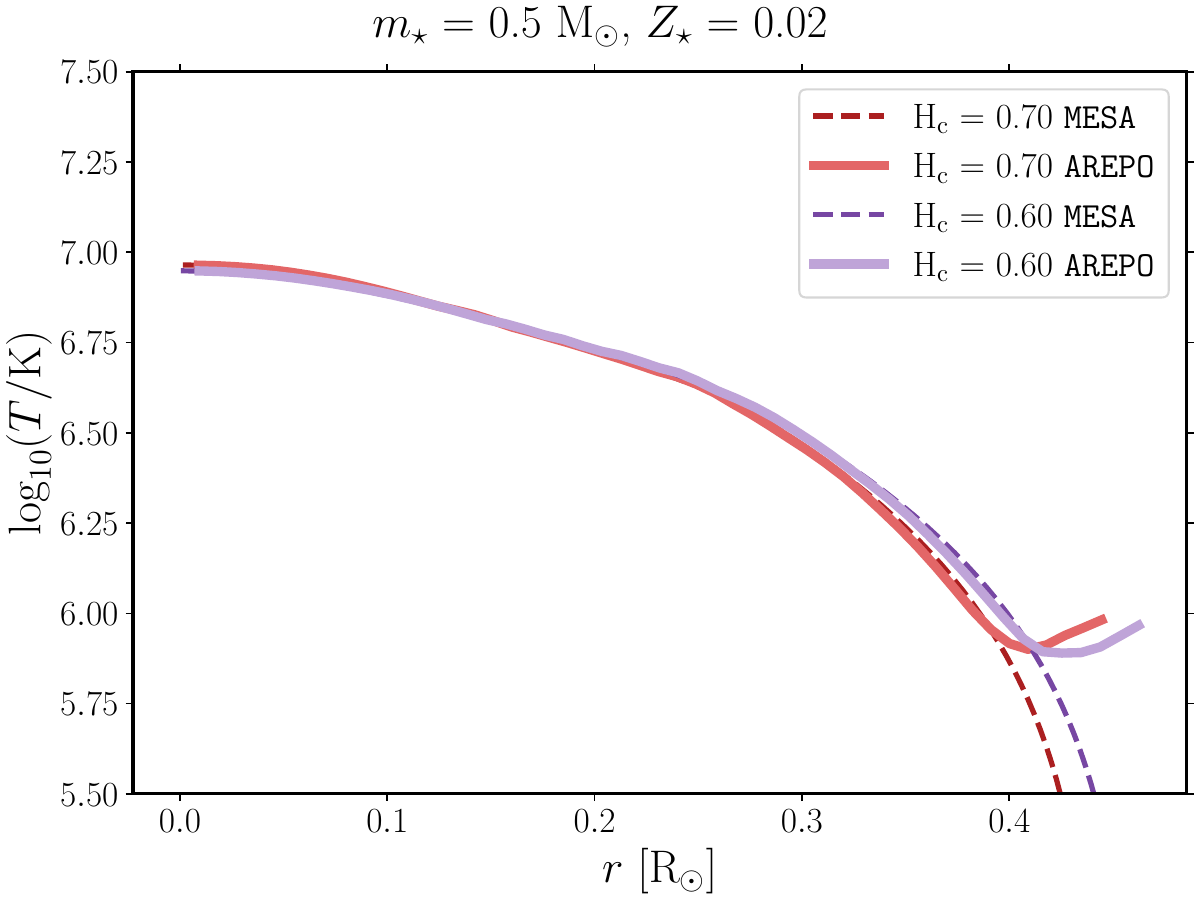}
        \caption{}
        \label{fig:temp_profile_0.5M}
    \end{subfigure}
    \centering
    \caption{Similar profiles as Figure \ref{fig:den_temp_profiles_1M}, but for a $0.5 \Msun$ MS star. A $0.5 \Msun$ star of age $\sim 13.5 \Gyr$ with $\Hc = 0.60$ (purple curves) is very similar to a ZAMS star with $\Hc = 0.70$ (red curves), albeit with a slightly larger radius.}
    \label{fig:den_temp_profiles_0.5M}
\end{figure*}

\subsection{Hydrodynamics}

We simulate the disruption of stars by black holes using the moving-mesh magnetohydrodynamics code \arepo{} \citep{2010MNRAS.401..791S,2016MNRAS.455.1134P,2020ApJS..248...32W}. \arepo{} is a massively parallel 3D magnetohydrodynamics code with gravity that inherits many advantages of the two popular schemes, Eulerian grid-based finite-volume codes and Lagrangian smoothed particle hydrodynamics (SPH). Although initially developed for cosmological simulations, \arepo{} has been successfully employed in phenomena involving stars, e.g., tidal disruptions and encounters \citep{2019MNRAS.487..981G,2023MNRAS.519.5787R,2023MNRAS.525.5752R,2024MNRAS.527.2734R}, TDEs in active galactic nuclei disks \citep{2024MNRAS.527.8103R}, collisions of main-sequence stars \citep{2019Natur.574..211S}, giant stars \citep{2024MNRAS.528.6193R}, white dwarfs \citep{2013ApJ...770L...8P,2021MNRAS.503.4734P,2022MNRAS.517.5260P,2023MNRAS.523..527B,2021A&A...649A.155G,2023arXiv230903300G} and neutron stars \citep{2024MNRAS.528.1906L}, and common envelope evolution \citep{2016ApJ...816L...9O,2016MNRAS.462L.121O,2020A&A...644A..60S,2020A&A...642A..97K,2022A&A...660L...8O}.

To solve the fluid equations, \arepo{} builds an unstructured Voronoi mesh with cells of varying volume and calculates fluxes between the cells in a finite-volume approach. The Voronoi mesh moves in time according to the approximate bulk velocities of the fluid elements. Gravity is handled in a tree-particle-mesh scheme (see \citealp{2002JApA...23..185B}),  with the minimum gravitational softening for the gas (star) cells set as one-tenth of the smallest gas cell for accuracy. \arepo{} allows for adaptation of spatial resolution according to arbitrary criteria on top of the default adaptivity to density, inherited from the near-Lagrangian nature of the scheme. It also includes particles that interact only gravitationally. We assume Newtonian gravity and do not include magnetic fields in our simulations.

\subsection{Stellar models}

We use the 1D stellar modeling code \mesa{} \citep{2011ApJS..192....3P} to generate main-sequence (MS) stellar models with initial masses $\ms=0.5 \Msun$ and $1.0 \Msun$, and metallicity $\Zs = 0.02$ (near-solar) at different ages. These masses are typical for globular clusters that predominantly consist of old, low-mass stars (e.g., \citealp{2002A&A...388..492S,2005AJ....130..116D,2009ApJ...694.1498M}), while the metallicity is higher than those of most clusters, which typically have a tenth (or less) of Solar metallicity (e.g., \citealp{2010arXiv1012.3224H})). However, the effect of metallicity is not expected to be significant and is quantified in Section \ref{sec:results}. \mesa{} solves the stellar structure equations to compute the evolution and provides us with stellar mass- and radial-profile parameters (e.g., densities, temperatures, and chemical abundances) at different evolutionary phases. We convert these 1D profiles to 3D \arepo{} initial conditions using the scheme adopted by \citet{2017A&A...599A...5O}, with the Helmholtz equation of state \citep{2000ApJS..126..501T}. After generating the star, we relax it for five stellar dynamical timescales $\tdyn = (\rs^3/G\ms)^{0.5}$, where $\rs$ is the stellar radius and $G$ is the gravitational constant. We then use this relaxed star for the disruption simulations.

\begin{table}
    \caption{Parameters of the $\Zs = 0.02$ MS star models.} 
    \label{tab:star_details}
    \centering
    \begin{tabular}{c c c c c c c}
    \hline
    $\ms$ [$\Msun$] & $\ts$ [$\Gyr$] & $\Hc$ & $\rs$ [$\Rsun$] & $\tdyn$ [$\hr$] & $\rhoconc^{-1}$ \TBstrut \\ 
    \hline
    0.5 & 0.01 & 0.70 & 0.44 & 0.19 & 0.46 \TBstrut \\
    \hline
     & 0.01 & 0.70 & 0.90 & 0.38 & 0.29 \Tstrut \\
    1.0 & 4.68 & 0.34 & 1.02 & 0.46 & 0.20 \\
     & 8.60 & 0.00 & 1.22 & 0.60 & 0.12 \Bstrut \\
    \hline
    \end{tabular}
    \tablefoot{The columns indicate stellar parameters - the mass $\ms$, the age $\ts$, the central hydrogen abundance $\Hc$, the radius $\rs$, the dynamical time $\tdyn$ and the inverse of the density concentration parameter $\rhoconc^{-1}$.}
\end{table}

Stellar density profiles are a major factor in determining whether a star undergoes a partial tidal disruption (PTDE) or a full tidal disruption (FTDE) for a given periapsis distance (e.g., \citealp{2020ApJ...904..100R}). To that end, we consider \mesa{} models of three evolutionary stages of $1.0 \Msun$ MS stars  with varying core hydrogen abundances $\Hc$, which correspond to distinct density profiles (see also \citealp{2018ApJ...857..109G,2019ApJ...882L..26G,2019ApJ...882L..25L,2019MNRAS.487..981G}). The first is close to the onset of hydrogen burning ($\Hc \approx 0.70$), i.e., zero-age main-sequence (hereafter ZAMS), the second is approximately midway through the main-sequence ($\Hc \approx 0.34$), i.e., middle-age main-sequence (hereafter MAMS), and the third is close to the depletion of core hydrogen ($\Hc \approx 0.00$), i.e., terminal-age main-sequence (hereafter TAMS). The total stellar mass remains nearly constant throughout the MS lifetime owing to the insignificant wind mass loss. In the case of the $0.5 \Msun$ stars, we examine only ZAMS profiles ($\Hc \approx 0.70$). This choice is motivated by the fact that their internal structure barely changes over time (see Figure \ref{fig:den_temp_profiles_0.5M}) owing to their large MS lifetimes. One could choose a $0.5\Msun$ star at an age comparable to a Hubble time \citep[e.g.,][]{2020ApJ...904...98R,2020ApJ...904..100R} but the results are expected to be very similar. Table \ref{tab:star_details} lists the stellar parameters of the MS star models for the stellar masses and ages we simulate.

 Furthermore, we performed resolution scaling tests to ensure that the PTDE simulations were robust. To this end, we varied the number of fluid (star) cells making up a $1 \Msun$ star -- $5\times10^4$, $8\times10^4$, $2\times10^5$, $5\times10^5$, and $8\times10^5$. We then performed PTDE simulations involving a $1 \Msun$ star and a $10 \Msun$ BH for each resolution. The masses lost due to tidal disruption plateaued close to the resolutions $2\times10^5$ and $5\times10^5$, with the mass loss values using $5\times10^5$ and $8\times10^5$ cells being essentially the same. Hence, we chose an initial resolution of $5\times10^5$ fluid cells for the stars in this study. It should be noted that the number of cells increases as the simulations progress since \arepo{} automatically adds extra cells when the density gradient is high.

\begin{table*}
    \caption{Initial parameters of our suite of simulations.}
    \label{tab:grid_suite}
    \centering
    \begin{tabular}{c c c c}
    \hline
    $\ms$ [$\Msun$] & $\mbh$ [$\Msun$] & $\Hc$ & $b = \rp/\rt$ \TBstrut \\ 
    \hline
    0.5 & 10.0, 40.0 & 0.70 & 0.25, 0.33, 0.50, 0.75, 1.00, 1.50, 2.00, 2.50 \Tstrut \\
    1.0 & 10.0, 40.0 & 0.70, 0.34, 0.00 & 0.25, 0.33, 0.50, 0.75, 1.00, 1.50, 2.00 \Bstrut \\
    \hline
    \end{tabular}
    \tablefoot{The columns indicate the initial parameters - the stellar mass $\ms$, the BH mass $\mbh$, the core hydrogen fraction $\Hc$, and the impact parameter $b$. In each row, every comma-separated combination of values in all columns is simulated, to give a total of 58 simulations.}
\end{table*}

Figure \ref{fig:den_temp_profiles_1M} shows that more evolved $1 \Msun$ MS stars have a denser (left panel) and hotter (right panel) core and a puffier outer layer than less evolved MS stars. Since the three stars have different radii $\rs$, they also have different $\tdyn$ and different tidal radii $\rt = (\mbh/\ms)^{1/3}\rs$ for the same BH mass $\mbh$. Figure \ref{fig:den_temp_profiles_0.5M} shows that $0.5 \Msun$ stars of different ages -- ZAMS ($\Hc = 0.70$) and $\sim 13.5 \Gyr$ ($\Hc = 0.60$) -- have very similar density and temperature profiles, which justifies excluding the latter from our simulation runs. It should be noted that the \arepo{} profiles, especially the temperature, diverge significantly from the \mesa{} ones near the surface of the star (corresponding to a fractional mass of  $\sim 10^{-4}$--$10^{-3}$). This amount of mass near the surface that deviates from the \mesa{} model sets the minimum amount of stripped mass in PTDEs that we can resolve. Therefore, we only present the results of PTDEs yielding a mass loss  $\gtrsim 10^{-4}\ms$ in this paper.

\subsection{Black hole}
The black holes are initialized in \arepo{} as non-rotating sink particles that interact gravitationally and grow in mass through accretion. We set the gravitational softening length of the BH to be ten times the minimum softening length of the  gas (star) cells. The scheme we use for accretion onto the BH, described here briefly, is the same as the one used in \citet{2022MNRAS.516.2204R,2023MNRAS.519.5787R,2023MNRAS.525.5752R,2024MNRAS.527.2734R}. Firstly, the cells around the BH within a radius of $10^4 \rg$, where gravitational radius $\rg = G \mbh/c^{2}$, are identified. Secondly, the accretion onto the BH is calculated using a weighted average of radial flux to account for the higher contribution of gas closer to the BH and vice-versa. Thirdly, mass is extracted from the neighboring cells of the BH, and momentum is inserted in the BH to conserve these quantities. Finally, the BH is spun up from accretion by adding angular momentum from the selected neighboring cells. The radiation feedback from accretion is not taken into account.  Although angular momentum is conserved through this scheme, there is no guarantee that the total energy is conserved. However, given that accretion onto the black hole is small in our simulations, the energy lost due to accretion is negligible.

We run simulations with SBH masses of $10 \Msun$ and $40 \Msun$, which are in proximity to the lower- and higher-mass peaks of observed LIGO BH masses (e.g., \citealp{2021ApJ...913L...7A,2023PhRvX..13a1048A}). Given the mass of the BH $\mbh$, the mass of the star $\ms$, and the radius of the star $\rs$, the classical tidal radius is defined as $\rt = (\mbh/\ms)^{1/3} \rs$.

\subsection{Initial conditions}

The distance of the closest approach is characterized by the impact parameter: $b \equiv \rp/\rt$ \footnote{$\beta \equiv b^{-1}$ is often introduced in the literature for TDEs by SMBHs as the `penetration factor'.}, where $\rp$ is the periapsis distance of the initially parabolic orbit. We vary the mass of the star: $0.5 \Msun$ and $1 \Msun$, and the mass of the black hole: $10 \Msun$ and $40 \Msun$. We then generate a suite of simulations of $b$ varying from $0.25$ to $2.00$ ($1 \Msun$ stars) or $2.50$ ($0.5 \Msun$ stars), and the three stellar ages (density profiles) to determine the post-disruption orbital, mass and spin parameter. The additional simulations for $0.5 \Msun$ and $b=2.50$ are performed to properly analyze the PTDE trends since $0.5 \Msun$ stars are fully disrupted for $b \lesssim 1.00$. 

 We also assume that the initial stars are non-rotating and that the star-BH encounters are parabolic. This can be justified for encounters in globular clusters because the velocity dispersions of Milky Way globular clusters are in the range $\sim 1$--$20 \kmsinv$ (e.g., \citealp{2018MNRAS.478.1520B}). For instance, given a velocity dispersion $\sigma\simeq 10 \kmsinv$, $|1-e| \simeq 10^{-4}$ (very close to parabolic) for the parameters considered in this paper.

 Each simulation is run for $\sim 70\ \tdyn$ after periapsis passage. This duration was chosen to ensure that the stellar masses reach steady values for proper analysis. Table \ref{tab:grid_suite} provides an overview of the parameters we use in our suite of simulations.

\section{Analysis} \label{sec:analysis}

\subsection{Calculation of bound and unbound mass}

To find the center of mass of a self-bound object (i.e., star before disruption, remnant after disruption) in each snapshot, we used an iterative procedure fairly similar to \citet{2013ApJ...767...25G}, with some differences to improve the accuracy of the identification of bound and unbound cells. For each snapshot, we chose the gas cell with the maximum density as the `initial' guess for the center of mass (CoM) position. Subsequently, we calculate the specific total energies of all the star cells relative to the star CoM, $\varepsilon_{\mathrm{cell,CoM}}$, and the BH, $\varepsilon_{\mathrm{cell,BH}}$, as the sum of their relative kinetic, potential and internal energies. We consider a cell to be bound to a self-gravitating object if $\varepsilon_{\mathrm{cell,CoM}} < 0$, $\varepsilon_{\mathrm{cell,CoM}} < \varepsilon_{\mathrm{cell,BH}}$. The last condition ensures that the cells bound to the star are `close' to the CoM. We compute a `new' CoM position (and velocity) using these bound cells and determine the `new' specific energy. We iterate this process until  there is no change in the relative position of the CoM. Subsequently, we calculate its `final' bound mass. In the case of an FTDE, we also visually inspect the surface density plots and assign the mass bound to the star to be zero. 

The mass bound to the BH is calculated in a similar fashion, though the position of the BH is trivially known from the simulation. Finally, we consider any cells with $\varepsilon_{\mathrm{cell,CoM}} \geq 0$ and $\varepsilon_{\mathrm{cell,CoM}} \geq 0$ unbound from the system.

\subsection{Calculation of orbital and spin parameters}

For each snapshot post-disruption, we compute the instantaneous Keplerian orbital parameters -- the semimajor axis $a$ and orbital eccentricity $e$ -- using the current values of the mass bound to the star $\ms$, the mass of the BH (plus the gas mass bound to it) $\mbh$, the positions and velocities of the star's CoM $(\vb{r}_{\mathrm{CoM}}, \vb{v}_{\mathrm{CoM}})$ and the BH $(\vb{r}_{\mathrm{BH}}, \vb{v}_{\mathrm{BH}})$. For completeness, with $\vb{r} \equiv \vb{r}_{\mathrm{CoM}} - \vb{r}_{\mathrm{BH}}$ and $\vb{v} \equiv \vb{v}_{\mathrm{CoM}} - \vb{v}_{\mathrm{BH}}$, the equations are as follows:
\begin{equation}
    \displaystyle a = \left( \frac{2}{r} - \frac{v^2}{G (\ms+\mbh)} \right) ^{-1}
    \label{eq:smaxis}
\end{equation}
\begin{equation}
    \displaystyle e = \left( 1 - \frac{(\vb{r} \times \vb{v})^2}{G (\ms+\mbh) a} \right) ^{0.5}
    \label{eq:ecc}
\end{equation}

We determine the spin angular momentum $\vb{L}_{\mathrm{\star}}$ about the star's CoM using the masses $m_i$, relative positions $\vb{r}_i$ and velocities $\vb{v}_i$ of the bound gas cells. In the cases of full disruption, we ignore the orbital and spin parameters.

The `final' post-disruption mass, orbital, and spin parameters reported in the following section are the means and standard deviations of these quantities during the last ten snapshots of each simulation (the time between consecutive snapshots was chosen to be approximately equal to the initial dynamical timescale of the star). In the cases where a partially disrupted star returns on a second passage within the simulation time, we chose snapshots close to the apoapsis of the first passage for the parameter calculations.

\section{Results} \label{sec:results}

\begin{figure*}
    \centering
    \includegraphics[width=\textwidth]{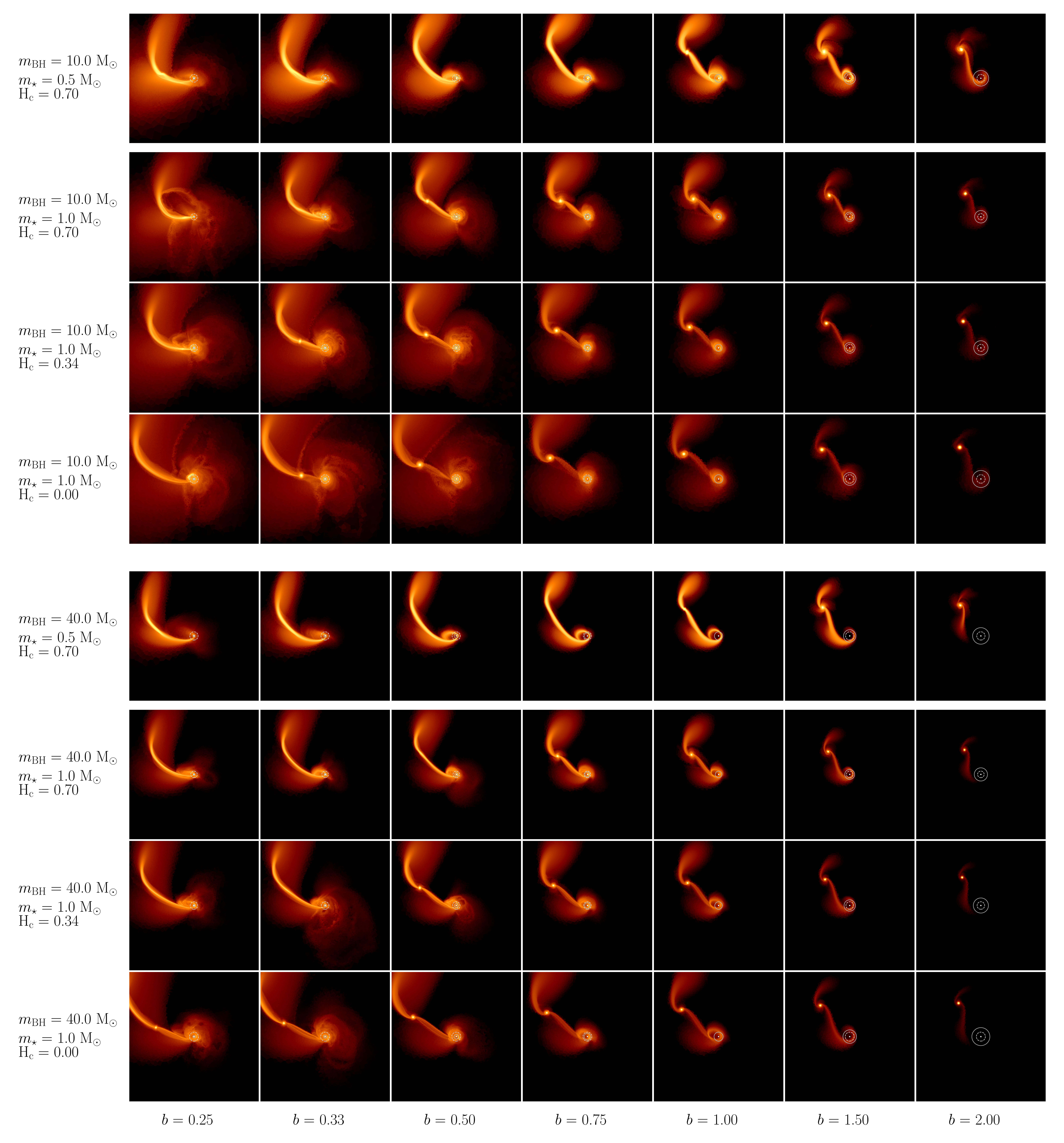}
    \caption{Grid of log density slices of stars $\sim 20\,\tdyn$ after undergoing $\mu$TDEs, with the BH at the center of each panel. Each row is for a different $\mbh$ (top and bottom half for $10 \Msun$ and $40 \Msun$ respectively), $\ms$ or $\Hc$, whereas the columns represent increasing $b$ (from left to right). The dashed and solid circles around the BH, most visible in the rightmost panels, denote the tidal radius, $\rt$, and the periapsis distance, $\rp$, respectively. Note that the spatial range of each panel is not the same.}
    \label{fig:snapshots}
\end{figure*}

Figure \ref{fig:snapshots} shows selected snapshots of our suite of simulations. It can be visually seen that the final outcomes (FTDE or PTDE) depend on all the parameters that we considered -- the stellar mass $\ms$, the core hydrogen fraction $\Hc$, the BH mass $\mbh$, and the impact parameter $b$.

The following sections detail the quantitative results. It should be noted that, henceforth, initial star parameters are given the subscript $\star$ (e.g. $\ms$, $\Ls$), while the post-disruption remnant parameters are given the subscript $\diamond$ (e.g. $\msr$, $\Lsr$). Table \ref{tab:param_values} provides detailed initial and post-disruption parameters for our 58 simulations.

\subsection{Effect of density concentration factor}

\begin{figure}
    \centering
    \includegraphics[width=\columnwidth]{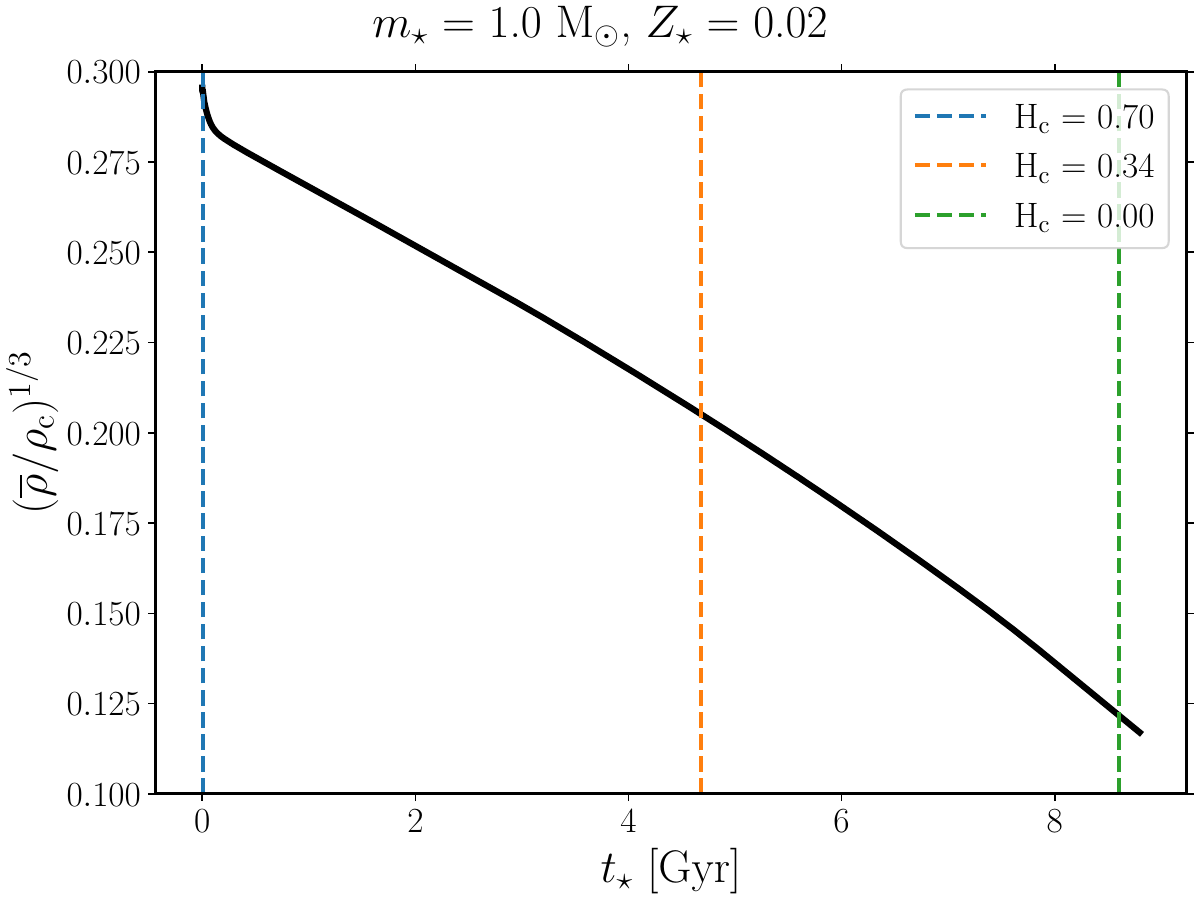}
    \caption{\mesa{}-derived inverses of density concentration factors $\rhoconc^{-1}$ for a $1 \Msun$ MS star as a function of stellar age. The three dashed vertical lines represent the ages of three chosen MS models in our study, corresponding to the profiles in Figure \ref{fig:den_temp_profiles_1M} and the values in Table \ref{tab:star_details}. The dependence with age is very close to linear.}
    \label{fig:concentration_factor_1Msun}
\end{figure}

\begin{figure}
    \centering
    \includegraphics[width=\columnwidth]{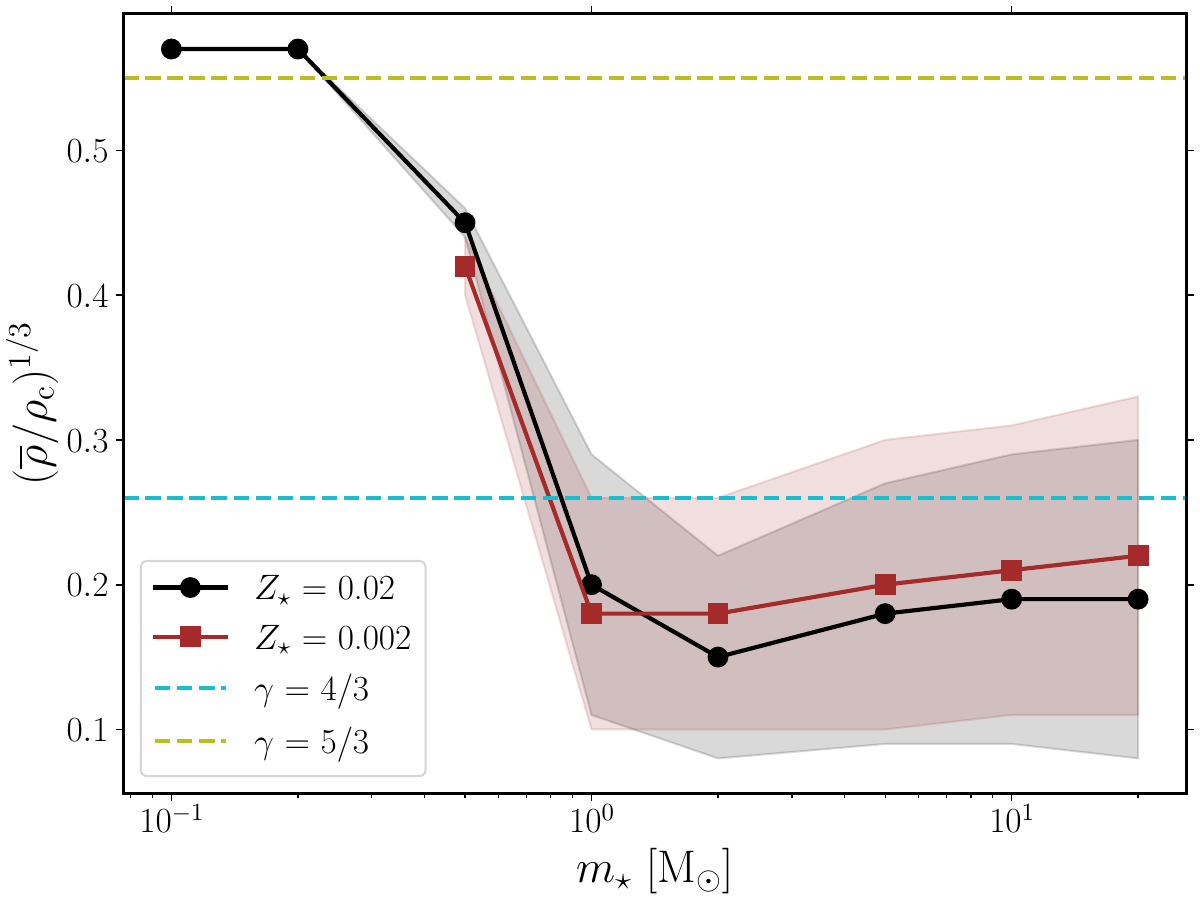}
    \caption{\mesa{}-derived inverses of density mean concentration factors $\rhoconc^{-1}$ for MS stars of a range of masses (0.1, 0.2, 0.5, 1.0, 2.0, 5.0, 10.0 and 20.0 in units of $\Msun$) and two metallicities -- $0.02$ and $0.002$. The shaded regions illustrate the variation in $\rhoconc^{-1}$ over the MS lifetime, with a TAMS star having the lowest $\rhoconc^{-1}$ value and a ZAMS star having the highest ( the values are for ZAMS stars only in the cases of stars of masses $< 1 \Msun$). The two dashed horizontal lines represent the corresponding values for polytropic stars, shown in Table \ref{tab:polytrope_stars}.}
    \label{fig:concentration_factor_mass}
\end{figure}

\begin{table}
    \caption{Evaluated values of the impact parameter for the full disruption of a star with uniform density, $\tilde{b}_{\mathrm{FTDE}} = b_{\mathrm{FTDE}}/\rhoconc$.}
    \label{tab:polytrope_stars}
    \centering
    \begin{tabular}{c c c c}
    \hline
    $\gamma$ & $b_{\mathrm{FTDE}}$ & $\rhoconc^{-1}$ & $\tilde{b}_{\mathrm{FTDE}}$ \TBstrut \\ 
    \hline
    4/3 & 0.50 & 0.26 & 1.92 \Tstrut \\
    5/3 & 1.08 & 0.55 & 1.96  \Bstrut \\
    \hline
    \end{tabular}
    \tablefoot{The columns indicate the polytropic index $\gamma$, the impact parameter for full disruption $b_{\mathrm{FTDE}}$, the inverse of the density concentration factor $\rhoconc^{-1}$ and the aforementioned parameter $\tilde{b}_{\mathrm{FTDE}}$. The $b_{\mathrm{FTDE}}$ values for polytropic stars are from \citet{2017A&A...600A.124M}.}
\end{table}

\citet{2020ApJ...904...99R} showed that, in the case of SMBHs, the critical impact parameter for full disruption, $b_{\mathrm{FTDE}}$, is inversely proportional to the density concentration factor $\rhoconc \equiv \rhoconcexp$, where $\overline{\rho}$ and $\rho_{\mathrm{c}}$ are average and central stellar density respectively (see also \citealp{2020ApJ...905..141L}). A larger value of $\rhoconc$ represents a denser core with a puffier outer layer, and vice-versa. This indicates that a star with higher core density requires a smaller impact parameter to fully disrupt. Quantitatively, $b_{\mathrm{FTDE}}$ can be written as:
\begin{equation}
    b_{\mathrm{FTDE}} = \rhoconc^{-1} \  \tilde{b}_{\mathrm{FTDE}}
    \label{eq:FTDE_b}
\end{equation}

Here, $\tilde{b}_{\mathrm{FTDE}}$ is the critical impact parameter for the full disruption of a hypothetical star with uniform density, i.e., $\rhoconc = 1$ (see also Equation 15 of \citealp{2020ApJ...904...99R}). Table \ref{tab:polytrope_stars} displays the values of $b_{\mathrm{FTDE}}$ of stars with polytropic indices $\gamma$ (where $P \propto \rho^{\gamma}$) 4/3 and 5/3 as calculated by \citet{2017A&A...600A.124M}. Shown also are the analytically computed $\rhoconc$ for these polytropes, and the subsequently evaluated (using Equation \ref{eq:FTDE_b}) $\tilde{b}_{\mathrm{FTDE}}$ values. We also estimate $\tilde{b}_{\mathrm{FTDE}}$ to be $1.95 \pm 0.20$ from our suite of simulations by fitting the remnant masses for given initial stellar and BH masses (see the following section for details), and find the values of $b_{\mathrm{FTDE}}$. Despite the significant difference in the masses of the black holes, our estimated values agree with those from \citet{2017A&A...600A.124M}. Estimates of this factor calculated by \citet{2020ApJ...904...99R}, \citealp{2020ApJ...905..141L} and \citet{2023ApJ...948...89K} all lie within the error range. Since the former two studies are on SMBHs and the latter is on IMBHs, we reiterate that $\tilde{b}_{\mathrm{FTDE}}$ is almost independent of the mass of the BH for a very large range of BH masses. The consensus between the previous works in spite of different numerical schemes (general relativistic grid-based hydrodynamics, adaptive-mesh refinement, and SPH respectively) and the current work strengthens this point. Therefore, we will use our computed value of $\tilde{b}_{\mathrm{FTDE}} = 1.95 \pm 0.20$ for the fit functions in the following sections.

For completeness, Figure \ref{fig:concentration_factor_1Msun} (and Table \ref{tab:star_details}) shows the trend $\rhoconc^{-1}$ for \mesa{} $1 \Msun$ MS stars, with our three stellar models highlighted. We see that a more evolved star has a smaller $\rhoconc^{-1}$ (harder to fully disrupt) compared to a less evolved star (easier to fully disrupt) and that this relation is close to linear as a function of stellar age $t_{\star}$. Figure \ref{fig:concentration_factor_mass} shows the averaged $\rhoconc^{-1}$ values for a range of masses of \mesa{}-generated MS stars ($0.1 \Msun$ -- $20.0 \Msun$) of two metallicities -- $0.02$ (near-Solar) and $0.002$ (sub-Solar). We see that higher mass MS stars have denser cores and puffier outer layers (hence lower $\rhoconc^{-1}$ values, which plateau for stars of masses $\gtrsim 1 \Msun$ at $\sim 1.6$--$2.0$), while the opposite is true for lower mass MS stars. Moreover, metallicity affects $\rhoconc^{-1}$ only slightly, with the metal-poor stars being slightly less (more) centrally concentrated when $\ms > 1.0 \Msun$ ($\ms \leq 1.0 \Msun$). The (approximate) values of $\rhoconc^{-1}$ are crucial to applying our fits (see the following sections).

\subsection{Post-disruption stellar and BH masses}

\begin{figure*}
    \begin{subfigure}{\columnwidth}
        \includegraphics[width=\columnwidth]{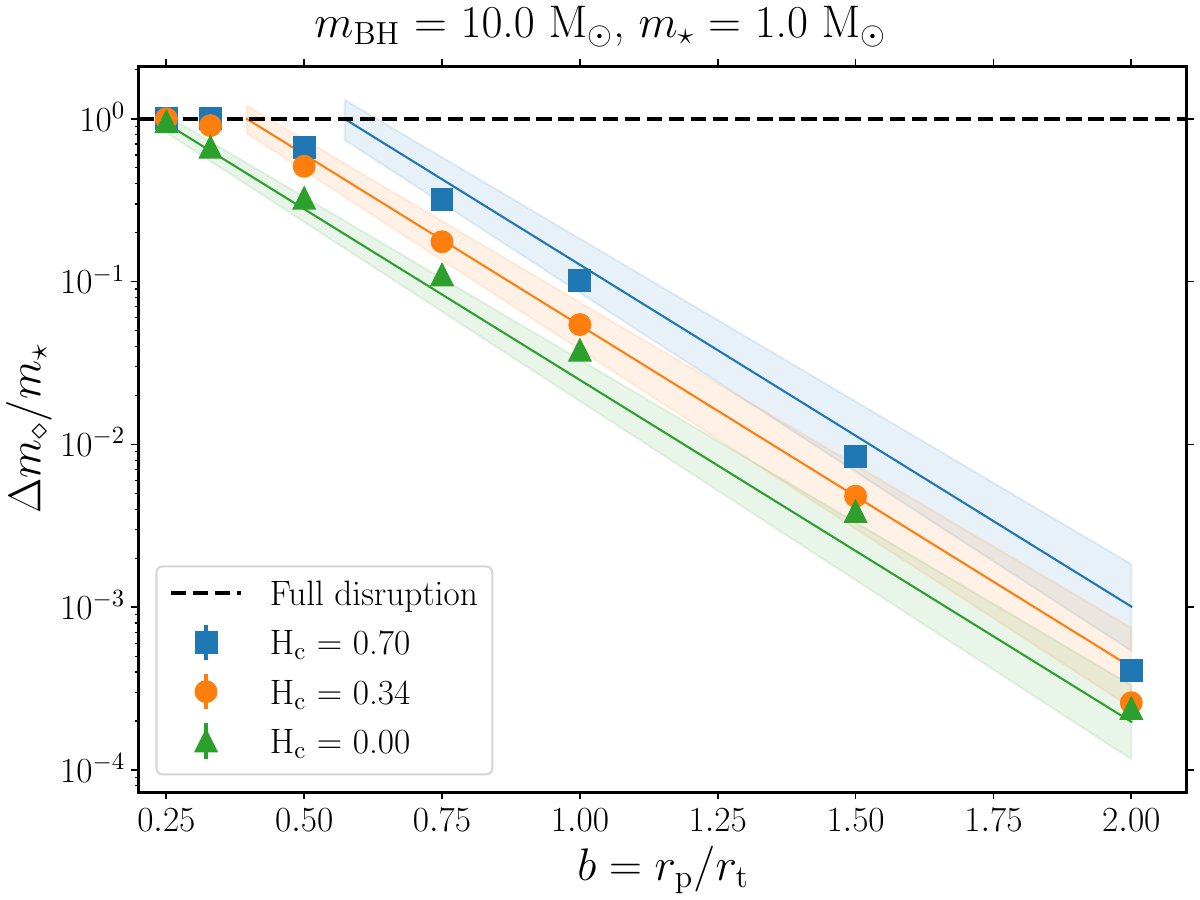}
        \caption{}
        \label{fig:mstar_trends_ms1mbh10}
    \end{subfigure}
    \hfill
    \begin{subfigure}{\columnwidth}
        \includegraphics[width=\columnwidth]{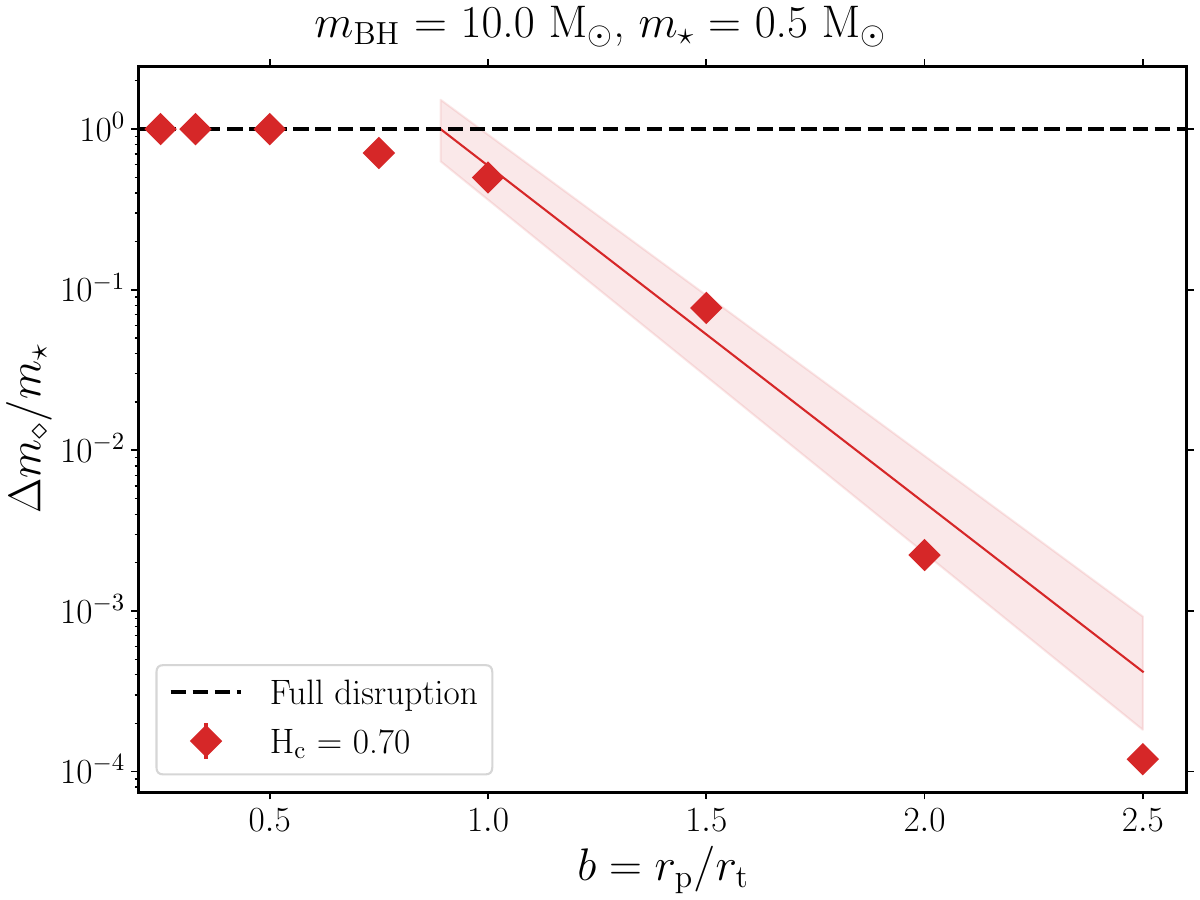}
        \caption{}
        \label{fig:mstar_trends_msp5mbh10}
    \end{subfigure}
    \begin{subfigure}{\columnwidth}
        \includegraphics[width=\columnwidth]{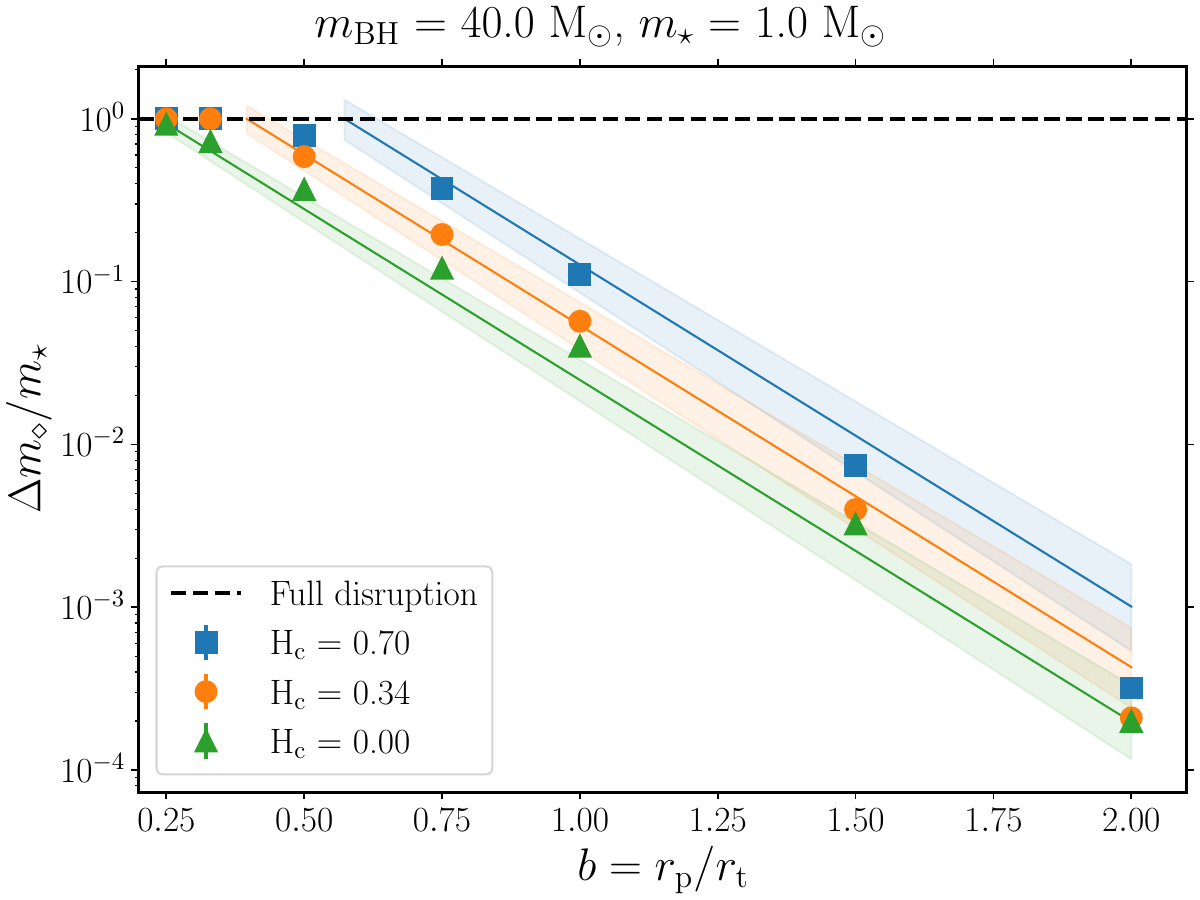}
        \caption{}
        \label{fig:mstar_trends_ms1mbh40}
    \end{subfigure}
    \hfill
    \begin{subfigure}{\columnwidth}
        \includegraphics[width=\columnwidth]{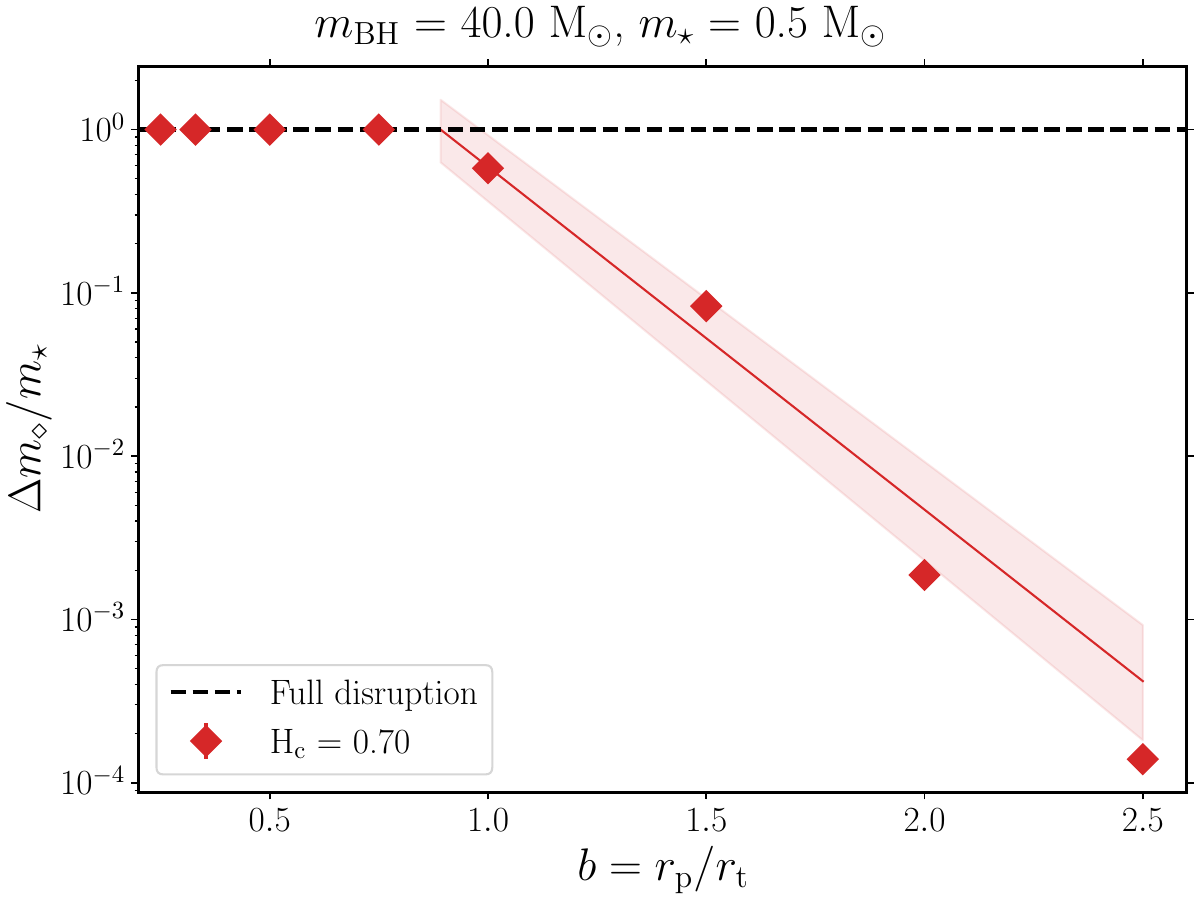}
        \caption{}
        \label{fig:mstar_trends_msp5mbh40}
    \end{subfigure}
    \centering
    \caption{Post-disruption fractional mass loss, $\Dmsr/\ms$, for a star of mass $1 \Msun$ (left) and $0.5 \Msun$  (right), due to a BH of mass $10 \Msun$ (top) and $40 \Msun$ (bottom), as a function of impact parameter $b$ and MS age ($\Hc$ abundance). The dashed line at the top signifies the full disruption of the star. The fractional mass losses are largely independent of BH mass. Mass loss is higher for lower $b$ values, and this decreases  roughly exponentially (best-fit lines and standard error regions shown) with increasing $b$. A less evolved star (higher $\Hc$) loses more mass for the same value of $b$. A $0.5 \Msun$ star has a higher mass loss than a $1 \Msun$ star for the same $b$.}
    \label{fig:mstar_trends_all}
\end{figure*}

After a disruption event, a star loses mass, some of which gets bound to (and eventually accreted onto) the BH while the rest of it becomes unbound from the system. Evidently, the closer the approach of the star is to the BH, the larger the mass lost from the star. 

Figure \ref{fig:mstar_trends_all} shows this trend for the case of a $1 \Msun$ (left) and a $0.5 \Msun$ (right) star that is disrupted by a $10 \Msun$ (top) and a $40 \Msun$ (bottom) BH. The post-disruption fractional mass loss $\Dmsr/\ms$, where $\Delta m_{\mathrm{\star}}$ is the difference between the star's initial mass $\ms$ and the remnant mass $\msr$. It decreases almost exponentially with the impact parameter $b$. Moreover, a more evolved MS star loses less mass compared to a less evolved star for the same value of $b$. This is because a more evolved star, with a higher central concentration, requires a stronger tidal force (or smaller $b$) to strip the same amount of mass when compared to a less evolved star. Consequently, $b_{\mathrm{FTDE}}$, corresponding to $\Dmsr/\ms = 1$, is also lower for more evolved stars.

Furthermore, we see that roughly half the mass lost from the star stays bound to the BH,  although this is not always true, especially when there is little mass loss (see Table \ref{tab:param_values}). This bound mass decreases through time due to continuous debris interaction resulting in mass unbinding, and hence is an upper limit for the mass that can be accreted onto the BH. Details of the accretion process and feedback from the BH might prevent some of the mass bound to the BH from being accreted, but following this process is beyond the scope of our paper. The other half of the stripped mass is unbound from the BH-star system.

We can fit  (with standard $1 \sigma$ errors) the fractional mass loss $\Delta m_{\mathrm{\star}}$ well by:
\begin{equation}
    \log_{10}\left( \frac{\Dmsr}{\ms} \right) = \min{\{(-2.10 \mbox{\tiny $\ \pm\ 0.10$})\ (b - b_{\mathrm{FTDE}})\ ,\ 0\}}
    \label{eq:mass_fit}
\end{equation}

Here, $b_{\mathrm{FTDE}}$ is as given in Equation \ref{eq:FTDE_b}. Since $\rhoconc$ is taken into account, this fit differs from those obtained by \citet{2013ApJ...767...25G} and \citet{2020ApJ...904..100R} for TDEs due to SMBHs. The remnant mass fit function does not explicitly depend on $\ms$ and $\mbh$, but only on $b$ and $\rhoconc$. Since $\rhoconc$ is higher for more evolved stars, they have a smaller $b_{\mathrm{FTDE}}$ and hence, less fractional mass loss. Figure \ref{fig:mstar_trends_all} plots these best-fit curves, with the error bars shaded.  It should be noted that this trend may not hold for larger values of $b$.

\subsection{Post-disruption stellar spins}

\begin{figure*}
    \begin{subfigure}{\columnwidth}
        \includegraphics[width=\columnwidth]{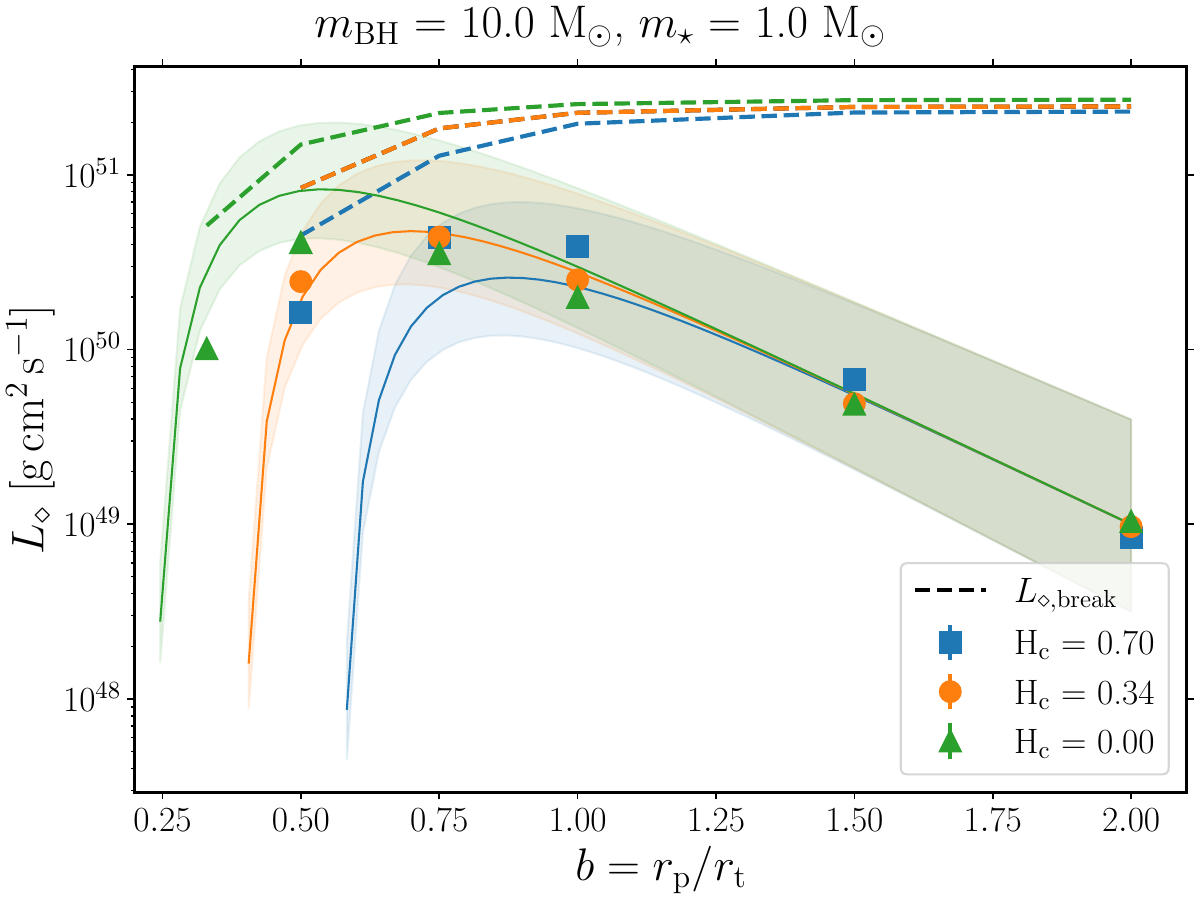}
        \caption{}
        \label{fig:Lstar_trends_ms1mbh10}
    \end{subfigure}
    \hfill
    \begin{subfigure}{\columnwidth}
        \includegraphics[width=\columnwidth]{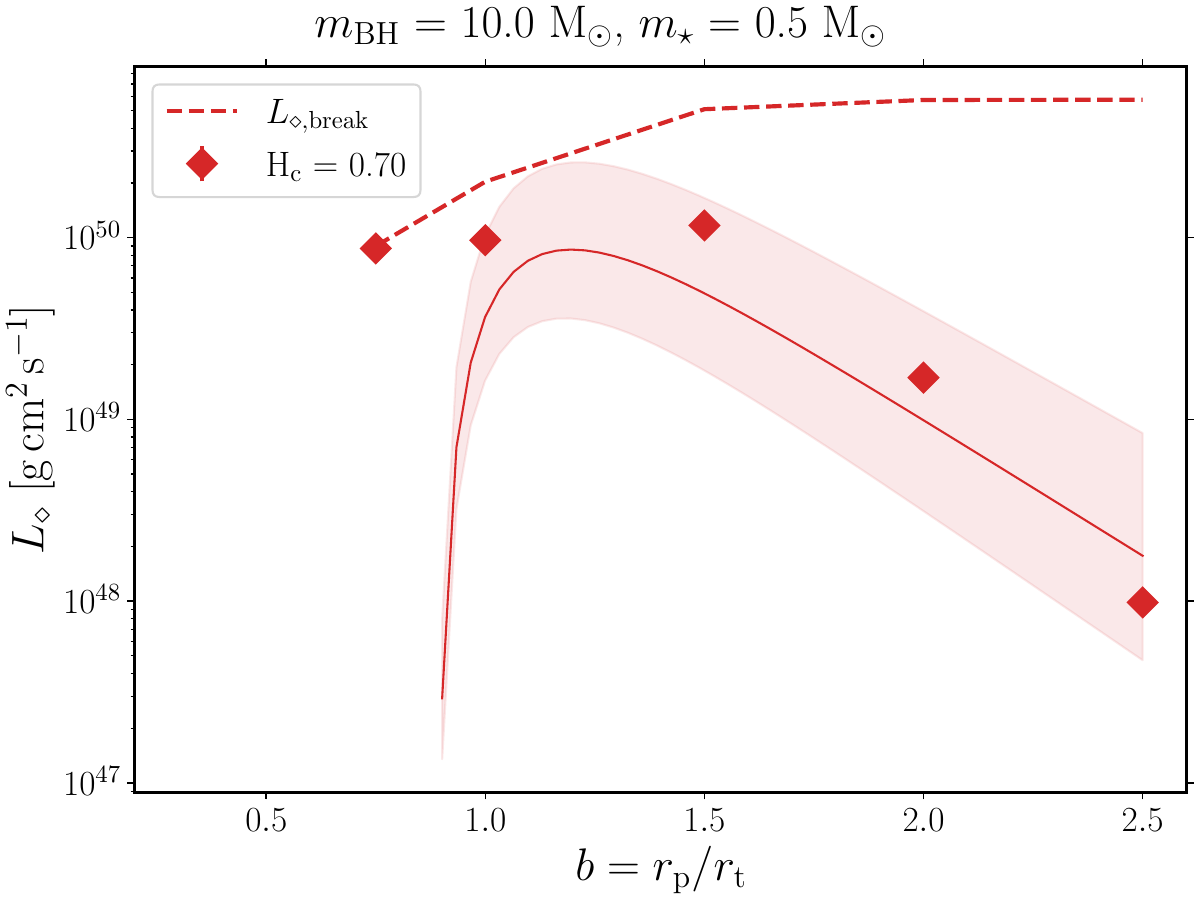}
        \caption{}
        \label{fig:Lstar_trends_msp5mbh10}
    \end{subfigure}
    \begin{subfigure}{\columnwidth}
        \includegraphics[width=\columnwidth]{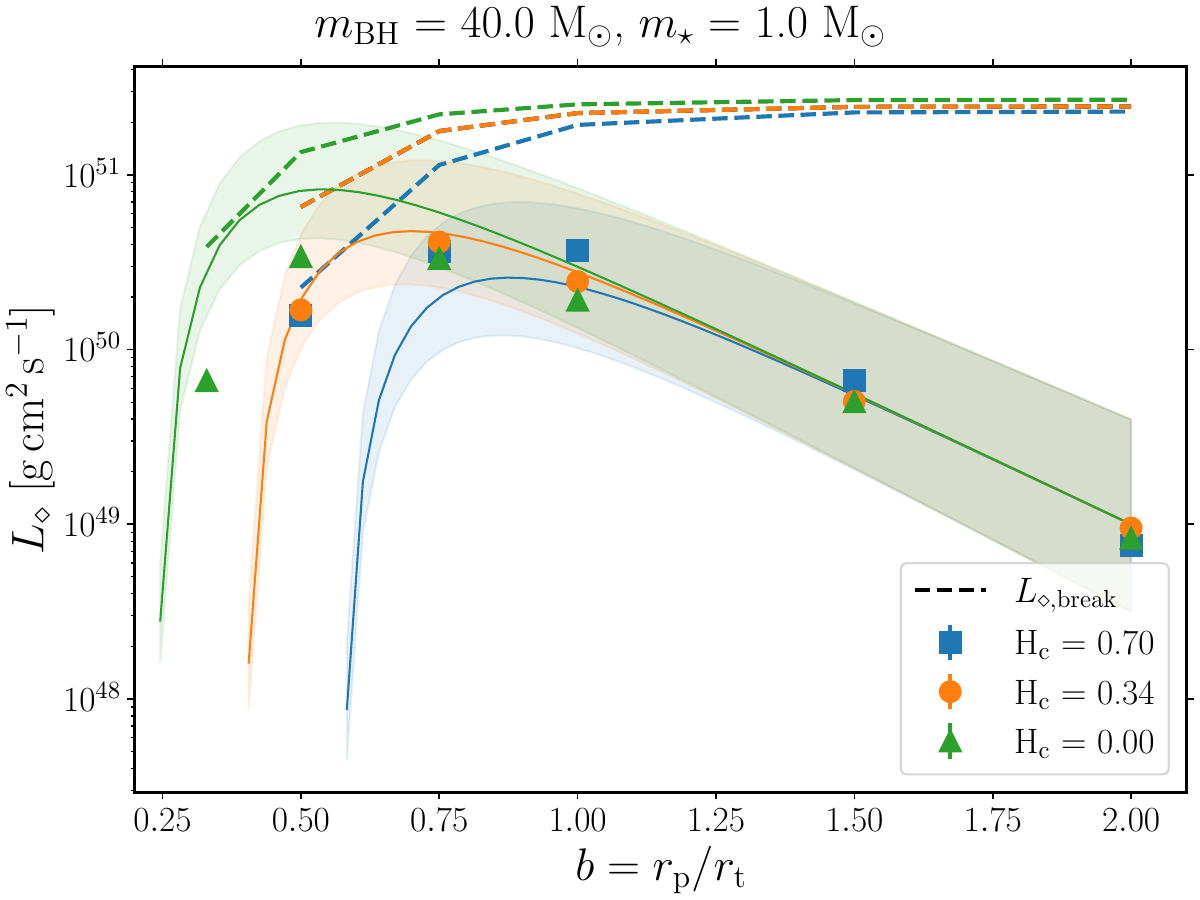}
        \caption{}
        \label{fig:Lstar_trends_ms1mbh40}
    \end{subfigure}
    \hfill
    \begin{subfigure}{\columnwidth}
        \includegraphics[width=\columnwidth]{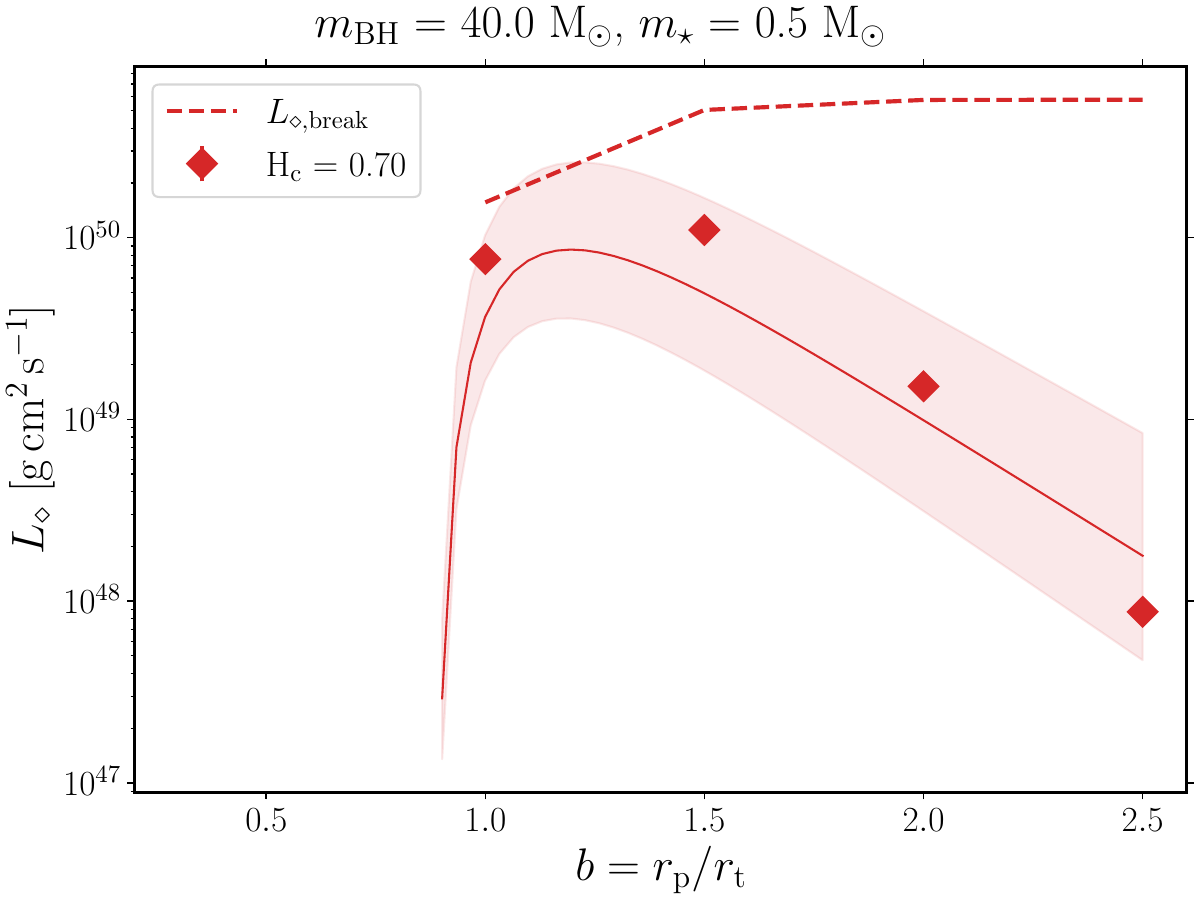}
        \caption{}
        \label{fig:Lstar_trends_msp5mbh40}
    \end{subfigure}
    \centering
    \caption{Post-disruption spin angular momentum, $\Lsr$, for an initially non-rotating star of mass $1 \Msun$ (left) and $0.5 \Msun$ (right), due to a BH of mass $10 \Msun$ (top) and $40 \Msun$  (bottom), as a function of impact parameter $b$ and MS age ($\Hc$ abundance). The dashed lines indicate the (approximate) estimated break-up angular momenta of the remnants $L_{\mathrm{\diamond,break}}$ respectively. The spin angular momenta are largely independent of BH mass. It decreases  almost exponentially for sufficiently high $b$ values, and it drops for very low $b$ values when the remnant mass is small (best-fit lines and standard error regions shown). The drop-off of angular momentum occurs at larger $b$ for a less evolved star.}
    \label{fig:Lstar_trends_all}
\end{figure*}

An initially non-rotating star gains spin after a tidal encounter due to tidal torques from the BH.  Although we do not delve into the details of the spin-up, this results in \textit{differential}, and not rigid, rotation of the star.  It should be noted that, due to spurious motions in our initial \arepo{} stellar models, the spin angular momenta $\Ls$ are not zero but $\sim 1$--$2 \times 10^{47} \gcmsqsinv$ ($\sim 2 \times 10^{46} \gcmsqsinv$) for $1 \Msun$ ($0.5 \Msun$) stars. Since these values correspond to negligible equatorial surface velocities\footnote{In contrast, the Sun, a very slow rotator, has an equatorial surface velocity of $\sim 2 \kmsinv$.} of $\sim 10 \msinv$, the stars are considered to be non-rotating. The spin angular momentum of the remnant $\Lsr$ depends on the internal structure of the original star and the impact parameter.

Figure \ref{fig:Lstar_trends_all} shows the trend of $\Lsr$ with respect to the impact parameter $b$ for the case of a $1 \Msun$ (left) and a $0.5 \Msun$ (right) star being \textit{partially} disrupted by a $10 \Msun$ (top) and a $40 \Msun$ (bottom) BH.  In addition, we indicate the  approximate, order-of-magnitude estimates of the break-up angular momenta  (assuming uniform density and rigid rotation) of the remnant stars,  $L_{\mathrm{\diamond,break}} = 0.4\ \msr r_{\mathrm{\diamond}} v_{\mathrm{\diamond,break}}$ \footnote{This is because hypothetical star (sphere) of uniform density has a moment of inertia of $I_{\mathrm{\star,uniform}} = 0.4\ \ms\rs^2$ about an axis of rotation passing through the center.}. Here,  we crudely estimate the remnant radius as $(r_{\mathrm{\diamond}}/\Rsun) = (\msr/\Msun)^{0.7}$ and assume the break-up velocity to be the Keplerian velocity at $\rs$, i.e., $v_{\mathrm{\diamond,break}} = \sqrt{G \msr / r_{\mathrm{\diamond}}}$.  In reality, the remnant is significantly puffed up, and the break-up velocity would depend on the true moment of inertia of the remnant (and thus, its density profile) and the differential rotation due to spin-up. For our range of simulation parameters, $\Lsr$ remains below  $L_{\mathrm{\diamond,break}}$, although significant mass loss can bring the remnant very close to break-up,  e.g., $\ms = 0.5 \Msun$ and $b \lesssim 1.00$. For higher values of $b$ (but still less than $b \sim 2$), $\Lsr$ decreases  almost exponentially as $b$ increases. On the other hand, when $b$ is less such that a fractional mass loss is $\geq 2 \times 10^{-1}$, the trend reverses until FTDE. This trend reflects the counter-balance between mass loss and spin-up: for stronger PTDEs, although the remnant is spun up more rapidly, the remnant mass is smaller.

The large spin-up for smaller $b$ indicates that the specific angular momentum monotonically increases as $b$ decreases. Motivated by this, we considered the \textit{scaled} (moment of inertia-adjusted) parameter $L_{\mathrm{\diamond,scaled}} \equiv \Lsr/(m_{\mathrm{\diamond,frac}}\ \tilde{r}_{\mathrm{\diamond,frac}}^{2})$. Here, the fractional remnant mass $m_{\mathrm{\diamond,frac}} \equiv \msr/\ms$ and the post-disruption estimated fractional radius $\tilde{r}_{\mathrm{\diamond,frac}}/\Rsun = m_{\mathrm{\diamond,frac}}^{0.7}$ is a crude estimate of the radius of an MS star.

We found the following exponential fit (with standard $1 \sigma$ errors) describes $L_{\mathrm{\diamond,scaled}}$ well:
\begin{equation}
    \log_{10}\left( \frac{L_{\mathrm{\diamond,scaled}}}{\gcmsqsinv} \right) = (-1.50 \mbox{\tiny $\ \pm\ 0.15$})\ b + (52.00 \mbox{\tiny $\ \pm\ 0.30$})
    \label{eq:spin_scaled}
\end{equation}

From the definition of $\tilde{r}_{\mathrm{\star,frac}}$ before, we have:
\begin{equation}
    \Lsr = L_{\mathrm{\diamond,scaled}} \left(\frac{\msr}{\ms}\right)^{2.4} = L_{\mathrm{\diamond,scaled}} \left(1 - \frac{\Dmsr}{\ms}\right)^{2.4}
    \label{eq:spin_fit}
\end{equation}

Here, $(\Dmsr/\ms)$ is estimated using Equation \ref{eq:mass_fit}. This equation implies that the post-disruption spin depends significantly on the fractional mass loss. Moreover, there is no dependence on $\mbh$. Figure \ref{fig:Lstar_trends_all} plots these best-fit curves, with the error bars shaded.

\subsection{Post-disruption orbital parameters}

\begin{figure*}
    \begin{subfigure}{\columnwidth}
        \includegraphics[width=\columnwidth]{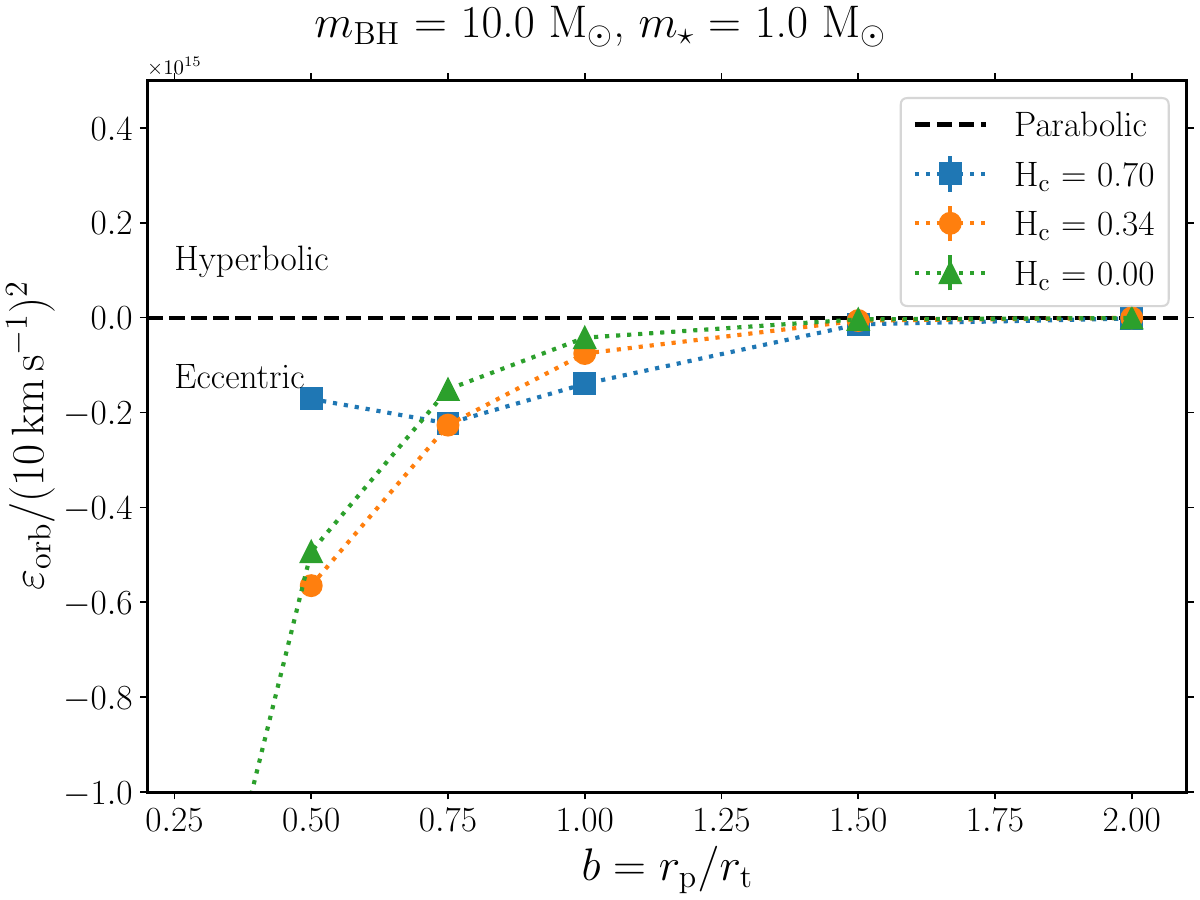}
        \caption{}
        \label{fig:Eorb_trends_ms1mbh10}
    \end{subfigure}
    \hfill
    \begin{subfigure}{\columnwidth}
        \includegraphics[width=\columnwidth]{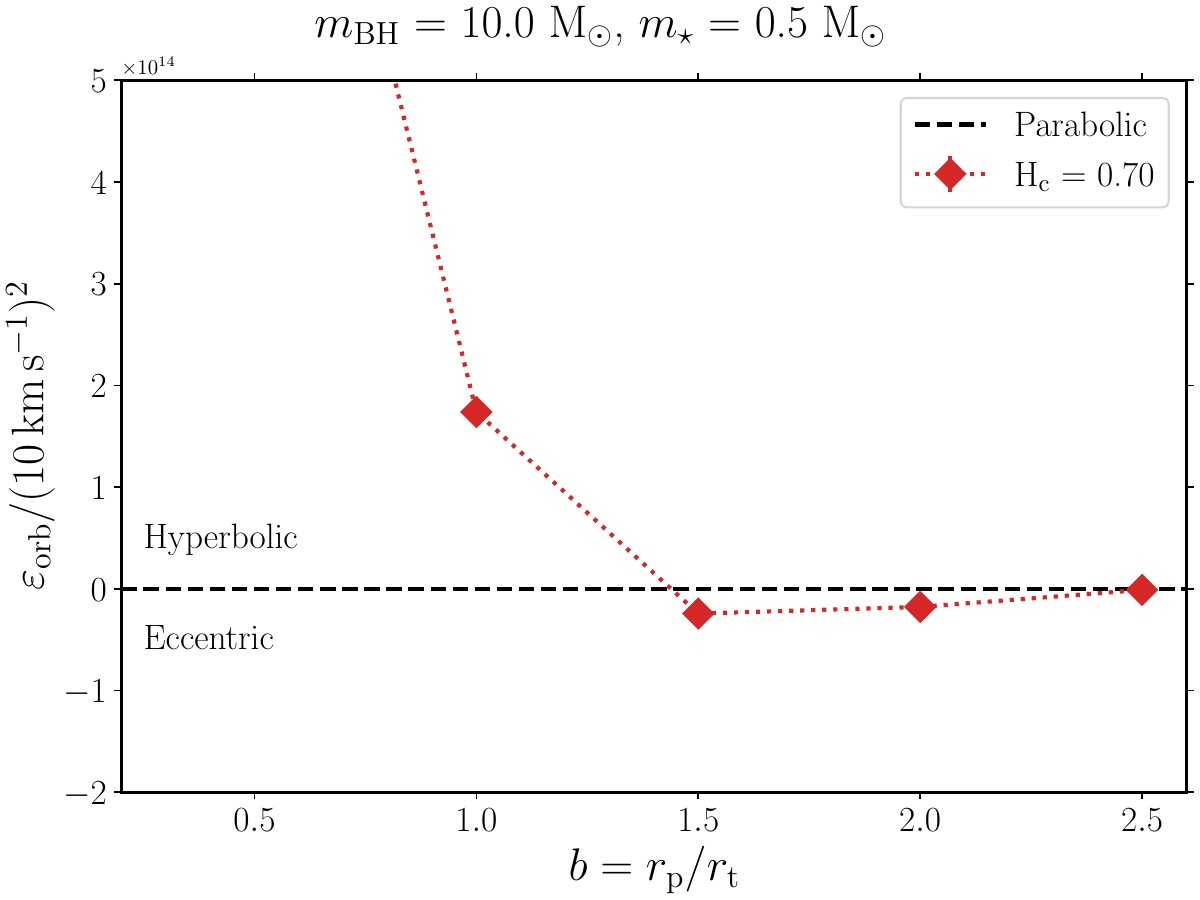}
        \caption{}
        \label{fig:Eorb_trends_msp5mbh10}
    \end{subfigure}
    \begin{subfigure}{\columnwidth}
        \includegraphics[width=\columnwidth]{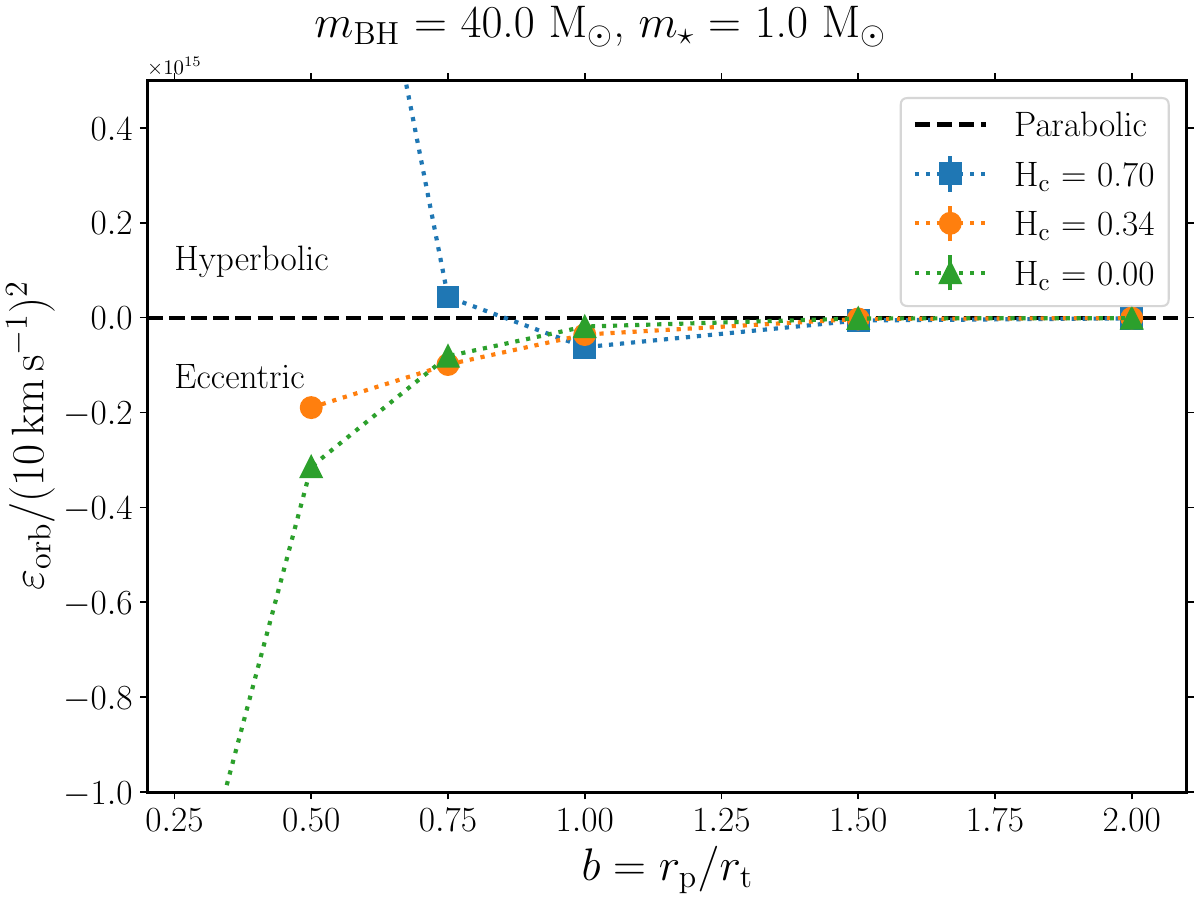}
        \caption{}
        \label{fig:Eorb_trends_ms1mbh40}
    \end{subfigure}
    \hfill
    \begin{subfigure}{\columnwidth}
        \includegraphics[width=\columnwidth]{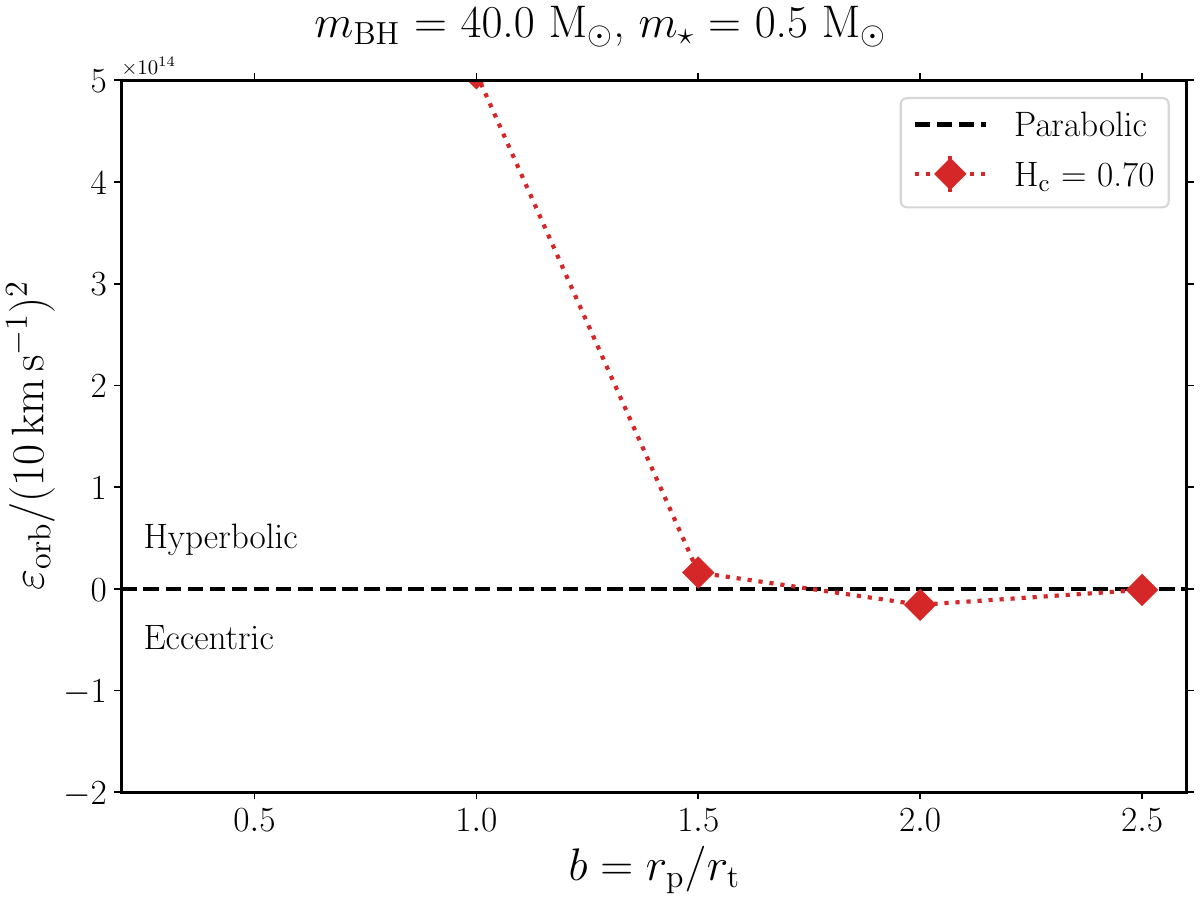}
        \caption{}
        \label{fig:Eorb_trends_msp5mbh40}
    \end{subfigure}
    \centering
    \caption{Post-disruption normalized specific orbital energies of the star-BH system, $\Eorb$, for a star of mass $1 \Msun$ (left) and $0.5 \Msun$ (right) initially in a parabolic orbit, due to a BH of mass $10 \Msun$ (top) and $40 \Msun$  (bottom), as a function of impact parameter $b$ and MS age ($\Hc$ abundance). The dashed line represents parabolic orbits and separates the parameter space into bound (eccentric) and unbound (hyperbolic) orbits. When mass loss is relatively low, a lower value of $b$ generally results in more negative $\Eorb$ in the case of $1 \Msun$ star. This trend continues for the TAMS star. However, for the ZAMS and MAMS stars, significant mass losses can reverse this trend, with hyperbolic orbits also being a possibility (e.g., $1 \Msun$ ZAMS star and $40 \Msun$ BH mass with $b < 1.00$). $0.5 \Msun$ stars, on the other hand, can become unbound for larger $b$, with boundedness also depending on the mass of the BH.}
    \label{fig:Eorb_trends_all}
\end{figure*}

\begin{figure*}
    \begin{subfigure}{\columnwidth}
        \includegraphics[width=\columnwidth]{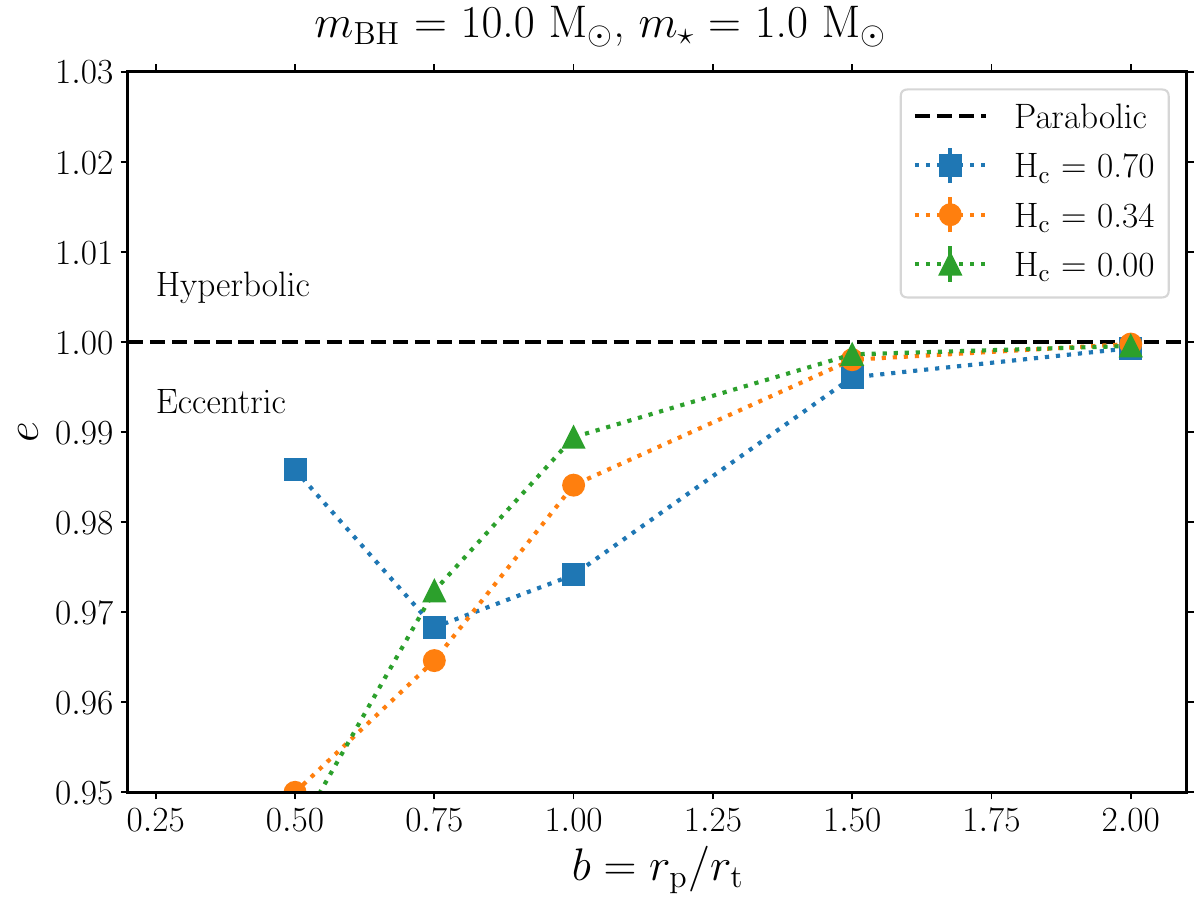}
        \caption{}
        \label{fig:ecc_trends_ms1mbh10}
    \end{subfigure}
    \hfill
    \begin{subfigure}{\columnwidth}
        \includegraphics[width=\columnwidth]{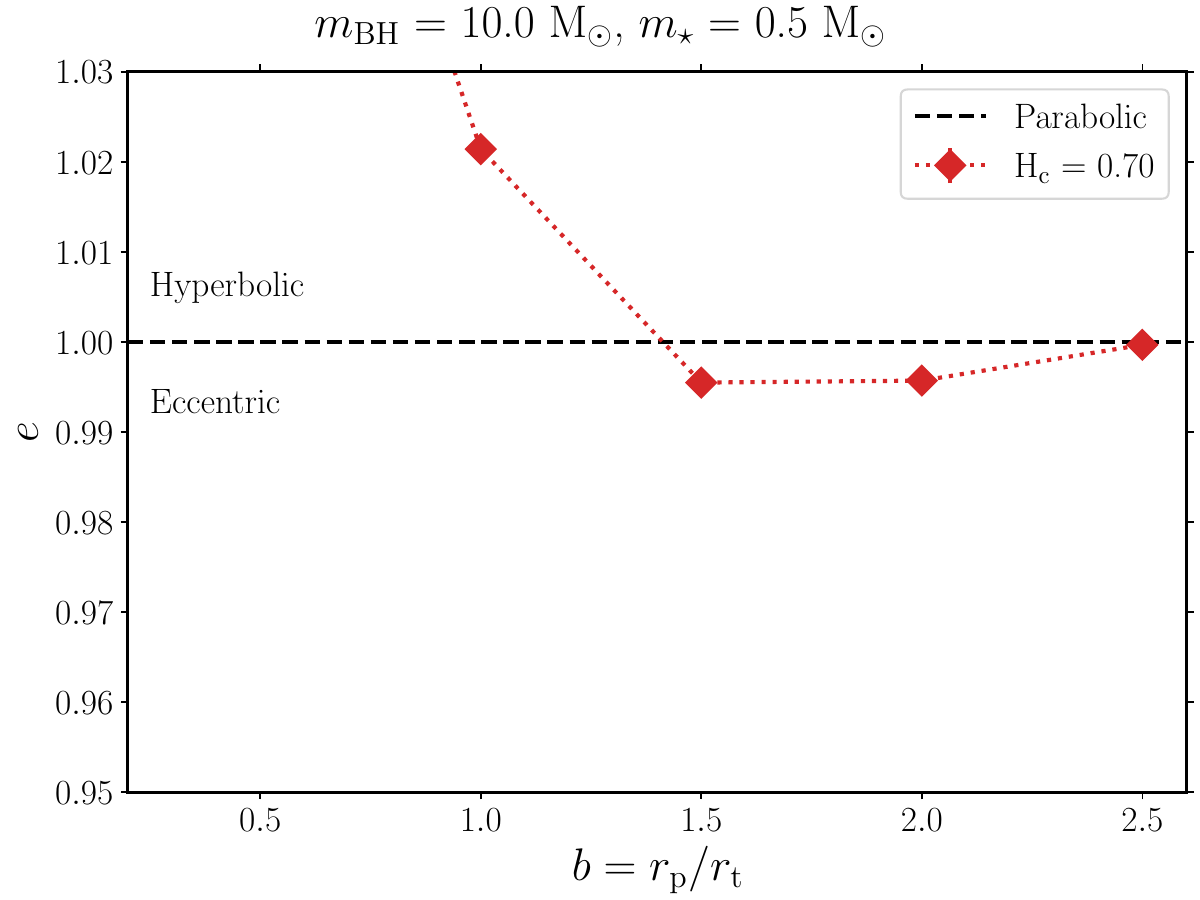}
        \caption{}
        \label{fig:ecc_trends_msp5mbh10}
    \end{subfigure}
    \begin{subfigure}{\columnwidth}
        \includegraphics[width=\columnwidth]{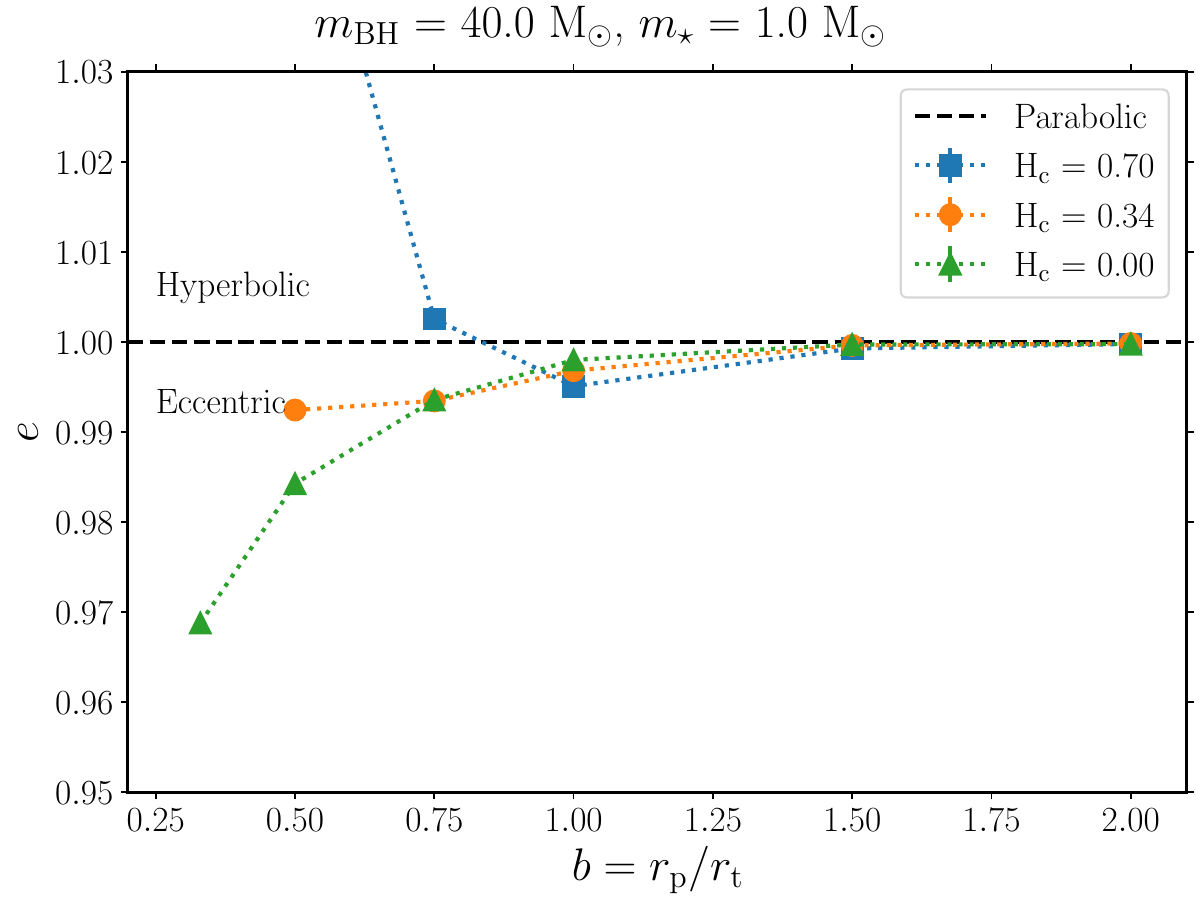}
        \caption{}
        \label{fig:ecc_trends_ms1mbh40}
    \end{subfigure}
    \hfill
    \begin{subfigure}{\columnwidth}
        \includegraphics[width=\columnwidth]{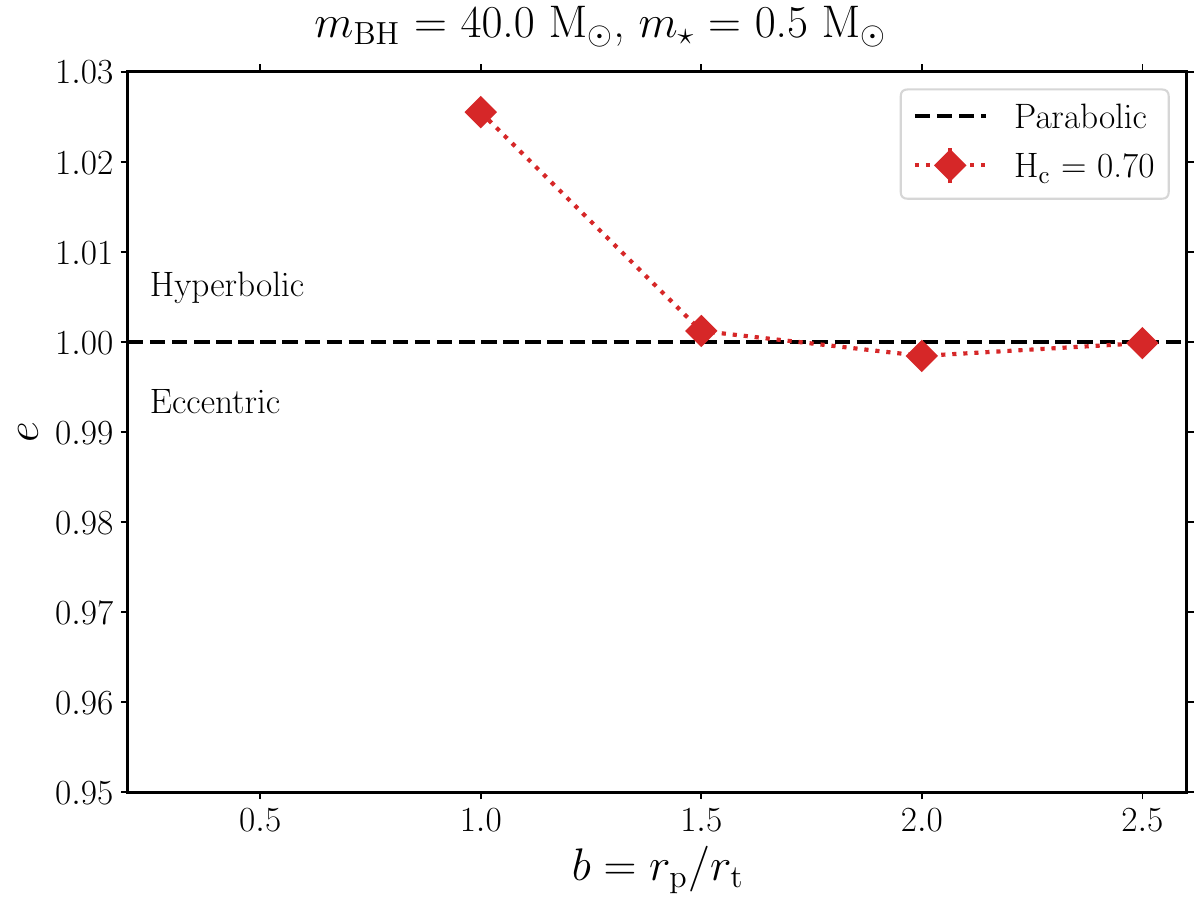}
        \caption{}
        \label{fig:ecc_trends_msp5mbh40}
    \end{subfigure}
    \centering
    \caption{Post-disruption orbital eccentricities, $e$, for a star of mass $1 \Msun$ (left) and $0.5 \Msun$ (right) initially in a parabolic orbit, due to a BH of mass $10 \Msun$ (top) and $40 \Msun$  (bottom), as a function of impact parameter $b$ and MS age ($\Hc$ abundance). The dashed line represents parabolic orbits and separates the parameter space into bound (eccentric) and unbound (hyperbolic) orbits. The reasoning for the trends is the same as that for Figure \ref{fig:Eorb_trends_all}.}
    \label{fig:ecc_trends_all}
\end{figure*}

Depending on the stellar structure and the BH mass, a star in an initially parabolic orbit ends up in an eccentric or even a hyperbolic orbit after a PTDE (see also \citealp{2020ApJ...904..100R,2022ApJ...933..203K,2023ApJ...948...89K}). The specific orbital energy of the star-BH system, $\Eorb$, is the sum of the specific energy injected into stellar tides, $\Etide$, the specific binding energy of the ejected material, $\Ebind$, and the specific `kick' energy, $\Ekick$  (see \citealp{2022ApJ...933..203K}). \citet{2013ApJ...771L..28M} estimated the potentially unbinding `kick' received by a star undergoing PTDE due to an SMBH from asymmetric mass loss and observed that it depends solely on the mass loss and not on the star-SMBH mass ratio. However, this is not true for $\mu$TDEs -- the mass ratio determines the kick, and hence the boundedness, of the star-SBH system (see also Figure 2 of \citealp{2023ApJ...948...89K} for TDEs due to IMBHs).

Figure \ref{fig:Eorb_trends_all} shows the post-disruption specific orbital energy of the star-BH system, $\Eorb$, for the disruption of a $1 \Msun$ and a $0.5 \Msun$ star by a $10 \Msun$ (top panel) and a $40 \Msun$ (bottom panel) BH. The values are normalized to the square of $\sigma = 10 \kmsinv$, the typical velocity dispersion of Milky Way globular clusters (e.g., \citealp{2018MNRAS.478.1520B}). More negative (positive) values represent more bound (unbound) orbits. In general, $\Eorb$ becomes more negative with decreasing $b$. If the density concentration factor $\rhoconc$ is high enough (as in the case of a TAMS $1 \Msun$ star; see Figure \ref{fig:concentration_factor_1Msun}), this trend continues till FTDE. However, in the case of a ZAMS $1 \Msun$ star (and less extremely in a MAMS star), there is a certain value of $b$ at which $\Eorb$ starts to increase with decreasing $b$. This is due to the momentum kick from asymmetric mass loss, which becomes significant when the mass loss is high. In the case of a PTDE of a $1 \Msun$ ZAMS star by a $40 \Msun$ BH, the post-disruption orbit can be hyperbolic (unbound) for $b$ values close to FTDE. This trend is more extreme for a $0.5 \Msun$ star, with hyperbolic orbits being possible even in the case of PTDE by a $10 \Msun$ BH. Moreover, their velocities at infinity are high enough ($\sim 100$--$300 \kmsinv$) to escape the cluster entirely. The two different BH masses result in quite different quantitative orbital parameters.   Similar to \citealp{2022ApJ...933..203K}, we can write each of the energy components as follows:

\begin{align}
   \Etide &= - \frac{G \mbh^2}{\rs \msr} \left[ \left(\frac{\rs}{\rp}\right)^6 T_2 + \left(\frac{\rs}{\rp}\right)^8 T_3 \right] \\
    \Ebind &= - \frac{G \Dmsr}{2 \rs} \\
    \Ekick &= \frac{G \mbh \Dmsr \rs^2}{\rp^3 \msr}
\end{align}

Here, $T_2$ and $T_3$ are terms which depend on the stellar mass, radius, and density structure, BH mass, and distance of approach \citep{1977ApJ...213..183P,1986ApJ...310..176L}. For the range of parameters considered, with $\mbh > \ms$ and $\rp \sim \rs$, the values of $T_2$ and $T_3$ lie in the range $0.01$ -- $0.1$. These values are systematically lower for stars with lower $\rhoconc$ values. As a result, in $0.5 \Msun$ stars (and ZAMS $1 \Msun$ stars), $\Etide$ is dominated by $\Ekick$, resulting in them becoming unbound from the BH. It should be noted that $\Etide$ depends very strongly on the distance of approach.

In addition, we find that the change in the specific orbital angular momentum of the star-BH system, $\horb$, is not significant. For orbits very close to being parabolic, which is the case for all our post-disruption orbits (see Figure \ref{fig:ecc_trends_all}), $\displaystyle \horb \approx \sqrt{2 G (\ms + \mbh) \rp} \approx \sqrt{2 G \mbh \rp}$. This implies that the post-disruption periapsis distance $r_{\mathrm{p}} = a (1 - e)$, where $a$ is the semimajor axis and $e$ is the eccentricity, is nearly the same as the initial periapsis distance $\rp \equiv b \rt$, which is also seen in PTDEs by SMBHs \citep{2020ApJ...904..100R}. Consequently, given that $a = - 0.5\,G (\ms + \mbh) / \Eorb$, the calculation of $e$ is straightforward. Figure \ref{fig:ecc_trends_all} shows the post-disruption $e$,  computed using Equations \ref{eq:smaxis} and \ref{eq:ecc}, as a function of $b$ which follows the trend in $\Eorb$.

\subsection{Rates of tidal encounters}
The rate of $\mu$TDEs per year per Milky Way-like galaxy (similar to \citealp{2016ApJ...823..113P}) can be expressed as:
\begin{equation}
    R_{\mu\mathrm{TDE}} \sim N_{\mathrm{clus}} N_{\mathrm{\star,c}}\,N_{\mathrm{BH,c}}\,\Sigma\,\sigma\,V_{\mathrm{c}}^{-1}
\end{equation}
Here, $N_{\mathrm{clus}} \sim 150 \MWGalinv$ is the number of globular clusters (in the Milky Way, e.g., \citealp{2010arXiv1012.3224H}), $N_{\mathrm{\star,c}}$ is the number of stars in a cluster core, $N_{\mathrm{BH,c}}$ is the number of BHs in a cluster core, $\Sigma = \pi \rp^2 (1 + v_{\mathrm{p}}^2/\sigma^2)$ is the gravitational focused encounter cross-section with periapsis distance $\rp = b \rt$ (see Section \ref{sec:methods}) and periapsis velocity $v_{\mathrm{p}} = \sqrt{G(\mbh+\ms)/\rp}$, and $\sigma$ is the velocity dispersion of the cluster. For globular clusters in the Milky Way Galaxy, the typical values of these parameters are $n_{\mathrm{\star,c}} = N_{\mathrm{\star,c}} V_{\mathrm{c}}^{-1} \sim 10^2$--$10^7 \pcinvcb$ (e.g., \citealp{2018MNRAS.478.1520B}), $N_{\mathrm{BH,c}} \sim 10$--$100$ (e.g., \citealp{2015ApJ...800....9M,2018MNRAS.478.1844A}), $\sigma \sim 1$--$20 \kmsinv$ (e.g., \citealp{2018MNRAS.478.1520B}). With these values, and assuming $v_{\mathrm{p}} >> \sigma$ in globular clusters, and typical star and BH masses of $1 \Msun$ and $10 \Msun$ respectively, the rate is:
\begin{multline}
    R_{\mu\mathrm{TDE}} \sim 10^{-6} \yrinv \MWGalinv \left(\frac{N_{\mathrm{clus}}}{150 \MWGalinv}\right) \left(\frac{n_{\mathrm{\star,c}}}{10^{-5} \pcinvcb}\right) \\ \left(\frac{N_{\mathrm{BH,c}}}{100}\right) \left(\frac{\sigma}{10 \kmsinv}\right) \left(\frac{\mbh+\ms}{11 \Msun}\right) \left(\frac{\rp/\rt}{1.0}\right)
    \label{eq:rate}
\end{multline}

This equation implies that PTDEs (with larger $\rp/\rt$) are more frequent than FTDEs. More specifically, from our simulations of TDEs of $1 \Msun$ stars by $10 \Msun$, encounters with $b \sim 1$--$2$ would result in eccentric bound orbits, which can circularize over time through tides. A more detailed rate calculation would involve integrating Equation \ref{eq:rate} over a range of star and black hole masses.

The above rates are derived for $\mu$TDEs produced in dense stellar systems such as globular clusters and nuclear star clusters. However, other channels also contribute to $\mu$TDEs (see discussion in \citealp{2016ApJ...823..113P}), such as ultra-wide binaries and triples in the field \citep{2016MNRAS.458.4188M,2020MNRAS.498.4924M} and post-natal-kicks leading to close encounter between a newly formed BH and a stellar companion \citep{2016arXiv161000593M}. These channels may give rise $\mu$TDEs in young environments and in the field.

\subsection{Summary of fits} \label{sec:results_summary}
Below we present the post-disruption fits detailed in previous sections for use in globular cluster codes. $m_{\mathrm{\star \rightarrow BH}}$ refers to the gas mass bound to the black hole after disruption,  including the accreted mass.

\subsubsection{Masses of remnant star and BH}

\begin{equation}
    \left( \frac{\Dmsr}{\ms} \right) = 10^{ \min{\{-2.10\,(b - 1.95 \rhoconc^{-1})\ ,\ 0\}} }
\end{equation}
\begin{equation}
    \left( \frac{m_{\mathrm{\star \rightarrow BH}}}{\ms} \right) \le 0.5 \,\left( \frac{\Dmsr}{\ms} \right)
\end{equation}

\subsubsection{Spin of remnant star}

\begin{equation}
    \left( \frac{\Lsr}{\gcmsqsinv} \right) = 10^{ (-1.50 b + 52.00) } \left(\frac{\ms}{\Msun}\right)^{-1} \left(1 - \frac{\Dmsr}{\ms}\right)^{2.4}
\end{equation}

\subsubsection{Validity of fits}
It should be noted that the density concentration factor, $\rhoconc$, only captures the full disruption of a star. Similarly defined factors for each radius $r$, i.e. $(\rho(r)/\overline{\rho})^{1/3}$, are required to calculate the distance at which mass beyond $r$ is lost (see \citealp{2020ApJ...904...99R}). Stars of higher masses tend to have much steeper density profiles when compared to $1 \Msun$ stars, thus Equation \ref{eq:mass_fit}, which fits the mass loss, is not universal. In particular, the mass loss fit for $2 \Msun$ stars would involve an exponent of $\sim 3$, unlike our derived exponent of $\sim 2.1$ in Equation \ref{eq:mass_fit}. Similar arguments follow for the fits of spins and orbits.

\section{Implications} \label{sec:implications}
In this section, we discuss the general observational implications and the caveats of our study. $\mu$TDEs can potentially produce both unique and peculiar transient events as well as longer-lived systems.

\subsection{Low-mass X-ray binaries}
Low-mass X-ray binaries (LMXBs) are interacting binary systems which consist of compact objects (neutrons stars or black holes) accreting mass from their low-mass ($\lesssim 1 \Msun$) stellar companions, thereby producing X-rays. A few hundred of LMXBs have been detected in the Milky Way Galaxy (e.g., \citealp{2007A&A...469..807L,2023A&A...675A.199A}), highlighting the prevalence of such systems. However, theoretical (e.g., \citealp{1997A&A...321..207P,1999ApJ...521..723K}) and population synthesis (e.g., \citealp{2003MNRAS.341..385P,2006MNRAS.369.1152K}) studies have not been able to match the rates of observed LMBXs through isolated binary evolution alone. The primary reason is that a low-mass companion has insufficient energy to expel a common envelope during the giant phase of the primary star, thereby resulting in the spiraling in of the companion and an eventual merger. Thus, a dynamical channel formation channel for LMXBs, where a low-mass star is tidally captured by a black hole in a globular/nuclear cluster, holds promise (e.g., \citealp{2007MNRAS.380.1685V,2016MNRAS.458.4188M,2017MNRAS.469.3088K}).

Some of our simulations of encounters of low-mass stars with black holes are not close enough for significant mass loss and can result in highly eccentric bound orbits. For example, in the parabolic encounter of a $1 \Msun$ star and a $10 \Msun$ BH, for $b \sim 1.0$--$1.5$ (see Section \ref{sec:results}), the fractional mass loss is $\sim 1$--$10$\%, and the post-disruption bound orbit has a period of $0.1$--$1 \yr$ and an eccentricity of $\sim 0.97$--$0.98$. However, tidal dissipation at subsequent periapsis passages can reduce the eccentricity and orbital period (e.g., \citealp{2017MNRAS.469.3088K}) such that the resulting bound system evolves into a compact LMXB. 

Another type of system that is potentially formed through dynamical encounters are non-interacting BH-star systems, e.g., \citet{2023MNRAS.521.4323E,2023MNRAS.518.1057E,2023AJ....166....6C}. These systems consist of $\sim 1 \Msun$ stars orbiting $\sim 10 \Msun$ black holes in moderately-eccentric orbits with periods of $\sim 1 \yr$. These parameters correspond quite well with our simulations with $b \sim 1.0$--$1.5$, which lends credence to dynamical formation.

\subsection{Intermediate-mass black hole growth}
Intermediate-mass black holes (IMBHs; see \citealp{2020ARA&A..58..257G} for a review), with masses greater than those of stellar-mass black holes ($\gtrsim 10^2 \Msun$) and less than those of supermassive black holes ($\lesssim 10^6 \Msun$), are expected to exist in the centers of dwarf galaxies (e.g., \citealp{1987AJ.....93...29K,1989ApJ...342L..11F,2022NatAs...6...26R,2022MNRAS.516.6123G}) and globular clusters (e.g., \citealp{2011ApJ...734..111D,2014MNRAS.437.1208F,2022ApJ...924...48P}). Among several IMBH formation mechanisms (see \citealp{2021NatRP...3..732V} for review), one possible avenue is runaway mass growth through tidal capture and disruption, which has been examined analytically (e.g., \citealp{2017MNRAS.467.4180S}) and numerically (e.g., \citealp{2023MNRAS.521.2930R,2023MNRAS.526..429A}). One of the most critical factors in this mechanism for BH growth is the amount of debris mass that ultimately accretes onto the BH. For example, \citet{2023MNRAS.521.2930R} assume that a star approaching a BH within the nominal tidal radius $\rt$ is fully destroyed and 50\% of the stellar mass is accreted onto the BH.

While these assumptions may be valid for a part of the parameter space, it is important to develop a prescription for the outcomes of TDEs that works over a broader parameter range to more accurately assess the possibility of BH mass growth through TDEs. Our simulations can provide such prescriptions that can improve the treatment of TDEs. Firstly, our simulations show, like previous work on TDEs by supermassive black holes (e.g., \citealp{2013ApJ...767...25G,2017A&A...600A.124M,2019MNRAS.487..981G,2020ApJ...904...98R,2020ApJ...904...99R,2020ApJ...904..100R,2020ApJ...904..101R,2020ApJ...905..141L}), that the periapis distances at which stars are fully or partially destroyed depend on the stellar internal structure. \citet{2020ApJ...904...99R} analytically demonstrated that the nominal tidal radius $\rt$ is a more relevant quantity for partial disruption events involving stars with $\ms \gtrsim 1 \Msun$. Only for lower-mass stars, where $\rhoconc \sim 1$, does $\rt$ come close to the genuine full disruption radius\footnote{The genuine full disruption radius and the partial disruption radius are proportional to $\rho_{\mathrm{c}}^{-1/3}$ and $\overline{\rho}^{-1/3}$ respectively (see Section 4 of \citep{2020ApJ...904...99R}).}. Our fitting formula for the fractional mass loss, Equation \ref{eq:mass_fit}, can provide a better prescription for determining the fate of the star (full or partial disruption) as well as the mass of the remnant.

The amount of debris accreted onto the BH remains highly uncertain. To the zeroth order, if the accretion rate is super-Eddington, the strong radiation pressure gradient would drive strong outflows, hindering continuous and steady accretion \citep{2014MNRAS.439..503S}. However, the accretion efficiency would be collectively affected by many factors, including magnetic fields, black hole spins, accretion flow structure, and jet formation (e.g., \citealp{2014MNRAS.439..503S,2019ApJ...880...67J, 2023MNRAS.518.3441C,2023ApJ...950...31K}). Our fitting formulae for the fractional mass loss cannot provide an accurate prescription for the accreted mass but can place constraints on the maximum mass that can be accreted onto the BH.

\subsection{Fast blue optical transients}
Fast blue optical transients (FBOTs) are a class of optical transients characterized by high peak luminosities $> 10^{43}\,\mathrm{erg}\,\mathrm{s}^{-1}$, rapid rise and decay times of the order of a few days, blue colors \citep{2011ApJ...730...89P,2014ApJ...794...23D}, and peak blackbody temperatures of a few $10^{4}\,\mathrm{K}$. Although a majority of these events can be explained by supernovae with low-mass ejecta \citep{2018MNRAS.481..894P}, a `luminous' subset, e.g., AT2018cow (`the Cow'; \citealp{2018ATel11727....1S,2018ApJ...865L...3P}), AT2018lqh \citep{2021ApJ...922..247O}, AT2020mrf \citep{2022ApJ...934..104Y}, CSS161010 \citep{2020ApJ...895L..23C},  ZTF18abvkwla (`the Koala'; \citealp{2020ApJ...895...49H}), AT2020xnd (`the Camel'; \citealp{2021MNRAS.508.5138P}), AT2022tsd (`the Tasmanian Devil'; \citealp{2023RNAAS...7..126M}), AT2023fhn (`the Finch'; \citealp{2024MNRAS.527L..47C}), are too bright at their peaks and fade too rapidly to be explained by supernovae. Proposed mechanisms involve compact objects, such as black hole accretion or magnetars formed in core-collapse supernovae \citep{2018ApJ...865L...3P}, mergers between Wolf-Rayet stars and compact objects accompanied by hyper-accretion \citep{2022ApJ...932...84M}, or tidal disruption events by intermediate-mass BHs \citep{2019MNRAS.487.2505K} or stellar-mass BHs \citep{2021ApJ...911..104K}.

Some luminous FBOTs also emit in radio or X-ray (e.g., \citealp{2019ApJ...872...18M,2020ApJ...895...49H}) which indicates the presence of a circumstellar medium that may result from partial TDEs \citep{2021ApJ...911..104K}. Our simulations support this notion by showing the presence of debris near the BH with occasional weak outflows from the BH. In addition, the remnants in our simulations are often bound to the BH and can subsequently undergo multiple partial disruptions, e.g., our simulations of TAMS $1 \Msun$ stars and $10 \Msun$ BHs for $b = 0.25$ or $b = 0.33$. While we cannot confirm a direct correlation between FBOTs and $\mu$TDEs, multiple disruption events of a star may fulfill some of the astrophysical conditions (e.g., the presence of a gas medium and rapid accretion onto a compact object) necessary to explain FBOTs with radio emission. For such cases, the detection of repeated bursts in FBOTs could provide valuable constraints on their formation mechanisms.

\subsection{Ultra-long gamma-ray bursts and X-ray flares}
The exact appearance of transients resulting from $\mu$TDEs is highly uncertain and depends on various assumptions. \citet{2016ApJ...823..113P} suggested that, if the accretion onto the BH is efficient and gives rise to jets, there is a possibility of producing ultra-long gamma-ray bursts (GRBs) and/or X-ray flares (XRFs). 

Assuming a proportional relation between the material accreted to the compact object and the luminosity produced by the jet, we can generally divide $\mu$TDEs into two regimes: (1) $t_{\mathrm{min}} \gg t_{\mathrm{acc}}$ and (2) $t_{\mathrm{min}} \ll t_{\mathrm{acc}}$, where $t_{\mathrm{min}}$ is the typical fallback time of the debris following the disruption, and $t_{\mathrm{acc}}$ is the typical viscous time of the accretion disk that can form around the BH (see \citealp{2016ApJ...823..113P} for more details). 

\begin{itemize}
\item $t_{\mathrm{min}} \gg t_{\mathrm{acc}}$: When the mass of the star is much smaller than that of the BH, the accretion evolution is dominated by the fallback rate, i.e., the light curve should generally follow the regular power-law (e.g.  $t^{-5/3}$ power-law for a full disruption). 
\item $t_{\mathrm{min}} \ll t_{\mathrm{acc}}$: When the masses of the star and the BH are comparable, the fallback material is expected to accumulate and form a disk on the fallback time, which then drains on the longer viscous time maintaining a low accretion rate. We would expect the flaring to begin only once the material is accreted onto the compact object. Therefore, we expect four stages in the light curve evolution: (1) A fast rise of the accretion flare once the disk material is processed and evolves to accrete on the compact object. (2) Accretion from the disk until the accumulated early fallback material is drained; if we \textit{assume} a steady state accretion until drainage, one might expect a relatively flat light curve. (3) Once the disk drains the accumulated early fall-back material the light curve should drop steeply. (4) The continuous fall-back of material would govern the accretion rate at times longer than the viscous time, and the accretion rate should then follow the $t^{-5/3}$ rate (or a different power-law). The exact light curve at the early stages is difficult to predict, but we do note a non-trivial signature of the $\mu$TDEs relating to the early and late stages.
\end{itemize}

The expected properties of $\mu$TDEs are consistent with and might explain the origins of ultra-long GRBs \citep{2014ApJ...781...13L}: long-lived ($\sim 10^{4}$ s) GRBs which show an initial plateau followed by a rapid decay. $\mu$TDEs may also explain the origin of some Swift TDE candidates \citep{2011Sci...333..203B,2011Natur.476..421B,2012ApJ...753...77C,2015MNRAS.452.4297B}, suggested to be produced through a TDE by a supermassive BH. The typical timescales for the latter are longer than the observed $10^{5}$ s, challenging the currently suggested origin, but quite consistent with a $\mu$TDE scenario. 

$\mu$TDEs producing ultra-long GRBs are also expected to produce afterglow emission if/when the strong outflows and jets launched by accretion propagate into the surrounding medium and the resulting shock interaction produces high energy particles which give rise to synchrotron and inverse Compton radiation. Such observational signature of $\mu$TDE has been little explored, though analogous observation and modeling have been suggested for TDEs by massive black holes (e.g., \citealp{2011MNRAS.416.2102G,2011Natur.476..425Z}). Detection of afterglows from $\mu$TDEs would provide information on the energetics and dynamics of the outflow, robust independent evidence for the jet collimation, as well as identification of the host galaxy.

\subsection{Remnants}
Partial $\mu$TDEs of stars by stellar BHs can spin up the stars significantly due to strong tidal interactions (see Fig.\ref{fig:Lstar_trends_all} and \citealp{2001ApJ...549..948A}). Though magnetic breaking can potentially spin down such stars over time, observations of highly spun-up low-mass old stars in globular clusters could provide potential signatures for partial $\mu$TDEs. In addition, the remnant undergoes violent chemical mixing during the first pericenter passage and has higher entropy than an ordinary star of the same mass and age \citep{2020ApJ...904..100R,2023MNRAS.519.5787R}. If the mass loss is significant, a unique chemical composition profile may persist even after the remnant has relaxed to a stable state. This suggests that the remnant could exhibit unique asteroseismic signatures that distinguish it from ordinary stars (\citealp{BellingerInPrep}, in prep).

\subsection{Caveats}
As mentioned in Section \ref{sec:methods}, we exclude the effects of relativity and magnetic fields in our study. The Newtonian approximation is justified for TDEs by SBHs because most of the debris stays outside $\rp$, which is approximately four orders of magnitude greater than the gravitational radius $\rg \equiv G \mbh/c^{2}$ \footnote{On the other hand, for TDEs with SMBHs, $\rg$ and $\rt$ can be comparable in scale.}. However, relativistic effects and proper treatment of radiation are more important for the accretion flow, which is beyond the scope of our paper. The effect of magnetic fields on the dynamics of the remnant is unclear, which we will leave for our future work.  A final point to note is that, unlike globular clusters, nuclear star clusters have typical velocity dispersions of $\sigma \gtrsim 100 \kmsinv$\citealp{2003ApJ...599.1139F,2009A&A...502...91S}) depending on the galaxy, and subsequently,  $|1-e| \gtrsim 10^{-2}$.  Thus, encounters are not always parabolic, and for a more thorough study, hyperbolic encounters must be taken into account.

\section{Summary and conclusion} \label{sec:summary}
We performed a grid of 58 hydrodynamics simulations, using the moving-mesh code \arepo{}, of partial (PTDE) and full (FTDE) tidal disruptions of main-sequence (MS) stars with stellar-mass black holes (SBH) on initially-parabolic orbits. Our varied parameters include stellar mass $\ms$ ($0.5 \Msun$ and $1.0 \Msun$), SBH mass ($10 \Msun$ and $40 \Msun$), core hydrogen abundance $\Hc$ (a proxy for MS age $\ts$) and impact parameter $b \equiv \rp/\rt$. Our stellar models are initialized in accordance with 1D detailed non-rotating stellar profiles from \mesa{}. We also define a star's density concentration parameter $\rhoconc  \equiv \rhoconcexp$. Our main results are summarized below:

\begin{itemize}
  \item The mass of a post-disruption stellar remnant $\msr$ decreases with decreasing $b$. A higher $\rhoconc$ (more centrally concentrated) results in a lesser mass loss for the same $b$. $\msr$ depends only on $b$ and $\rhoconc$, and not on $\mbh$. Roughly half of the disrupted mass remains bound to the black hole, while the other half is unbound from the system.
  \item The spin angular momentum of a stellar remnant $\Lsr$ increases  after periapsis passage. For large $b$ (when the mass loss is small), $\Lsr$ increases with decreasing $b$, but this trend reverses for very small $b$ (close to FTDE) due to significant mass loss. This reversal occurs for lower $b$ when $\rhoconc$ is higher. The final spin is also independent of $\mbh$.  For low $b$ values, the rotational velocities can be very close to break-up velocities.
  \item Unlike mass and spin, the orbit of the remnant star also depends on $\mbh$. For large $b$, eccentricity $e$ and semimajor axis $a$ decrease with decreasing $b$ keeping a constant periapsis distance, i.e., the remnant tends to be more `bound'. Depending on $\mbh$ and $\rhoconc$, this no longer holds for approaches close to FTDE -- $e$ can increase and the remnant can become unbound. In general, a higher $\mbh/\ms$ ratio and a lower $\rhoconc$ tend to be unbinding.
  \item We provide relatively simple and accurate fitting formulae for the masses and spins of the remnant stars. These analytical fits are listed in  Section \ref{sec:results_summary} for easy access.
\end{itemize}

We discussed the implications that $\mu$TDEs can have on low-mass X-ray binary formation, intermediate-mass black hole formation, and transient searches. Given the limited range of stellar masses and ages that we covered in this work, we will extend our suite of simulations in the future to cover a wider parameter space, including massive stars. It is necessary to explore this regime to model stellar dynamics in dense stellar clusters.

\begin{acknowledgements}
The authors thank Thorsten Naab for useful discussions regarding star-black hole encounters in star clusters. PV acknowledges the computational resources provided by the Max Planck Institute for Astrophysics to carry out this research.
\end{acknowledgements}

%
  \bibliographystyle{aa} 
  \bibliography{microTDE} 

\begin{thebibliography}{126}
\expandafter\ifx\csname natexlab\endcsname\relax\def\natexlab#1{#1}\fi

\bibitem[{{Abbott} {et~al.}(2021){Abbott}, {Abbott}, {Abraham}, {Acernese},
  {Ackley}, {Adams}, {Adams}, {Adhikari}, {Adya}, {Affeldt}, {Agathos},
  {Agatsuma}, {Aggarwal}, {Aguiar}, {Aiello}, {Ain}, {Ajith}, {Allen},
  {Allocca}, {Altin}, {Amato}, {Anand}, {Ananyeva}, {Anderson}, {Anderson},
  {Angelova}, {Ansoldi}, {Antelis}, {Antier}, {Appert}, {Arai}, {Araya},
  {Areeda}, {Ar{\`e}ne}, {Arnaud}, {Aronson}, {Arun}, {Asali}, {Ascenzi},
  {Ashton}, {Aston}, {Astone}, {Aubin}, {Aufmuth}, {AultONeal}, {Austin},
  {Avendano}, {Babak}, {Badaracco}, {Bader}, {Bae}, {Baer}, {Bagnasco},
  {Baird}, {Ball}, {Ballardin}, {Ballmer}, {Bals}, {Balsamo}, {Baltus},
  {Banagiri}, {Bankar}, {Bankar}, {Barayoga}, {Barbieri}, {Barish}, {Barker},
  {Barneo}, {Barnum}, {Barone}, {Barr}, {Barsotti}, {Barsuglia}, {Barta},
  {Bartlett}, {Bartos}, {Bassiri}, {Basti}, {Bawaj}, {Bayley}, {Bazzan},
  {Becher}, {B{\'e}csy}, {Bedakihale}, {Bejger}, {Belahcene}, {Beniwal},
  {Benjamin}, {Bennett}, {Bentley}, {Bergamin}, {Berger}, {Bergmann},
  {Bernuzzi}, {Berry}, {Bersanetti}, {Bertolini}, {Betzwieser}, {Bhandare},
  {Bhandari}, {Bhattacharjee}, {Bidler}, {Bilenko}, {Billingsley}, {Birney},
  {Birnholtz}, {Biscans}, {Bischi}, {Biscoveanu}, {Bisht}, {Bitossi},
  {Bizouard}, {Blackburn}, {Blackman}, {Blair}, {Blair}, {Blair}, {Blanch},
  {Bobba}, {Bode}, {Boer}, {Boetzel}, {Bogaert}, {Boldrini}, {Bondu},
  {Bonilla}, {Bonnand}, {Booker}, {Boom}, {Bork}, {Boschi}, {Bose},
  {Bossilkov}, {Boudart}, {Bouffanais}, {Bozzi}, {Bradaschia}, {Brady},
  {Bramley}, {Branchesi}, {Brau}, {Breschi}, {Briant}, {Briggs}, {Brighenti},
  {Brillet}, {Brinkmann}, {Brockill}, {Brooks}, {Brooks}, {Brown}, {Brunett},
  {Bruno}, {Bruntz}, {Buikema}, {Bulik}, {Bulten}, {Buonanno}, {Buscicchio},
  {Buskulic}, {Byer}, {Cabero}, {Cadonati}, {Caesar}, {Cagnoli}, {Cahillane},
  {Calder{\'o}n Bustillo}, {Callaghan}, {Callister}, {Calloni}, {Camp},
  {Canepa}, {Cannon}, {Cao}, {Cao}, {Carapella}, {Carbognani}, {Carney},
  {Carpinelli}, {Carullo}, {Carver}, {Casanueva Diaz}, {Casentini}, {Caudill},
  {Cavagli{\`a}}, {Cavalier}, {Cavalieri}, {Cella}, {Cerd{\'a}-Dur{\'a}n},
  {Cesarini}, {Chaibi}, {Chakravarti}, {Chan}, {Chan}, {Chandra}, {Chanial},
  {Chao}, {Charlton}, {Chase}, {Chassande-Mottin}, {Chatterjee},
  {Chattopadhyay}, {Chaturvedi}, {Chatziioannou}, {Chen}, {Chen}, {Chen},
  {Chen}, {Cheng}, {Cheong}, {Chia}, {Chiadini}, {Chierici}, {Chincarini},
  {Chiummo}, {Cho}, {Cho}, {Cho}, {Choate}, {Christensen}, {Chu}, {Chua},
  {Chung}, {Chung}, {Ciani}, {Ciecielag}, {Cie{\'s}lar}, {Cifaldi}, {Ciobanu},
  {Ciolfi}, {Cipriano}, {Cirone}, {Clara}, {Clark}, {Clark}, {Clarke},
  {Clearwater}, {Clesse}, {Cleva}, {Coccia}, {Cohadon}, {Cohen}, {Colleoni},
  {Collette}, {Collins}, {Colpi}, {Constancio}, {Conti}, {Cooper}, {Corban},
  {Corbitt}, {Cordero-Carri{\'o}n}, {Corezzi}, {Corley}, {Cornish}, {Corre},
  {Corsi}, {Cortese}, {Costa}, {Cotesta}, {Coughlin}, {Coughlin}, {Coulon},
  {Countryman}, {Couvares}, {Covas}, {Coward}, {Cowart}, {Coyne}, {Coyne},
  {Creighton}, {Creighton}, {Croquette}, {Crowder}, {Cudell}, {Cullen},
  {Cumming}, {Cummings}, {Cunningham}, {Cuoco}, {Curylo}, {Dal Canton},
  {D{\'a}lya}, {Dana}, {DaneshgaranBajastani}, {D'Angelo}, {Danilishin},
  {D'Antonio}, {Danzmann}, {Darsow-Fromm}, {Dasgupta}, {Datrier}, {Dattilo},
  {Dave}, {Davier}, {Davies}, {Davis}, {Daw}, {Dean}, {DeBra}, {Deenadayalan},
  {Degallaix}, {De Laurentis}, {Del{\'e}glise}, {Del Favero}, {De Lillo}, {De
  Lillo}, {Del Pozzo}, {DeMarchi}, {De Matteis}, {D'Emilio}, {Demos}, {Denker},
  {Dent}, {Depasse}, {De Pietri}, {De Rosa}, {De Rossi}, {DeSalvo}, {de
  Varona}, {Dhurandhar}, {D{\'\i}az}, {Diaz-Ortiz}, {Didio}, {Dietrich}, {Di
  Fiore}, {DiFronzo}, {Di Giorgio}, {Di Giovanni}, {Di Giovanni}, {Di
  Girolamo}, {Di Lieto}, {Ding}, {Di Pace}, {Di Palma}, {Di Renzo},
  {Divakarla}, {Dmitriev}, {Doctor}, {D'Onofrio}, {Donovan}, {Dooley},
  {Doravari}, {Dorrington}, {Downes}, {Drago}, {Driggers}, {Du}, {Ducoin},
  {Dupej}, {Durante}, {D'Urso}, {Duverne}, {Dwyer}, {Easter}, {Eddolls},
  {Edelman}, {Edo}, {Edy}, {Effler}, {Eichholz}, {Eikenberry}, {Eisenmann},
  {Eisenstein}, {Ejlli}, {Errico}, {Essick}, {Estell{\'e}s}, {Estevez},
  {Etienne}, {Etzel}, {Evans}, {Evans}, {Ewing}, {Fafone}, {Fair}, {Fairhurst},
  {Fan}, {Farah}, {Farinon}, {Farr}, {Farr}, {Fauchon-Jones}, {Favata}, {Fays},
  {Fazio}, {Feicht}, {Fejer}, {Feng}, {Fenyvesi}, {Ferguson},
  {Fernandez-Galiana}, {Ferrante}, {Ferreira}, {Fidecaro}, {Figura}, {Fiori},
  {Fiorucci}, {Fishbach}, {Fisher}, {Fishner}, {Fittipaldi}, {Fitz-Axen},
  {Fiumara}, {Flaminio}, {Floden}, {Flynn}, {Fong}, {Font}, {Forsyth},
  {Fournier}, {Frasca}, {Frasconi}, {Frei}, {Freise}, {Frey}, {Frey},
  {Fritschel}, {Frolov}, {Fronz{\'e}}, {Fulda}, {Fyffe}, {Gabbard}, {Gadre},
  {Gaebel}, {Gair}, {Gais}, {Galaudage}, {Gamba}, {Ganapathy}, {Ganguly},
  {Gaonkar}, {Garaventa}, {Garc{\'\i}a-Quir{\'o}s}, {Garufi}, {Gateley},
  {Gaudio}, {Gayathri}, {Gemme}, {Gennai}, {George}, {George}, {Gergely},
  {Ghonge}, {Ghosh}, {Ghosh}, {Ghosh}, {Giacomazzo}, {Giacoppo}, {Giaime},
  {Giardina}, {Gibson}, {Gier}, {Gill}, {Giri}, {Glanzer}, {Gleckl}, {Godwin},
  {Goetz}, {Goetz}, {Gohlke}, {Goncharov}, {Gonz{\'a}lez}, {Gopakumar},
  {Gossan}, {Gosselin}, {Gouaty}, {Grace}, {Grado}, {Granata}, {Granata},
  {Grant}, {Gras}, {Grassia}, {Gray}, {Gray}, {Greco}, {Green}, {Green},
  {Gretarsson}, {Griggs}, {Grignani}, {Grimaldi}, {Grimes}, {Grimm}, {Grote},
  {Grunewald}, {Gruning}, {Guerrero}, {Guidi}, {Guimaraes}, {Guix{\'e}},
  {Gulati}, {Guo}, {Gupta}, {Gupta}, {Gupta}, {Gustafson}, {Gustafson},
  {Guzman}, {Haegel}, {Halim}, {Hall}, {Hamilton}, {Hammond}, {Haney}, {Hanke},
  {Hanks}, {Hanna}, {Hannuksela}, {Hannuksela}, {Hansen}, {Hansen}, {Hanson},
  {Harder}, {Hardwick}, {Haris}, {Harms}, {Harry}, {Harry}, {Hartwig},
  {Hasskew}, {Haster}, {Haughian}, {Hayes}, {Healy}, {Heidmann}, {Heintze},
  {Heinze}, {Heinzel}, {Heitmann}, {Hellman}, {Hello}, {Helmling-Cornell},
  {Hemming}, {Hendry}, {Heng}, {Hennes}, {Hennig}, {Hennig}, {Hernandez
  Vivanco}, {Heurs}, {Hild}, {Hill}, {Hines}, {Hochheim}, {Hofgard}, {Hofman},
  {Hohmann}, {Holgado}, {Holland}, {Hollows}, {Holmes}, {Holt}, {Holz},
  {Hopkins}, {Horst}, {Hough}, {Howell}, {Hoy}, {Hoyland}, {Huang},
  {H{\"u}bner}, {Huddart}, {Huerta}, {Hughey}, {Hui}, {Husa}, {Huttner},
  {Hutzler}, {Huxford}, {Huynh-Dinh}, {Idzkowski}, {Iess}, {Imperato},
  {Inchauspe}, {Ingram}, {Intini}, {Isi}, {Iyer}, {JaberianHamedan}, {Jacqmin},
  {Jadhav}, {Jadhav}, {James}, {Jani}, {Janssens}, {Janthalur}, {Jaranowski},
  {Jariwala}, {Jaume}, {Jenkins}, {Jeunon}, {Jiang}, {Johns}, {Jones}, {Jones},
  {Jones}, {Jones}, {Jones}, {Jonker}, {Ju}, {Junker}, {Kalaghatgi},
  {Kalogera}, {Kamai}, {Kandhasamy}, {Kang}, {Kanner}, {Kapadia}, {Kapasi},
  {Karathanasis}, {Karki}, {Kashyap}, {Kasprzack}, {Kastaun}, {Katsanevas},
  {Katsavounidis}, {Katzman}, {Kawabe}, {K{\'e}f{\'e}lian}, {Keitel}, {Key},
  {Khadka}, {Khalili}, {Khan}, {Khan}, {Khazanov}, {Khetan}, {Khursheed},
  {Kijbunchoo}, {Kim}, {Kim}, {Kim}, {Kim}, {Kim}, {Kim}, {Kimball}, {King},
  {Kinley-Hanlon}, {Kirchhoff}, {Kissel}, {Kleybolte}, {Klimenko}, {Knowles},
  {Knyazev}, {Koch}, {Koehlenbeck}, {Koekoek}, {Koley}, {Kolstein}, {Komori},
  {Kondrashov}, {Kontos}, {Koper}, {Korobko}, {Korth}, {Kovalam}, {Kozak},
  {Kr{\"a}mer}, {Kringel}, {Krishnendu}, {Kr{\'o}lak}, {Kuehn}, {Kumar},
  {Kumar}, {Kumar}, {Kumar}, {Kuns}, {Kwang}, {Lackey}, {Laghi}, {Lalande},
  {Lam}, {Lamberts}, {Landry}, {Lane}, {Lang}, {Lange}, {Lantz}, {Lanza}, {La
  Rosa}, {Lartaux-Vollard}, {Lasky}, {Laxen}, {Lazzarini}, {Lazzaro}, {Leaci},
  {Leavey}, {Lecoeuche}, {Lee}, {Lee}, {Lee}, {Lee}, {Lehmann}, {Leon},
  {Leroy}, {Letendre}, {Levin}, {Li}, {Li}, {Li}, {Li}, {Li}, {Linde},
  {Linker}, {Linley}, {Littenberg}, {Liu}, {Liu}, {Llorens-Monteagudo}, {Lo},
  {Lockwood}, {London}, {Longo}, {Lorenzini}, {Loriette}, {Lormand}, {Losurdo},
  {Lough}, {Lousto}, {Lovelace}, {L{\"u}ck}, {Lumaca}, {Lundgren}, {Ma},
  {Macas}, {MacInnis}, {Macleod}, {MacMillan}, {Macquet}, {Maga{\~n}a
  Hernandez}, {Maga{\~n}a-Sandoval}, {Magazz{\`u}}, {Magee}, {Majorana},
  {Maksimovic}, {Maliakal}, {Malik}, {Man}, {Mandic}, {Mangano}, {Mansell},
  {Manske}, {Mantovani}, {Mapelli}, {Marchesoni}, {Marion}, {M{\'a}rka},
  {M{\'a}rka}, {Markakis}, {Markosyan}, {Markowitz}, {Maros}, {Marquina},
  {Marsat}, {Martelli}, {Martin}, {Martin}, {Martinez}, {Martinez}, {Martynov},
  {Masalehdan}, {Mason}, {Massera}, {Masserot}, {Massinger}, {Masso-Reid},
  {Mastrogiovanni}, {Matas}, {Mateu-Lucena}, {Matichard}, {Matiushechkina},
  {Mavalvala}, {Maynard}, {McCann}, {McCarthy}, {McClelland}, {McCormick},
  {McCuller}, {McGuire}, {McIsaac}, {McIver}, {McManus}, {McRae}, {McWilliams},
  {Meacher}, {Meadors}, {Mehmet}, {Mehta}, {Melatos}, {Melchor}, {Mendell},
  {Menendez-Vazquez}, {Mercer}, {Mereni}, {Merfeld}, {Merilh}, {Merritt},
  {Merzougui}, {Meshkov}, {Messenger}, {Messick}, {Metzdorff}, {Meyers},
  {Meylahn}, {Mhaske}, {Miani}, {Miao}, {Michaloliakos}, {Michel}, {Middleton},
  {Milano}, {Miller}, {Miller}, {Millhouse}, {Mills}, {Milotti},
  {Milovich-Goff}, {Minazzoli}, {Minenkov}, {Mir}, {Mishkin}, {Mishra},
  {Mistry}, {Mitra}, {Mitrofanov}, {Mitselmakher}, {Mittleman}, {Mo},
  {Mogushi}, {Mohapatra}, {Mohite}, {Molina}, {Molina-Ruiz}, {Mondin},
  {Montani}, {Moore}, {Moraru}, {Morawski}, {Moreno}, {Morisaki}, {Mours},
  {Mow-Lowry}, {Mozzon}, {Muciaccia}, {Mukherjee}, {Mukherjee}, {Mukherjee},
  {Mukherjee}, {Mukund}, {Mullavey}, {Munch}, {Mu{\~n}iz}, {Murray}, {Nadji},
  {Nagar}, {Nardecchia}, {Naticchioni}, {Nayak}, {Neil}, {Neilson}, {Nelemans},
  {Nelson}, {Nery}, {Neunzert}, {Ng}, {Ng}, {Nguyen}, {Nguyen}, {Nguyen},
  {Nichols}, {Nissanke}, {Nocera}, {Noh}, {North}, {Nothard}, {Nuttall},
  {Oberling}, {O'Brien}, {O'Dell}, {Oganesyan}, {Ogin}, {Oh}, {Oh}, {Ohme},
  {Ohta}, {Okada}, {Olivetto}, {Oppermann}, {Oram}, {O'Reilly}, {Ormiston},
  {Ormsby}, {Ortega}, {O'Shaughnessy}, {Ossokine}, {Osthelder}, {Ottaway},
  {Overmier}, {Owen}, {Pace}, {Pagano}, {Page}, {Pagliaroli}, {Pai}, {Pai},
  {Palamos}, {Palashov}, {Palomba}, {Pan}, {Panda}, {Pang}, {Pankow},
  {Pannarale}, {Pant}, {Paoletti}, {Paoli}, {Paolone}, {Parker}, {Pascucci},
  {Pasqualetti}, {Passaquieti}, {Passuello}, {Patel}, {Patricelli}, {Payne},
  {Pechsiri}, {Pedraza}, {Pegoraro}, {Pele}, {Penn}, {Perego}, {Perez},
  {P{\'e}rigois}, {Perreca}, {Perri{\`e}s}, {Petermann}, {Petterson},
  {Pfeiffer}, {Pham}, {Phukon}, {Piccinni}, {Pichot}, {Piendibene},
  {Piergiovanni}, {Pierini}, {Pierro}, {Pillant}, {Pilo}, {Pinard}, {Pinto},
  {Piotrzkowski}, {Pirello}, {Pitkin}, {Placidi}, {Plastino}, {Pluchar},
  {Poggiani}, {Polini}, {Pong}, {Ponrathnam}, {Popolizio}, {Porter},
  {Poverman}, {Powell}, {Pracchia}, {Prajapati}, {Prasai}, {Prasanna},
  {Pratten}, {Prestegard}, {Principe}, {Prodi}, {Prokhorov}, {Prosposito},
  {Puecher}, {Punturo}, {Puosi}, {Puppo}, {P{\"u}rrer}, {Qi}, {Quetschke},
  {Quinonez}, {Quitzow-James}, {Raab}, {Raaijmakers}, {Radkins}, {Radulesco},
  {Raffai}, {Rafferty}, {Rail}, {Raja}, {Rajan}, {Rajbhandari}, {Rakhmanov},
  {Ramirez}, {Ramirez}, {Ramos-Buades}, {Rana}, {Rao}, {Rapagnani}, {Rapol},
  {Ratto}, {Raymond}, {Razzano}, {Read}, {Regimbau}, {Rei}, {Reid}, {Reitze},
  {Rettegno}, {Ricci}, {Richardson}, {Richardson}, {Richardson}, {Ricker},
  {Riemenschneider}, {Riles}, {Rizzo}, {Robertson}, {Robinet}, {Rocchi},
  {Rocha}, {Rodriguez}, {Rodriguez-Soto}, {Rolland}, {Rollins}, {Roma},
  {Romanelli}, {Romano}, {Romel}, {Romero}, {Romero-Shaw}, {Romie}, {Ronchini},
  {Rose}, {Rose}, {Rose}, {Rosell}, {Rosi{\'n}ska}, {Rosofsky}, {Ross},
  {Rowan}, {Rowlinson}, {Roy}, {Roy}, {Ruggi}, {Ryan}, {Sachdev}, {Sadecki},
  {Sakellariadou}, {Salafia}, {Salconi}, {Saleem}, {Samajdar}, {Sanchez},
  {Sanchez}, {Sanchez}, {Sanchis-Gual}, {Sanders}, {Santiago}, {Santos},
  {Saravanan}, {Sarin}, {Sassolas}, {Sathyaprakash}, {Sauter}, {Savage},
  {Savant}, {Sawant}, {Sayah}, {Schaetzl}, {Schale}, {Scheel}, {Scheuer},
  {Schindler-Tyka}, {Schmidt}, {Schnabel}, {Schofield}, {Sch{\"o}nbeck},
  {Schreiber}, {Schulte}, {Schutz}, {Schwarm}, {Schwartz}, {Scott}, {Scott},
  {Seglar-Arroyo}, {Seidel}, {Sellers}, {Sengupta}, {Sennett}, {Sentenac},
  {Sequino}, {Sergeev}, {Setyawati}, {Shaffer}, {Shahriar}, {Sharifi},
  {Sharma}, {Sharma}, {Shawhan}, {Shen}, {Shikauchi}, {Shink}, {Shoemaker},
  {Shoemaker}, {Shukla}, {ShyamSundar}, {Sieniawska}, {Sigg}, {Singer},
  {Singh}, {Singh}, {Singha}, {Singhal}, {Sintes}, {Sipala}, {Skliris},
  {Slagmolen}, {Slaven-Blair}, {Smetana}, {Smith}, {Smith}, {Somala}, {Son},
  {Soni}, {Sorazu}, {Sordini}, {Sorrentino}, {Sorrentino}, {Soulard},
  {Souradeep}, {Sowell}, {Spencer}, {Spera}, {Srivastava}, {Srivastava},
  {Staats}, {Stachie}, {Steer}, {Steinke}, {Steinlechner}, {Steinlechner},
  {Steinmeyer}, {Stevenson}, {Stolle-McAllister}, {Stops}, {Stover}, {Strain},
  {Stratta}, {Strunk}, {Sturani}, {Stuver}, {S{\"u}dbeck}, {Sudhagar},
  {Sudhir}, {Suh}, {Summerscales}, {Sun}, {Sun}, {Sunil}, {Sur}, {Suresh},
  {Sutton}, {Swinkels}, {Szczepa{\'n}czyk}, {Tacca}, {Tait}, {Talbot},
  {Tanasijczuk}, {Tanner}, {Tao}, {Tapia}, {Tapia San Martin}, {Tasson},
  {Taylor}, {Tenorio}, {Terkowski}, {Thirugnanasambandam}, {Thomas}, {Thomas},
  {Thomas}, {Thompson}, {Thondapu}, {Thorne}, {Thrane}, {Tiwari}, {Tiwari},
  {Tiwari}, {Toland}, {Tolley}, {Tonelli}, {Tornasi}, {Torres-Forn{\'e}},
  {Torrie}, {Tosta e Melo}, {T{\"o}yr{\"a}}, {Tran}, {Trapananti}, {Travasso},
  {Traylor}, {Tringali}, {Tripathee}, {Trovato}, {Trudeau}, {Tsai}, {Tsang},
  {Tse}, {Tso}, {Tsukada}, {Tsuna}, {Tsutsui}, {Turconi}, {Ubhi}, {Udall},
  {Ueno}, {Ugolini}, {Unnikrishnan}, {Urban}, {Usman}, {Utina}, {Vahlbruch},
  {Vajente}, {Vajpeyi}, {Valdes}, {Valentini}, {Valsan}, {van Bakel},
  {Beuzekom}, {van den Brand}, {Van Den Broeck}, {Vander-Hyde}, {van der
  Schaaf}, {van Heijningen}, {Vardaro}, {Vargas}, {Varma}, {Vass},
  {Vas{\'u}th}, {Vecchio}, {Vedovato}, {Veitch}, {Veitch}, {Venkateswara},
  {Venneberg}, {Venugopalan}, {Verkindt}, {Verma}, {Veske}, {Vetrano},
  {Vicer{\'e}}, {Viets}, {Villa-Ortega}, {Vinet}, {Vitale}, {Vo}, {Vocca},
  {Vorvick}, {Vyatchanin}, {Wade}, {Wade}, {Wade}, {Walet}, {Walker},
  {Wallace}, {Wallace}, {Walsh}, {Wang}, {Wang}, {Wang}, {Wang}, {Ward},
  {Warner}, {Was}, {Washington}, {Watchi}, {Weaver}, {Wei}, {Weinert},
  {Weinstein}, {Weiss}, {Wellmann}, {Wen}, {We{\ss}els}, {Westhouse}, {Wette},
  {Whelan}, {White}, {White}, {Whiting}, {Whittle}, {Wilken}, {Williams},
  {Williams}, {Williamson}, {Willis}, {Willke}, {Wilson}, {Wimmer}, {Winkler},
  {Wipf}, {Woan}, {Woehler}, {Wofford}, {Wong}, {Wrangel}, {Wright}, {Wu},
  {Wysocki}, {Xiao}, {Yamamoto}, {Yang}, {Yang}, {Yang}, {Yap}, {Yeeles},
  {Yoon}, {Yu}, {Yu}, {Yuen}, {Zadro{\.z}ny}, {Zanolin}, {Zelenova}, {Zendri},
  {Zevin}, {Zhang}, {Zhang}, {Zhang}, {Zhang}, {Zhao}, {Zhao}, {Zhou}, {Zhou},
  {Zhu}, {Zimmerman}, {Zucker}, {Zweizig}, {LIGO Scientific Collaboration}, \&
  {Virgo Collaboration}}]{2021ApJ...913L...7A}
{Abbott}, R., {Abbott}, T.~D., {Abraham}, S., {et~al.} 2021, \apjl, 913, L7

\bibitem[{{Abbott} {et~al.}(2023){Abbott}, {Abbott}, {Acernese}, {Ackley},
  {Adams}, {Adhikari}, {Adhikari}, {Adya}, {Affeldt}, {Agarwal}, {Agathos},
  {Agatsuma}, {Aggarwal}, {Aguiar}, {Aiello}, {Ain}, {Ajith}, {Akutsu}, {de
  Alarc{\'o}n}, {Akcay}, {Albanesi}, {Allocca}, {Altin}, {Amato}, {Anand},
  {Anand}, {Ananyeva}, {Anderson}, {Anderson}, {Ando}, {Andrade}, {Andres},
  {Andri{\'c}}, {Angelova}, {Ansoldi}, {Antelis}, {Antier}, {Antonini},
  {Appert}, {Arai}, {Arai}, {Arai}, {Araki}, {Araya}, {Araya}, {Areeda},
  {Ar{\`e}ne}, {Aritomi}, {Arnaud}, {Arogeti}, {Aronson}, {Arun}, {Asada},
  {Asali}, {Ashton}, {Aso}, {Assiduo}, {Aston}, {Astone}, {Aubin}, {Austin},
  {Babak}, {Badaracco}, {Bader}, {Badger}, {Bae}, {Bae}, {Baer}, {Bagnasco},
  {Bai}, {Baiotti}, {Baird}, {Bajpai}, {Ball}, {Ballardin}, {Ballmer},
  {Balsamo}, {Baltus}, {Banagiri}, {Bankar}, {Barayoga}, {Barbieri}, {Barish},
  {Barker}, {Barneo}, {Barone}, {Barr}, {Barsotti}, {Barsuglia}, {Barta},
  {Bartlett}, {Barton}, {Bartos}, {Bassiri}, {Basti}, {Bawaj}, {Bayley},
  {Baylor}, {Bazzan}, {B{\'e}csy}, {Bedakihale}, {Bejger}, {Belahcene},
  {Benedetto}, {Beniwal}, {Bennett}, {Bentley}, {Benyaala}, {Bergamin},
  {Berger}, {Bernuzzi}, {Berry}, {Bersanetti}, {Bertolini}, {Betzwieser},
  {Beveridge}, {Bhandare}, {Bhardwaj}, {Bhattacharjee}, {Bhaumik}, {Bilenko},
  {Billingsley}, {Bini}, {Birney}, {Birnholtz}, {Biscans}, {Bischi},
  {Biscoveanu}, {Bisht}, {Biswas}, {Bitossi}, {Bizouard}, {Blackburn}, {Blair},
  {Blair}, {Blair}, {Bobba}, {Bode}, {Boer}, {Bogaert}, {Boldrini}, {Bonavena},
  {Bondu}, {Bonilla}, {Bonnand}, {Booker}, {Boom}, {Bork}, {Boschi}, {Bose},
  {Bose}, {Bossilkov}, {Boudart}, {Bouffanais}, {Bozzi}, {Bradaschia}, {Brady},
  {Bramley}, {Branch}, {Branchesi}, {Brandt}, {Brau}, {Breschi}, {Briant},
  {Briggs}, {Brillet}, {Brinkmann}, {Brockill}, {Brooks}, {Brooks}, {Brown},
  {Brunett}, {Bruno}, {Bruntz}, {Bryant}, {Bulik}, {Bulten}, {Buonanno},
  {Buscicchio}, {Buskulic}, {Buy}, {Byer}, {Cadonati}, {Cagnoli}, {Cahillane},
  {Bustillo}, {Callaghan}, {Callister}, {Calloni}, {Cameron}, {Camp}, {Canepa},
  {Canevarolo}, {Cannavacciuolo}, {Cannon}, {Cao}, {Cao}, {Capocasa}, {Capote},
  {Carapella}, {Carbognani}, {Carlin}, {Carney}, {Carpinelli}, {Carrillo},
  {Carullo}, {Carver}, {Diaz}, {Casentini}, {Castaldi}, {Caudill},
  {Cavagli{\`a}}, {Cavalier}, {Cavalieri}, {Ceasar}, {Cella},
  {Cerd{\'a}-Dur{\'a}n}, {Cesarini}, {Chaibi}, {Chakravarti}, {Subrahmanya},
  {Champion}, {Chan}, {Chan}, {Chan}, {Chan}, {Chan}, {Chandra}, {Chanial},
  {Chao}, {Chapman-Bird}, {Charlton}, {Chase}, {Chassande-Mottin},
  {Chatterjee}, {Chatterjee}, {Chatterjee}, {Chaturvedi}, {Chaty},
  {Chatziioannou}, {Chen}, {Chen}, {Chen}, {Chen}, {Chen}, {Chen}, {Chen},
  {Chen}, {Cheng}, {Cheong}, {Cheung}, {Chia}, {Chiadini}, {Chiang},
  {Chiarini}, {Chierici}, {Chincarini}, {Chiofalo}, {Chiummo}, {Cho}, {Cho},
  {Choudhary}, {Choudhary}, {Christensen}, {Chu}, {Chu}, {Chu}, {Chua},
  {Chung}, {Ciani}, {Ciecielag}, {Cie{\'s}lar}, {Cifaldi}, {Ciobanu}, {Ciolfi},
  {Cipriano}, {Cirone}, {Clara}, {Clark}, {Clark}, {Clarke}, {Clearwater},
  {Clesse}, {Cleva}, {Coccia}, {Codazzo}, {Cohadon}, {Cohen}, {Cohen},
  {Colleoni}, {Collette}, {Colombo}, {Colpi}, {Compton}, {Constancio}, {Conti},
  {Cooper}, {Corban}, {Corbitt}, {Cordero-Carri{\'o}n}, {Corezzi}, {Corley},
  {Cornish}, {Corre}, {Corsi}, {Cortese}, {Costa}, {Cotesta}, {Coughlin},
  {Coulon}, {Countryman}, {Cousins}, {Couvares}, {Coward}, {Cowart}, {Coyne},
  {Coyne}, {Creighton}, {Creighton}, {Criswell}, {Croquette}, {Crowder},
  {Cudell}, {Cullen}, {Cumming}, {Cummings}, {Cunningham}, {Cuoco},
  {Cury{\l}o}, {Dabadie}, {Canton}, {Dall'Osso}, {D{\'a}lya}, {Dana},
  {Daneshgaranbajastani}, {D'Angelo}, {Danila}, {Danilishin}, {D'Antonio},
  {Danzmann}, {Darsow-Fromm}, {Dasgupta}, {Datrier}, {Datta}, {Dattilo},
  {Dave}, {Davier}, {Davies}, {Davis}, {Davis}, {Daw}, {Dean}, {Debra},
  {Deenadayalan}, {Degallaix}, {de Laurentis}, {Del{\'e}glise}, {Del Favero},
  {de Lillo}, {de Lillo}, {Del Pozzo}, {Demarchi}, {de Matteis}, {D'Emilio},
  {Demos}, {Dent}, {Depasse}, {de Pietri}, {De Rosa}, {de Rossi}, {Desalvo},
  {de Simone}, {Dhurandhar}, {D{\'\i}az}, {Diaz-Ortiz}, {Didio}, {Dietrich},
  {di Fiore}, {di Fronzo}, {di Giorgio}, {di Giovanni}, {di Giovanni}, {di
  Girolamo}, {di Lieto}, {Ding}, {di Pace}, {di Palma}, {di Renzo},
  {Divakarla}, {Dmitriev}, {Doctor}, {D'Onofrio}, {Donovan}, {Dooley},
  {Doravari}, {Dorrington}, {Drago}, {Driggers}, {Drori}, {Ducoin}, {Dupej},
  {Durante}, {D'Urso}, {Duverne}, {Dwyer}, {Eassa}, {Easter}, {Ebersold},
  {Eckhardt}, {Eddolls}, {Edelman}, {Edo}, {Edy}, {Effler}, {Eguchi},
  {Eichholz}, {Eikenberry}, {Eisenmann}, {Eisenstein}, {Ejlli}, {Engelby},
  {Enomoto}, {Errico}, {Essick}, {Estell{\'e}s}, {Estevez}, {Etienne}, {Etzel},
  {Evans}, {Evans}, {Ewing}, {Fafone}, {Fair}, {Fairhurst}, {Farah}, {Farinon},
  {Farr}, {Farr}, {Farrow}, {Fauchon-Jones}, {Favaro}, {Favata}, {Fays},
  {Fazio}, {Feicht}, {Fejer}, {Fenyvesi}, {Ferguson}, {Fernandez-Galiana},
  {Ferrante}, {Ferreira}, {Fidecaro}, {Figura}, {Fiori}, {Fishbach}, {Fisher},
  {Fittipaldi}, {Fiumara}, {Flaminio}, {Floden}, {Fong}, {Font}, {Fornal},
  {Forsyth}, {Franke}, {Frasca}, {Frasconi}, {Frederick}, {Freed}, {Frei},
  {Freise}, {Frey}, {Fritschel}, {Frolov}, {Fronz{\'e}}, {Fujii}, {Fujikawa},
  {Fukunaga}, {Fukushima}, {Fulda}, {Fyffe}, {Gabbard}, {Gadre}, {Gair},
  {Gais}, {Galaudage}, {Gamba}, {Ganapathy}, {Ganguly}, {Gao}, {Gaonkar},
  {Garaventa}, {Garc{\'\i}a}, {Garc{\'\i}a-N{\'u}{\~n}ez},
  {Garc{\'\i}a-Quir{\'o}s}, {Garufi}, {Gateley}, {Gaudio}, {Gayathri}, {Ge},
  {Gemme}, {Gennai}, {George}, {George}, {Gerberding}, {Gergely}, {Gewecke},
  {Ghonge}, {Ghosh}, {Ghosh}, {Ghosh}, {Ghosh}, {Giacomazzo}, {Giacoppo},
  {Giaime}, {Giardina}, {Gibson}, {Gier}, {Giesler}, {Giri}, {Gissi},
  {Glanzer}, {Gleckl}, {Godwin}, {Golomb}, {Goetz}, {Goetz}, {Gohlke},
  {Goncharov}, {Gonz{\'a}lez}, {Gopakumar}, {Gosselin}, {Gouaty}, {Gould},
  {Grace}, {Grado}, {Granata}, {Granata}, {Grant}, {Gras}, {Grassia}, {Gray},
  {Gray}, {Greco}, {Green}, {Green}, {Gretarsson}, {Gretarsson}, {Griffith},
  {Griffiths}, {Griggs}, {Grignani}, {Grimaldi}, {Grimm}, {Grote}, {Grunewald},
  {Gruning}, {Guerra}, {Guidi}, {Guimaraes}, {Guix{\'e}}, {Gulati}, {Guo},
  {Guo}, {Gupta}, {Gupta}, {Gupta}, {Gustafson}, {Gustafson}, {Guzman}, {Ha},
  {Haegel}, {Hagiwara}, {Haino}, {Halim}, {Hall}, {Hamilton}, {Hammond}, {Han},
  {Haney}, {Hanks}, {Hanna}, {Hannam}, {Hannuksela}, {Hansen}, {Hansen},
  {Hanson}, {Harder}, {Hardwick}, {Haris}, {Harms}, {Harry}, {Harry},
  {Hartwig}, {Hasegawa}, {Haskell}, {Hasskew}, {Haster}, {Hattori}, {Haughian},
  {Hayakawa}, {Hayama}, {Hayes}, {Healy}, {Heidmann}, {Heidt}, {Heintze},
  {Heinze}, {Heinzel}, {Heitmann}, {Hellman}, {Hello}, {Helmling-Cornell},
  {Hemming}, {Hendry}, {Heng}, {Hennes}, {Hennig}, {Hennig}, {Hernandez},
  {Vivanco}, {Heurs}, {Hild}, {Hill}, {Himemoto}, {Hines}, {Hiranuma},
  {Hirata}, {Hirose}, {Hochheim}, {Hofman}, {Hohmann}, {Holcomb}, {Holland},
  {Hollows}, {Holmes}, {Holt}, {Holz}, {Hong}, {Hopkins}, {Hough}, {Hourihane},
  {Howell}, {Hoy}, {Hoyland}, {Hreibi}, {Hsieh}, {Hsu}, {Huang}, {Huang},
  {Huang}, {Huang}, {Huang}, {Huang}, {H{\"u}bner}, {Huddart}, {Hughey}, {Hui},
  {Hui}, {Husa}, {Huttner}, {Huxford}, {Huynh-Dinh}, {Ide}, {Idzkowski},
  {Iess}, {Ikenoue}, {Imam}, {Inayoshi}, {Ingram}, {Inoue}, {Ioka}, {Isi},
  {Isleif}, {Ito}, {Itoh}, {Iyer}, {Izumi}, {Jaberianhamedan}, {Jacqmin},
  {Jadhav}, {Jadhav}, {James}, {Jan}, {Jani}, {Janquart}, {Janssens},
  {Janthalur}, {Jaranowski}, {Jariwala}, {Jaume}, {Jenkins}, {Jenner}, {Jeon},
  {Jeunon}, {Jia}, {Jin}, {Johns}, {Jones}, {Jones}, {Jones}, {Jones}, {Jones},
  {Jonker}, {Ju}, {Jung}, {Jung}, {Junker}, {Juste}, {Kaihotsu}, {Kajita},
  {Kakizaki}, {Kalaghatgi}, {Kalogera}, {Kamai}, {Kamiizumi}, {Kanda},
  {Kandhasamy}, {Kang}, {Kanner}, {Kao}, {Kapadia}, {Kapasi}, {Karat},
  {Karathanasis}, {Karki}, {Kashyap}, {Kasprzack}, {Kastaun}, {Katsanevas},
  {Katsavounidis}, {Katzman}, {Kaur}, {Kawabe}, {Kawaguchi}, {Kawai},
  {Kawasaki}, {K{\'e}f{\'e}lian}, {Keitel}, {Key}, {Khadka}, {Khalili}, {Khan},
  {Khazanov}, {Khetan}, {Khursheed}, {Kijbunchoo}, {Kim}, {Kim}, {Kim}, {Kim},
  {Kim}, {Kim}, {Kimball}, {Kimura}, {Kinley-Hanlon}, {Kirchhoff}, {Kissel},
  {Kita}, {Kitazawa}, {Kleybolte}, {Klimenko}, {Knee}, {Knowles}, {Knyazev},
  {Koch}, {Koekoek}, {Kojima}, {Kokeyama}, {Koley}, {Kolitsidou}, {Kolstein},
  {Komori}, {Kondrashov}, {Kong}, {Kontos}, {Koper}, {Korobko}, {Kotake},
  {Kovalam}, {Kozak}, {Kozakai}, {Kozu}, {Kringel}, {Krishnendu}, {Kr{\'o}lak},
  {Kuehn}, {Kuei}, {Kuijer}, {Kulkarni}, {Kumar}, {Kumar}, {Kumar}, {Kumar},
  {Kume}, {Kuns}, {Kuo}, {Kuo}, {Kuromiya}, {Kuroyanagi}, {Kusayanagi},
  {Kuwahara}, {Kwak}, {Lagabbe}, {Laghi}, {Lalande}, {Lam}, {Lamberts},
  {Landry}, {Landry}, {Lane}, {Lang}, {Lange}, {Lantz}, {La Rosa},
  {Lartaux-Vollard}, {Lasky}, {Laxen}, {Lazzarini}, {Lazzaro}, {Leaci},
  {Leavey}, {Lecoeuche}, {Lee}, {Lee}, {Lee}, {Lee}, {Lee}, {Lee}, {Lehmann},
  {Lema{\^\i}tre}, {Leonardi}, {Leroy}, {Letendre}, {Levesque}, {Levin},
  {Leviton}, {Leyde}, {Li}, {Li}, {Li}, {Li}, {Li}, {Li}, {Lin}, {Lin}, {Lin},
  {Lin}, {Lin}, {Linde}, {Linker}, {Linley}, {Littenberg}, {Liu}, {Liu}, {Liu},
  {Liu}, {Llamas}, {Llorens-Monteagudo}, {Lo}, {Lockwood}, {Loh}, {London},
  {Longo}, {Lopez}, {Portilla}, {Lorenzini}, {Loriette}, {Lormand}, {Losurdo},
  {Lott}, {Lough}, {Lousto}, {Lovelace}, {Lucaccioni}, {L{\"u}ck}, {Lumaca},
  {Lundgren}, {Luo}, {Lynam}, {Macas}, {Macinnis}, {MacLeod}, {MacMillan},
  {Macquet}, {Hernandez}, {Magazz{\`u}}, {Magee}, {Maggiore}, {Magnozzi},
  {Mahesh}, {Majorana}, {Makarem}, {Maksimovic}, {Maliakal}, {Malik}, {Man},
  {Mandic}, {Mangano}, {Mango}, {Mansell}, {Manske}, {Mantovani}, {Mapelli},
  {Marchesoni}, {Marchio}, {Marion}, {Mark}, {M{\'a}rka}, {M{\'a}rka},
  {Markakis}, {Markosyan}, {Markowitz}, {Maros}, {Marquina}, {Marsat},
  {Martelli}, {Martin}, {Martin}, {Martinez}, {Martinez}, {Martinez},
  {Martinovic}, {Martynov}, {Marx}, {Masalehdan}, {Mason}, {Massera},
  {Masserot}, {Massinger}, {Masso-Reid}, {Mastrogiovanni}, {Matas},
  {Mateu-Lucena}, {Matichard}, {Matiushechkina}, {Mavalvala}, {McCann},
  {McCarthy}, {McClelland}, {McClincy}, {McCormick}, {McCuller}, {McGhee},
  {McGuire}, {McIsaac}, {McIver}, {McRae}, {McWilliams}, {Meacher}, {Mehmet},
  {Mehta}, {Meijer}, {Melatos}, {Melchor}, {Mendell}, {Menendez-Vazquez},
  {Menoni}, {Mercer}, {Mereni}, {Merfeld}, {Merilh}, {Merritt}, {Merzougui},
  {Meshkov}, {Messenger}, {Messick}, {Meyers}, {Meylahn}, {Mhaske}, {Miani},
  {Miao}, {Michaloliakos}, {Michel}, {Michimura}, {Middleton}, {Milano},
  {Miller}, {Miller}, {Miller}, {Miller}, {Millhouse}, {Mills}, {Milotti},
  {Minazzoli}, {Minenkov}, {Mio}, {Mir}, {Miravet-Ten{\'e}s}, {Mishra},
  {Mishra}, {Mistry}, {Mitra}, {Mitrofanov}, {Mitselmakher}, {Mittleman},
  {Miyakawa}, {Miyamoto}, {Miyazaki}, {Miyo}, {Miyoki}, {Mo}, {Modafferi},
  {Moguel}, {Mogushi}, {Mohapatra}, {Mohite}, {Molina}, {Molina-Ruiz},
  {Mondin}, {Montani}, {Moore}, {Moraru}, {Morawski}, {More}, {Moreno},
  {Moreno}, {Mori}, {Morisaki}, {Moriwaki}, {Morr{\'a}s}, {Mours}, {Mow-Lowry},
  {Mozzon}, {Muciaccia}, {Mukherjee}, {Mukherjee}, {Mukherjee}, {Mukherjee},
  {Mukherjee}, {Mukund}, {Mullavey}, {Munch}, {Mu{\~n}iz}, {Murray},
  {Musenich}, {Muusse}, {Nadji}, {Nagano}, {Nagano}, {Nagar}, {Nakamura},
  {Nakano}, {Nakano}, {Nakashima}, {Nakayama}, {Napolano}, {Nardecchia},
  {Narikawa}, {Naticchioni}, {Nayak}, {Nayak}, {Negishi}, {Neil}, {Neilson},
  {Nelemans}, {Nelson}, {Nery}, {Neubauer}, {Neunzert}, {Ng}, {Ng}, {Nguyen},
  {Nguyen}, {Nguyen}, {Quynh}, {Ni}, {Nichols}, {Nishizawa}, {Nissanke},
  {Nitoglia}, {Nocera}, {Norman}, {North}, {Nozaki}, {Siles}, {Nuttall},
  {Oberling}, {O'Brien}, {Obuchi}, {O'Dell}, {Oelker}, {Ogaki}, {Oganesyan},
  {Oh}, {Oh}, {Oh}, {Ohashi}, {Ohishi}, {Ohkawa}, {Ohme}, {Ohta}, {Okada},
  {Okutani}, {Okutomi}, {Olivetto}, {Oohara}, {Ooi}, {Oram}, {O'Reilly},
  {Ormiston}, {Ormsby}, {Ortega}, {O'Shaughnessy}, {O'Shea}, {Oshino},
  {Ossokine}, {Osthelder}, {Otabe}, {Ottaway}, {Overmier}, {Pace}, {Pagano},
  {Page}, {Pagliaroli}, {Pai}, {Pai}, {Palamos}, {Palashov}, {Palomba}, {Pan},
  {Pan}, {Panda}, {Pang}, {Pang}, {Pankow}, {Pannarale}, {Pant}, {Panther},
  {Paoletti}, {Paoli}, {Paolone}, {Parisi}, {Park}, {Park}, {Parker},
  {Pascucci}, {Pasqualetti}, {Passaquieti}, {Passuello}, {Patel}, {Pathak},
  {Patricelli}, {Patron}, {Paul}, {Payne}, {Pedraza}, {Pegoraro}, {Pele},
  {Arellano}, {Penn}, {Perego}, {Pereira}, {Pereira}, {Perez}, {P{\'e}rigois},
  {Perkins}, {Perreca}, {Perri{\`e}s}, {Petermann}, {Petterson}, {Pfeiffer},
  {Pham}, {Phukon}, {Piccinni}, {Pichot}, {Piendibene}, {Piergiovanni},
  {Pierini}, {Pierro}, {Pillant}, {Pillas}, {Pilo}, {Pinard}, {Pinto}, {Pinto},
  {Piotrzkowski}, {Piotrzkowski}, {Pirello}, {Pitkin}, {Placidi}, {Planas},
  {Plastino}, {Pluchar}, {Poggiani}, {Polini}, {Pong}, {Ponrathnam},
  {Popolizio}, {Porter}, {Poulton}, {Powell}, {Pracchia}, {Pradier},
  {Prajapati}, {Prasai}, {Prasanna}, {Pratten}, {Principe}, {Prodi},
  {Prokhorov}, {Prosposito}, {Prudenzi}, {Puecher}, {Punturo}, {Puosi},
  {Puppo}, {P{\"u}rrer}, {Qi}, {Quetschke}, {Quitzow-James}, {Raab},
  {Raaijmakers}, {Radkins}, {Radulesco}, {Raffai}, {Rail}, {Raja}, {Rajan},
  {Ramirez}, {Ramirez}, {Ramos-Buades}, {Rana}, {Rapagnani}, {Rapol}, {Ray},
  {Raymond}, {Raza}, {Razzano}, {Read}, {Rees}, {Regimbau}, {Rei}, {Reid},
  {Reid}, {Reitze}, {Relton}, {Renzini}, {Rettegno}, {Reza}, {Rezac}, {Ricci},
  {Richards}, {Richardson}, {Richardson}, {Riemenschneider}, {Riles},
  {Rinaldi}, {Rink}, {Rizzo}, {Robertson}, {Robie}, {Robinet}, {Rocchi},
  {Rodriguez}, {Rolland}, {Rollins}, {Romanelli}, {Romano}, {Romel},
  {Romero-Rodr{\'\i}guez}, {Romero-Shaw}, {Romie}, {Ronchini}, {Rosa}, {Rose},
  {Rosi{\'n}ska}, {Ross}, {Rowan}, {Rowlinson}, {Roy}, {Roy}, {Roy}, {Rozza},
  {Ruggi}, {Ryan}, {Sachdev}, {Sadecki}, {Sadiq}, {Sago}, {Saito}, {Saito},
  {Sakai}, {Sakai}, {Sakellariadou}, {Sakuno}, {Salafia}, {Salconi}, {Saleem},
  {Salemi}, {Samajdar}, {Sanchez}, {Sanchez}, {Sanchez}, {Sanchis-Gual},
  {Sanders}, {Sanuy}, {Saravanan}, {Sarin}, {Sassolas}, {Satari},
  {Sathyaprakash}, {Sato}, {Sato}, {Sauter}, {Savage}, {Sawada}, {Sawant},
  {Sawant}, {Sayah}, {Schaetzl}, {Scheel}, {Scheuer}, {Schiworski}, {Schmidt},
  {Schmidt}, {Schnabel}, {Schneewind}, {Schofield}, {Sch{\"o}nbeck}, {Schulte},
  {Schutz}, {Schwartz}, {Scott}, {Scott}, {Seglar-Arroyo}, {Sekiguchi},
  {Sekiguchi}, {Sellers}, {Sengupta}, {Sentenac}, {Seo}, {Sequino}, {Sergeev},
  {Setyawati}, {Shaffer}, {Shahriar}, {Shams}, {Shao}, {Sharma}, {Sharma},
  {Shawhan}, {Shcheblanov}, {Shibagaki}, {Shikauchi}, {Shimizu}, {Shimoda},
  {Shimode}, {Shinkai}, {Shishido}, {Shoda}, {Shoemaker}, {Shoemaker},
  {Shyamsundar}, {Sieniawska}, {Sigg}, {Singer}, {Singh}, {Singh}, {Singha},
  {Sintes}, {Sipala}, {Skliris}, {Slagmolen}, {Slaven-Blair}, {Smetana},
  {Smith}, {Smith}, {Soldateschi}, {Somala}, {Somiya}, {Son}, {Soni}, {Soni},
  {Sordini}, {Sorrentino}, {Sorrentino}, {Sotani}, {Soulard}, {Souradeep},
  {Sowell}, {Spagnuolo}, {Spencer}, {Spera}, {Srinivasan}, {Srivastava},
  {Srivastava}, {Staats}, {Stachie}, {Steer}, {Steinhoff}, {Steinlechner},
  {Steinlechner}, {Stevenson}, {Stops}, {Stover}, {Strain}, {Strang},
  {Stratta}, {Strunk}, {Sturani}, {Stuver}, {Sudhagar}, {Sudhir}, {Sugimoto},
  {Suh}, {Sullivan}, {Summerscales}, {Sun}, {Sun}, {Sunil}, {Sur}, {Suresh},
  {Sutton}, {Suzuki}, {Suzuki}, {Swinkels}, {Szczepa{\'n}czyk}, {Szewczyk},
  {Tacca}, {Tagoshi}, {Tait}, {Takahashi}, {Takahashi}, {Takamori}, {Takano},
  {Takeda}, {Takeda}, {Talbot}, {Talbot}, {Tanaka}, {Tanaka}, {Tanaka},
  {Tanaka}, {Tanaka}, {Tanasijczuk}, {Tanioka}, {Tanner}, {Tao}, {Tao},
  {Mart{\'\i}n}, {Taranto}, {Tasson}, {Telada}, {Tenorio}, {Terhune},
  {Terkowski}, {Thirugnanasambandam}, {Thomas}, {Thomas}, {Thomas}, {Thompson},
  {Thondapu}, {Thorne}, {Thrane}, {Tiwari}, {Tiwari}, {Tiwari}, {Toivonen},
  {Toland}, {Tolley}, {Tomaru}, {Tomigami}, {Tomura}, {Tonelli},
  {Torres-Forn{\'e}}, {Torrie}, {E Melo}, {T{\"o}yr{\"a}}, {Trapananti},
  {Travasso}, {Traylor}, {Trevor}, {Tringali}, {Tripathee}, {Troiano},
  {Trovato}, {Trozzo}, {Trudeau}, {Tsai}, {Tsai}, {Tsang}, {Tsang}, {Tsao},
  {Tse}, {Tso}, {Tsubono}, {Tsuchida}, {Tsukada}, {Tsuna}, {Tsutsui},
  {Tsuzuki}, {Turbang}, {Turconi}, {Tuyenbayev}, {Ubhi}, {Uchikata},
  {Uchiyama}, {Udall}, {Ueda}, {Uehara}, {Ueno}, {Ueshima}, {Unnikrishnan},
  {Uraguchi}, {Urban}, {Ushiba}, {Utina}, {Vahlbruch}, {Vajente}, {Vajpeyi},
  {Valdes}, {Valentini}, {Valsan}, {van Bakel}, {van Beuzekom}, {van den
  Brand}, {van den Broeck}, {Vander-Hyde}, {van der Schaaf}, {van Heijningen},
  {Vanosky}, {van Putten}, {van Remortel}, {Vardaro}, {Vargas}, {Varma},
  {Vas{\'u}th}, {Vecchio}, {Vedovato}, {Veitch}, {Veitch}, {Venneberg},
  {Venugopalan}, {Verkindt}, {Verma}, {Verma}, {Veske}, {Vetrano},
  {Vicer{\'e}}, {Vidyant}, {Viets}, {Vijaykumar}, {Villa-Ortega}, {Vinet},
  {Virtuoso}, {Vitale}, {Vo}, {Vocca}, {von Reis}, {von Wrangel}, {Vorvick},
  {Vyatchanin}, {Wade}, {Wade}, {Wagner}, {Walet}, {Walker}, {Wallace},
  {Wallace}, {Walsh}, {Wang}, {Wang}, {Wang}, {Ward}, {Warner}, {Was},
  {Washimi}, {Washington}, {Watchi}, {Weaver}, {Webster}, {Weinert},
  {Weinstein}, {Weiss}, {Weller}, {Wellmann}, {Wen}, {We{\ss}els}, {Wette},
  {Whelan}, {White}, {Whiting}, {Whittle}, {Wilken}, {Williams}, {Williams},
  {Williamson}, {Willis}, {Willke}, {Wilson}, {Winkler}, {Wipf}, {Wlodarczyk},
  {Woan}, {Woehler}, {Wofford}, {Wong}, {Wu}, {Wu}, {Wu}, {Wu}, {Wysocki},
  {Xiao}, {Xu}, {Yamada}, {Yamamoto}, {Yamamoto}, {Yamamoto}, {Yamamoto},
  {Yamashita}, {Yamazaki}, {Yang}, {Yang}, {Yang}, {Yang}, {Yang}, {Yap},
  {Yeeles}, {Yelikar}, {Ying}, {Yokogawa}, {Yokoyama}, {Yokozawa}, {Yoo},
  {Yoshioka}, {Yu}, {Yu}, {Yuzurihara}, {Zadro{\.z}ny}, {Zanolin}, {Zeidler},
  {Zelenova}, {Zendri}, {Zevin}, {Zhan}, {Zhang}, {Zhang}, {Zhang}, {Zhang},
  {Zhang}, {Zhao}, {Zhao}, {Zhao}, {Zhao}, {Zheng}, {Zhou}, {Zhou}, {Zhu},
  {Zhu}, {Zimmerman}, {Zlochower}, {Zucker}, {Zweizig}, {LIGO Scientific
  Collaboration}, {VIRGO Collaboration}, \& {KAGRA
  Collaboration}}]{2023PhRvX..13a1048A}
{Abbott}, R., {Abbott}, T.~D., {Acernese}, F., {et~al.} 2023, Physical Review
  X, 13, 011048

\bibitem[{{Alexander} \& {Kumar}(2001)}]{2001ApJ...549..948A}
{Alexander}, T. \& {Kumar}, P. 2001, \apj, 549, 948

\bibitem[{{Arca Sedda} {et~al.}(2023){Arca Sedda}, {Kamlah}, {Spurzem},
  {Rizzuto}, {Naab}, {Giersz}, \& {Berczik}}]{2023MNRAS.526..429A}
{Arca Sedda}, M., {Kamlah}, A. W.~H., {Spurzem}, R., {et~al.} 2023, \mnras,
  526, 429

\bibitem[{{Askar} {et~al.}(2018){Askar}, {Arca Sedda}, \&
  {Giersz}}]{2018MNRAS.478.1844A}
{Askar}, A., {Arca Sedda}, M., \& {Giersz}, M. 2018, \mnras, 478, 1844

\bibitem[{{Avakyan} {et~al.}(2023){Avakyan}, {Neumann}, {Zainab}, {Doroshenko},
  {Wilms}, \& {Santangelo}}]{2023A&A...675A.199A}
{Avakyan}, A., {Neumann}, M., {Zainab}, A., {et~al.} 2023, \aap, 675, A199

\bibitem[{{Bagla}(2002)}]{2002JApA...23..185B}
{Bagla}, J.~S. 2002, Journal of Astrophysics and Astronomy, 23, 185

\bibitem[{{Baumgardt} \& {Hilker}(2018)}]{2018MNRAS.478.1520B}
{Baumgardt}, H. \& {Hilker}, M. 2018, \mnras, 478, 1520

\bibitem[{{Bellinger} {et~al.}(2023){Bellinger}, {Ryu}, \&
  {Spruit}}]{BellingerInPrep}
{Bellinger}, E., {Ryu}, T., \& {Spruit}, H. 2023, (in prep)

\bibitem[{{Bellm} {et~al.}(2019){Bellm}, {Kulkarni}, {Graham}, {Dekany},
  {Smith}, {Riddle}, {Masci}, {Helou}, {Prince}, {Adams}, {Barbarino},
  {Barlow}, {Bauer}, {Beck}, {Belicki}, {Biswas}, {Blagorodnova}, {Bodewits},
  {Bolin}, {Brinnel}, {Brooke}, {Bue}, {Bulla}, {Burruss}, {Cenko}, {Chang},
  {Connolly}, {Coughlin}, {Cromer}, {Cunningham}, {De}, {Delacroix}, {Desai},
  {Duev}, {Eadie}, {Farnham}, {Feeney}, {Feindt}, {Flynn}, {Franckowiak},
  {Frederick}, {Fremling}, {Gal-Yam}, {Gezari}, {Giomi}, {Goldstein},
  {Golkhou}, {Goobar}, {Groom}, {Hacopians}, {Hale}, {Henning}, {Ho}, {Hover},
  {Howell}, {Hung}, {Huppenkothen}, {Imel}, {Ip}, {Ivezi{\'c}}, {Jackson},
  {Jones}, {Juric}, {Kasliwal}, {Kaspi}, {Kaye}, {Kelley}, {Kowalski},
  {Kramer}, {Kupfer}, {Landry}, {Laher}, {Lee}, {Lin}, {Lin}, {Lunnan},
  {Giomi}, {Mahabal}, {Mao}, {Miller}, {Monkewitz}, {Murphy}, {Ngeow},
  {Nordin}, {Nugent}, {Ofek}, {Patterson}, {Penprase}, {Porter}, {Rauch},
  {Rebbapragada}, {Reiley}, {Rigault}, {Rodriguez}, {van Roestel}, {Rusholme},
  {van Santen}, {Schulze}, {Shupe}, {Singer}, {Soumagnac}, {Stein}, {Surace},
  {Sollerman}, {Szkody}, {Taddia}, {Terek}, {Van Sistine}, {van Velzen},
  {Vestrand}, {Walters}, {Ward}, {Ye}, {Yu}, {Yan}, \&
  {Zolkower}}]{2019PASP..131a8002B}
{Bellm}, E.~C., {Kulkarni}, S.~R., {Graham}, M.~J., {et~al.} 2019, \pasp, 131,
  018002

\bibitem[{{Bloom} {et~al.}(2011){Bloom}, {Giannios}, {Metzger}, {Cenko},
  {Perley}, {Butler}, {Tanvir}, {Levan}, {O'Brien}, {Strubbe}, {De Colle},
  {Ramirez-Ruiz}, {Lee}, {Nayakshin}, {Quataert}, {King}, {Cucchiara},
  {Guillochon}, {Bower}, {Fruchter}, {Morgan}, \& {van der
  Horst}}]{2011Sci...333..203B}
{Bloom}, J.~S., {Giannios}, D., {Metzger}, B.~D., {et~al.} 2011, Science, 333,
  203

\bibitem[{{Brown} {et~al.}(2015){Brown}, {Levan}, {Stanway}, {Tanvir}, {Cenko},
  {Berger}, {Chornock}, \& {Cucchiaria}}]{2015MNRAS.452.4297B}
{Brown}, G.~C., {Levan}, A.~J., {Stanway}, E.~R., {et~al.} 2015, \mnras, 452,
  4297

\bibitem[{{Burmester} {et~al.}(2023){Burmester}, {Ferrario}, {Pakmor},
  {Seitenzahl}, {Ruiter}, \& {Hole}}]{2023MNRAS.523..527B}
{Burmester}, U.~P., {Ferrario}, L., {Pakmor}, R., {et~al.} 2023, \mnras, 523,
  527

\bibitem[{{Burrows} {et~al.}(2011){Burrows}, {Kennea}, {Ghisellini}, {Mangano},
  {Zhang}, {Page}, {Eracleous}, {Romano}, {Sakamoto}, {Falcone}, {Osborne},
  {Campana}, {Beardmore}, {Breeveld}, {Chester}, {Corbet}, {Covino},
  {Cummings}, {D'Avanzo}, {D'Elia}, {Esposito}, {Evans}, {Fugazza}, {Gelbord},
  {Hiroi}, {Holland}, {Huang}, {Im}, {Israel}, {Jeon}, {Jeon}, {Jun}, {Kawai},
  {Kim}, {Krimm}, {Marshall}, {P. M{\'e}sz{\'a}ros}, {Negoro}, {Omodei},
  {Park}, {Perkins}, {Sugizaki}, {Sung}, {Tagliaferri}, {Troja}, {Ueda},
  {Urata}, {Usui}, {Antonelli}, {Barthelmy}, {Cusumano}, {Giommi}, {Melandri},
  {Perri}, {Racusin}, {Sbarufatti}, {Siegel}, \&
  {Gehrels}}]{2011Natur.476..421B}
{Burrows}, D.~N., {Kennea}, J.~A., {Ghisellini}, G., {et~al.} 2011, \nat, 476,
  421

\bibitem[{{Cenko} {et~al.}(2012){Cenko}, {Krimm}, {Horesh}, {Rau}, {Frail},
  {Kennea}, {Levan}, {Holland}, {Butler}, {Quimby}, {Bloom}, {Filippenko},
  {Gal-Yam}, {Greiner}, {Kulkarni}, {Ofek}, {Olivares E.}, {Schady},
  {Silverman}, {Tanvir}, \& {Xu}}]{2012ApJ...753...77C}
{Cenko}, S.~B., {Krimm}, H.~A., {Horesh}, A., {et~al.} 2012, \apj, 753, 77

\bibitem[{{Chakrabarti} {et~al.}(2023){Chakrabarti}, {Simon}, {Craig},
  {Reggiani}, {Brandt}, {Guhathakurta}, {Dalba}, {Kirby}, {Chang}, {Hey},
  {Savino}, {Geha}, \& {Thompson}}]{2023AJ....166....6C}
{Chakrabarti}, S., {Simon}, J.~D., {Craig}, P.~A., {et~al.} 2023, \aj, 166, 6

\bibitem[{{Chambers} {et~al.}(2016){Chambers}, {Magnier}, {Metcalfe},
  {Flewelling}, {Huber}, {Waters}, {Denneau}, {Draper}, {Farrow}, {Finkbeiner},
  {Holmberg}, {Koppenhoefer}, {Price}, {Rest}, {Saglia}, {Schlafly}, {Smartt},
  {Sweeney}, {Wainscoat}, {Burgett}, {Chastel}, {Grav}, {Heasley}, {Hodapp},
  {Jedicke}, {Kaiser}, {Kudritzki}, {Luppino}, {Lupton}, {Monet}, {Morgan},
  {Onaka}, {Shiao}, {Stubbs}, {Tonry}, {White}, {Ba{\~n}ados}, {Bell},
  {Bender}, {Bernard}, {Boegner}, {Boffi}, {Botticella}, {Calamida},
  {Casertano}, {Chen}, {Chen}, {Cole}, {Deacon}, {Frenk}, {Fitzsimmons},
  {Gezari}, {Gibbs}, {Goessl}, {Goggia}, {Gourgue}, {Goldman}, {Grant},
  {Grebel}, {Hambly}, {Hasinger}, {Heavens}, {Heckman}, {Henderson}, {Henning},
  {Holman}, {Hopp}, {Ip}, {Isani}, {Jackson}, {Keyes}, {Koekemoer}, {Kotak},
  {Le}, {Liska}, {Long}, {Lucey}, {Liu}, {Martin}, {Masci}, {McLean}, {Mindel},
  {Misra}, {Morganson}, {Murphy}, {Obaika}, {Narayan}, {Nieto-Santisteban},
  {Norberg}, {Peacock}, {Pier}, {Postman}, {Primak}, {Rae}, {Rai}, {Riess},
  {Riffeser}, {Rix}, {R{\"o}ser}, {Russel}, {Rutz}, {Schilbach}, {Schultz},
  {Scolnic}, {Strolger}, {Szalay}, {Seitz}, {Small}, {Smith}, {Soderblom},
  {Taylor}, {Thomson}, {Taylor}, {Thakar}, {Thiel}, {Thilker}, {Unger},
  {Urata}, {Valenti}, {Wagner}, {Walder}, {Walter}, {Watters}, {Werner},
  {Wood-Vasey}, \& {Wyse}}]{2016arXiv161205560C}
{Chambers}, K.~C., {Magnier}, E.~A., {Metcalfe}, N., {et~al.} 2016, arXiv
  e-prints, arXiv:1612.05560

\bibitem[{{Chrimes} {et~al.}(2024){Chrimes}, {Jonker}, {Levan}, {Coppejans},
  {Gaspari}, {Gompertz}, {Groot}, {Malesani}, {Mummery}, {Stanway}, \&
  {Wiersema}}]{2024MNRAS.527L..47C}
{Chrimes}, A.~A., {Jonker}, P.~G., {Levan}, A.~J., {et~al.} 2024, \mnras, 527,
  L47

\bibitem[{{Coppejans} {et~al.}(2020){Coppejans}, {Margutti}, {Terreran},
  {Nayana}, {Coughlin}, {Laskar}, {Alexander}, {Bietenholz}, {Caprioli},
  {Chandra}, {Drout}, {Frederiks}, {Frohmaier}, {Hurley}, {Kochanek},
  {MacLeod}, {Meisner}, {Nugent}, {Ridnaia}, {Sand}, {Svinkin}, {Ward}, {Yang},
  {Baldeschi}, {Chilingarian}, {Dong}, {Esquivia}, {Fong}, {Guidorzi},
  {Lundqvist}, {Milisavljevic}, {Paterson}, {Reichart}, {Shappee}, {Stroh},
  {Valenti}, {Zauderer}, \& {Zhang}}]{2020ApJ...895L..23C}
{Coppejans}, D.~L., {Margutti}, R., {Terreran}, G., {et~al.} 2020, \apjl, 895,
  L23

\bibitem[{{Curd} \& {Narayan}(2023)}]{2023MNRAS.518.3441C}
{Curd}, B. \& {Narayan}, R. 2023, \mnras, 518, 3441

\bibitem[{{Davis} {et~al.}(2011){Davis}, {Narayan}, {Zhu}, {Barret}, {Farrell},
  {Godet}, {Servillat}, \& {Webb}}]{2011ApJ...734..111D}
{Davis}, S.~W., {Narayan}, R., {Zhu}, Y., {et~al.} 2011, \apj, 734, 111

\bibitem[{{De Angeli} {et~al.}(2005){De Angeli}, {Piotto}, {Cassisi}, {Busso},
  {Recio-Blanco}, {Salaris}, {Aparicio}, \& {Rosenberg}}]{2005AJ....130..116D}
{De Angeli}, F., {Piotto}, G., {Cassisi}, S., {et~al.} 2005, \aj, 130, 116

\bibitem[{{Drout} {et~al.}(2014){Drout}, {Chornock}, {Soderberg}, {Sanders},
  {McKinnon}, {Rest}, {Foley}, {Milisavljevic}, {Margutti}, {Berger},
  {Calkins}, {Fong}, {Gezari}, {Huber}, {Kankare}, {Kirshner}, {Leibler},
  {Lunnan}, {Mattila}, {Marion}, {Narayan}, {Riess}, {Roth}, {Scolnic},
  {Smartt}, {Tonry}, {Burgett}, {Chambers}, {Hodapp}, {Jedicke}, {Kaiser},
  {Magnier}, {Metcalfe}, {Morgan}, {Price}, \& {Waters}}]{2014ApJ...794...23D}
{Drout}, M.~R., {Chornock}, R., {Soderberg}, A.~M., {et~al.} 2014, \apj, 794,
  23

\bibitem[{{El-Badry} {et~al.}(2023{\natexlab{a}}){El-Badry}, {Rix}, {Cendes},
  {Rodriguez}, {Conroy}, {Quataert}, {Hawkins}, {Zari}, {Hobson}, {Breivik},
  {Rau}, {Berger}, {Shahaf}, {Seeburger}, {Burdge}, {Latham}, {Buchhave},
  {Bieryla}, {Bashi}, {Mazeh}, \& {Faigler}}]{2023MNRAS.521.4323E}
{El-Badry}, K., {Rix}, H.-W., {Cendes}, Y., {et~al.} 2023{\natexlab{a}},
  \mnras, 521, 4323

\bibitem[{{El-Badry} {et~al.}(2023{\natexlab{b}}){El-Badry}, {Rix}, {Quataert},
  {Howard}, {Isaacson}, {Fuller}, {Hawkins}, {Breivik}, {Wong}, {Rodriguez},
  {Conroy}, {Shahaf}, {Mazeh}, {Arenou}, {Burdge}, {Bashi}, {Faigler}, {Weisz},
  {Seeburger}, {Almada Monter}, \& {Wojno}}]{2023MNRAS.518.1057E}
{El-Badry}, K., {Rix}, H.-W., {Quataert}, E., {et~al.} 2023{\natexlab{b}},
  \mnras, 518, 1057

\bibitem[{{Farrell} {et~al.}(2014){Farrell}, {Servillat}, {Gladstone}, {Webb},
  {Soria}, {Maccarone}, {Wiersema}, {Hau}, {Pforr}, {Hakala}, {Knigge},
  {Barret}, {Maraston}, \& {Kong}}]{2014MNRAS.437.1208F}
{Farrell}, S.~A., {Servillat}, M., {Gladstone}, J.~C., {et~al.} 2014, \mnras,
  437, 1208

\bibitem[{{Figer} {et~al.}(2003){Figer}, {Gilmore}, {Kim}, {Morris}, {Becklin},
  {McLean}, {Gilbert}, {Graham}, {Larkin}, {Levenson}, \&
  {Teplitz}}]{2003ApJ...599.1139F}
{Figer}, D.~F., {Gilmore}, D., {Kim}, S.~S., {et~al.} 2003, \apj, 599, 1139

\bibitem[{{Filippenko} \& {Sargent}(1989)}]{1989ApJ...342L..11F}
{Filippenko}, A.~V. \& {Sargent}, W. L.~W. 1989, \apjl, 342, L11

\bibitem[{{Gallegos-Garcia} {et~al.}(2018){Gallegos-Garcia}, {Law-Smith}, \&
  {Ramirez-Ruiz}}]{2018ApJ...857..109G}
{Gallegos-Garcia}, M., {Law-Smith}, J., \& {Ramirez-Ruiz}, E. 2018, \apj, 857,
  109

\bibitem[{{Gehrels} {et~al.}(2004){Gehrels}, {Chincarini}, {Giommi}, {Mason},
  {Nousek}, {Wells}, {White}, {Barthelmy}, {Burrows}, {Cominsky}, {Hurley},
  {Marshall}, {M{\'e}sz{\'a}ros}, {Roming}, {Angelini}, {Barbier}, {Belloni},
  {Campana}, {Caraveo}, {Chester}, {Citterio}, {Cline}, {Cropper}, {Cummings},
  {Dean}, {Feigelson}, {Fenimore}, {Frail}, {Fruchter}, {Garmire}, {Gendreau},
  {Ghisellini}, {Greiner}, {Hill}, {Hunsberger}, {Krimm}, {Kulkarni}, {Kumar},
  {Lebrun}, {Lloyd-Ronning}, {Markwardt}, {Mattson}, {Mushotzky}, {Norris},
  {Osborne}, {Paczynski}, {Palmer}, {Park}, {Parsons}, {Paul}, {Rees},
  {Reynolds}, {Rhoads}, {Sasseen}, {Schaefer}, {Short}, {Smale}, {Smith},
  {Stella}, {Tagliaferri}, {Takahashi}, {Tashiro}, {Townsley}, {Tueller},
  {Turner}, {Vietri}, {Voges}, {Ward}, {Willingale}, {Zerbi}, \&
  {Zhang}}]{2004ApJ...611.1005G}
{Gehrels}, N., {Chincarini}, G., {Giommi}, P., {et~al.} 2004, \apj, 611, 1005

\bibitem[{{Gezari}(2021)}]{2021ARA&A..59...21G}
{Gezari}, S. 2021, \araa, 59, 21

\bibitem[{{Giannios} \& {Metzger}(2011)}]{2011MNRAS.416.2102G}
{Giannios}, D. \& {Metzger}, B.~D. 2011, \mnras, 416, 2102

\bibitem[{{Glanz} {et~al.}(2023){Glanz}, {Perets}, \&
  {Pakmor}}]{2023arXiv230903300G}
{Glanz}, H., {Perets}, H.~B., \& {Pakmor}, R. 2023, arXiv e-prints,
  arXiv:2309.03300

\bibitem[{{Goicovic} {et~al.}(2019){Goicovic}, {Springel}, {Ohlmann}, \&
  {Pakmor}}]{2019MNRAS.487..981G}
{Goicovic}, F.~G., {Springel}, V., {Ohlmann}, S.~T., \& {Pakmor}, R. 2019,
  \mnras, 487, 981

\bibitem[{{Golightly} {et~al.}(2019){Golightly}, {Nixon}, \&
  {Coughlin}}]{2019ApJ...882L..26G}
{Golightly}, E.~C.~A., {Nixon}, C.~J., \& {Coughlin}, E.~R. 2019, \apjl, 882,
  L26

\bibitem[{{Greene} {et~al.}(2020){Greene}, {Strader}, \&
  {Ho}}]{2020ARA&A..58..257G}
{Greene}, J.~E., {Strader}, J., \& {Ho}, L.~C. 2020, \araa, 58, 257

\bibitem[{{Gronow} {et~al.}(2021){Gronow}, {Collins}, {Sim}, \&
  {R{\"o}pke}}]{2021A&A...649A.155G}
{Gronow}, S., {Collins}, C.~E., {Sim}, S.~A., \& {R{\"o}pke}, F.~K. 2021, \aap,
  649, A155

\bibitem[{{Guillochon} \& {Ramirez-Ruiz}(2013)}]{2013ApJ...767...25G}
{Guillochon}, J. \& {Ramirez-Ruiz}, E. 2013, \apj, 767, 25

\bibitem[{{G{\"u}ltekin} {et~al.}(2022){G{\"u}ltekin}, {Nyland}, {Gray},
  {Fehmer}, {Huang}, {Sparkman}, {Reines}, {Greene}, {Cackett}, \&
  {Baldassare}}]{2022MNRAS.516.6123G}
{G{\"u}ltekin}, K., {Nyland}, K., {Gray}, N., {et~al.} 2022, \mnras, 516, 6123

\bibitem[{{Harris}(2010)}]{2010arXiv1012.3224H}
{Harris}, W.~E. 2010, arXiv e-prints, arXiv:1012.3224

\bibitem[{{Ho} {et~al.}(2020){Ho}, {Perley}, {Kulkarni}, {Dong}, {De},
  {Chandra}, {Andreoni}, {Bellm}, {Burdge}, {Coughlin}, {Dekany}, {Feeney},
  {Frederiks}, {Fremling}, {Golkhou}, {Graham}, {Hale}, {Helou}, {Horesh},
  {Kasliwal}, {Laher}, {Masci}, {Miller}, {Porter}, {Ridnaia}, {Rusholme},
  {Shupe}, {Soumagnac}, \& {Svinkin}}]{2020ApJ...895...49H}
{Ho}, A. Y.~Q., {Perley}, D.~A., {Kulkarni}, S.~R., {et~al.} 2020, \apj, 895,
  49

\bibitem[{{Jansen} {et~al.}(2001){Jansen}, {Lumb}, {Altieri}, {Clavel}, {Ehle},
  {Erd}, {Gabriel}, {Guainazzi}, {Gondoin}, {Much}, {Munoz}, {Santos},
  {Schartel}, {Texier}, \& {Vacanti}}]{2001A&A...365L...1J}
{Jansen}, F., {Lumb}, D., {Altieri}, B., {et~al.} 2001, \aap, 365, L1

\bibitem[{{Jiang} {et~al.}(2019){Jiang}, {Stone}, \&
  {Davis}}]{2019ApJ...880...67J}
{Jiang}, Y.-F., {Stone}, J.~M., \& {Davis}, S.~W. 2019, \apj, 880, 67

\bibitem[{{Kaaz} {et~al.}(2023){Kaaz}, {Murguia-Berthier}, {Chatterjee},
  {Liska}, \& {Tchekhovskoy}}]{2023ApJ...950...31K}
{Kaaz}, N., {Murguia-Berthier}, A., {Chatterjee}, K., {Liska}, M. T.~P., \&
  {Tchekhovskoy}, A. 2023, \apj, 950, 31

\bibitem[{{Kalogera}(1999)}]{1999ApJ...521..723K}
{Kalogera}, V. 1999, \apj, 521, 723

\bibitem[{{Kiel} \& {Hurley}(2006)}]{2006MNRAS.369.1152K}
{Kiel}, P.~D. \& {Hurley}, J.~R. 2006, \mnras, 369, 1152

\bibitem[{{K{\i}ro{\u{g}}lu} {et~al.}(2023){K{\i}ro{\u{g}}lu}, {Lombardi},
  {Kremer}, {Fragione}, {Fogarty}, \& {Rasio}}]{2023ApJ...948...89K}
{K{\i}ro{\u{g}}lu}, F., {Lombardi}, J.~C., {Kremer}, K., {et~al.} 2023, \apj,
  948, 89

\bibitem[{{Klencki} {et~al.}(2017){Klencki}, {Wiktorowicz}, {G{\l}adysz}, \&
  {Belczynski}}]{2017MNRAS.469.3088K}
{Klencki}, J., {Wiktorowicz}, G., {G{\l}adysz}, W., \& {Belczynski}, K. 2017,
  \mnras, 469, 3088

\bibitem[{{Kochanek} {et~al.}(2017){Kochanek}, {Shappee}, {Stanek}, {Holoien},
  {Thompson}, {Prieto}, {Dong}, {Shields}, {Will}, {Britt}, {Perzanowski}, \&
  {Pojma{\'n}ski}}]{2017PASP..129j4502K}
{Kochanek}, C.~S., {Shappee}, B.~J., {Stanek}, K.~Z., {et~al.} 2017, \pasp,
  129, 104502

\bibitem[{{Kollmeier} {et~al.}(2019){Kollmeier}, {Anderson}, {Blanc},
  {Blanton}, {Covey}, {Crane}, {Drory}, {Frinchaboy}, {Froning}, {Johnson},
  {Kneib}, {Kreckel}, {Merloni}, {Pellegrini}, {Pogge}, {Ramirez}, {Rix},
  {Sayres}, {S{\'a}nchez-Gallego}, {Shen}, {Tkachenko}, {Trump}, {Tuttle},
  {Weijmans}, {Zasowski}, {Barbuy}, {Beaton}, {Bergemann}, {Bochanski},
  {Brandt}, {Casey}, {Cherinka}, {Eracleous}, {Fan}, {Garc{\'\i}a}, {Green},
  {Hekker}, {Lane}, {Longa-Pe{\~n}a}, {Mathur}, {Meza}, {Minchev}, {Myers},
  {Nidever}, {Nitschelm}, {O'Connell}, {Price-Whelan}, {Raddick}, {Rossi},
  {Sankrit}, {Simon}, {Stutz}, {Ting}, {Trakhtenbrot}, {Weaver}, {Willmer}, \&
  {Weinberg}}]{2019BAAS...51g.274K}
{Kollmeier}, J., {Anderson}, S.~F., {Blanc}, G.~A., {et~al.} 2019, in Bulletin
  of the American Astronomical Society, Vol.~51, 274

\bibitem[{{Kramer} {et~al.}(2020){Kramer}, {Schneider}, {Ohlmann}, {Geier},
  {Schaffenroth}, {Pakmor}, \& {R{\"o}pke}}]{2020A&A...642A..97K}
{Kramer}, M., {Schneider}, F.~R.~N., {Ohlmann}, S.~T., {et~al.} 2020, \aap,
  642, A97

\bibitem[{{Kremer} {et~al.}(2022){Kremer}, {Lombardi}, {Lu}, {Piro}, \&
  {Rasio}}]{2022ApJ...933..203K}
{Kremer}, K., {Lombardi}, J.~C., {Lu}, W., {Piro}, A.~L., \& {Rasio}, F.~A.
  2022, \apj, 933, 203

\bibitem[{{Kremer} {et~al.}(2021){Kremer}, {Lu}, {Piro}, {Chatterjee}, {Rasio},
  \& {Ye}}]{2021ApJ...911..104K}
{Kremer}, K., {Lu}, W., {Piro}, A.~L., {et~al.} 2021, \apj, 911, 104

\bibitem[{{Kuin} {et~al.}(2019){Kuin}, {Wu}, {Oates}, {Lien}, {Emery},
  {Kennea}, {de Pasquale}, {Han}, {Brown}, {Tohuvavohu}, {Breeveld}, {Burrows},
  {Cenko}, {Campana}, {Levan}, {Markwardt}, {Osborne}, {Page}, {Page},
  {Sbarufatti}, {Siegel}, \& {Troja}}]{2019MNRAS.487.2505K}
{Kuin}, N. P.~M., {Wu}, K., {Oates}, S., {et~al.} 2019, \mnras, 487, 2505

\bibitem[{{Kunth} {et~al.}(1987){Kunth}, {Sargent}, \&
  {Bothun}}]{1987AJ.....93...29K}
{Kunth}, D., {Sargent}, W.~L.~W., \& {Bothun}, G.~D. 1987, \aj, 93, 29

\bibitem[{{Law} {et~al.}(2009){Law}, {Kulkarni}, {Dekany}, {Ofek}, {Quimby},
  {Nugent}, {Surace}, {Grillmair}, {Bloom}, {Kasliwal}, {Bildsten}, {Brown},
  {Cenko}, {Ciardi}, {Croner}, {Djorgovski}, {van Eyken}, {Filippenko}, {Fox},
  {Gal-Yam}, {Hale}, {Hamam}, {Helou}, {Henning}, {Howell}, {Jacobsen},
  {Laher}, {Mattingly}, {McKenna}, {Pickles}, {Poznanski}, {Rahmer}, {Rau},
  {Rosing}, {Shara}, {Smith}, {Starr}, {Sullivan}, {Velur}, {Walters}, \&
  {Zolkower}}]{2009PASP..121.1395L}
{Law}, N.~M., {Kulkarni}, S.~R., {Dekany}, R.~G., {et~al.} 2009, \pasp, 121,
  1395

\bibitem[{{Law-Smith} {et~al.}(2019){Law-Smith}, {Guillochon}, \&
  {Ramirez-Ruiz}}]{2019ApJ...882L..25L}
{Law-Smith}, J., {Guillochon}, J., \& {Ramirez-Ruiz}, E. 2019, \apjl, 882, L25

\bibitem[{{Law-Smith} {et~al.}(2020){Law-Smith}, {Coulter}, {Guillochon},
  {Mockler}, \& {Ramirez-Ruiz}}]{2020ApJ...905..141L}
{Law-Smith}, J. A.~P., {Coulter}, D.~A., {Guillochon}, J., {Mockler}, B., \&
  {Ramirez-Ruiz}, E. 2020, \apj, 905, 141

\bibitem[{{Lee} \& {Ostriker}(1986)}]{1986ApJ...310..176L}
{Lee}, H.~M. \& {Ostriker}, J.~P. 1986, \apj, 310, 176

\bibitem[{{Levan} {et~al.}(2014){Levan}, {Tanvir}, {Starling}, {Wiersema},
  {Page}, {Perley}, {Schulze}, {Wynn}, {Chornock}, {Hjorth}, {Cenko},
  {Fruchter}, {O'Brien}, {Brown}, {Tunnicliffe}, {Malesani}, {Jakobsson},
  {Watson}, {Berger}, {Bersier}, {Cobb}, {Covino}, {Cucchiara}, {de Ugarte
  Postigo}, {Fox}, {Gal-Yam}, {Goldoni}, {Gorosabel}, {Kaper}, {Kr{\"u}hler},
  {Karjalainen}, {Osborne}, {Pian}, {S{\'a}nchez-Ram{\'\i}rez}, {Schmidt},
  {Skillen}, {Tagliaferri}, {Th{\"o}ne}, {Vaduvescu}, {Wijers}, \&
  {Zauderer}}]{2014ApJ...781...13L}
{Levan}, A.~J., {Tanvir}, N.~R., {Starling}, R.~L.~C., {et~al.} 2014, \apj,
  781, 13

\bibitem[{{Lioutas} {et~al.}(2024){Lioutas}, {Bauswein}, {Soultanis}, {Pakmor},
  {Springel}, \& {R{\"o}pke}}]{2024MNRAS.528.1906L}
{Lioutas}, G., {Bauswein}, A., {Soultanis}, T., {et~al.} 2024, \mnras, 528,
  1906

\bibitem[{{Liu} {et~al.}(2007){Liu}, {van Paradijs}, \& {van den
  Heuvel}}]{2007A&A...469..807L}
{Liu}, Q.~Z., {van Paradijs}, J., \& {van den Heuvel}, E.~P.~J. 2007, \aap,
  469, 807

\bibitem[{{Lopez} {et~al.}(2019){Lopez}, {Batta}, {Ramirez-Ruiz}, {Martinez},
  \& {Samsing}}]{2019ApJ...877...56L}
{Lopez}, Martin, J., {Batta}, A., {Ramirez-Ruiz}, E., {Martinez}, I., \&
  {Samsing}, J. 2019, \apj, 877, 56

\bibitem[{{Mainetti} {et~al.}(2017){Mainetti}, {Lupi}, {Campana}, {Colpi},
  {Coughlin}, {Guillochon}, \& {Ramirez-Ruiz}}]{2017A&A...600A.124M}
{Mainetti}, D., {Lupi}, A., {Campana}, S., {et~al.} 2017, \aap, 600, A124

\bibitem[{{Manukian} {et~al.}(2013){Manukian}, {Guillochon}, {Ramirez-Ruiz}, \&
  {O'Leary}}]{2013ApJ...771L..28M}
{Manukian}, H., {Guillochon}, J., {Ramirez-Ruiz}, E., \& {O'Leary}, R.~M. 2013,
  \apjl, 771, L28

\bibitem[{{Margutti} {et~al.}(2019){Margutti}, {Metzger}, {Chornock}, {Vurm},
  {Roth}, {Grefenstette}, {Savchenko}, {Cartier}, {Steiner}, {Terreran},
  {Margalit}, {Migliori}, {Milisavljevic}, {Alexander}, {Bietenholz},
  {Blanchard}, {Bozzo}, {Brethauer}, {Chilingarian}, {Coppejans}, {Ducci},
  {Ferrigno}, {Fong}, {G{\"o}tz}, {Guidorzi}, {Hajela}, {Hurley}, {Kuulkers},
  {Laurent}, {Mereghetti}, {Nicholl}, {Patnaude}, {Ubertini}, {Banovetz},
  {Bartel}, {Berger}, {Coughlin}, {Eftekhari}, {Frederiks}, {Kozlova},
  {Laskar}, {Svinkin}, {Drout}, {MacFadyen}, \&
  {Paterson}}]{2019ApJ...872...18M}
{Margutti}, R., {Metzger}, B.~D., {Chornock}, R., {et~al.} 2019, \apj, 872, 18

\bibitem[{{Mar{\'\i}n-Franch} {et~al.}(2009){Mar{\'\i}n-Franch}, {Aparicio},
  {Piotto}, {Rosenberg}, {Chaboyer}, {Sarajedini}, {Siegel}, {Anderson},
  {Bedin}, {Dotter}, {Hempel}, {King}, {Majewski}, {Milone}, {Paust}, \&
  {Reid}}]{2009ApJ...694.1498M}
{Mar{\'\i}n-Franch}, A., {Aparicio}, A., {Piotto}, G., {et~al.} 2009, \apj,
  694, 1498

\bibitem[{{Martin} {et~al.}(2005){Martin}, {Fanson}, {Schiminovich},
  {Morrissey}, {Friedman}, {Barlow}, {Conrow}, {Grange}, {Jelinsky},
  {Milliard}, {Siegmund}, {Bianchi}, {Byun}, {Donas}, {Forster}, {Heckman},
  {Lee}, {Madore}, {Malina}, {Neff}, {Rich}, {Small}, {Surber}, {Szalay},
  {Welsh}, \& {Wyder}}]{2005ApJ...619L...1M}
{Martin}, D.~C., {Fanson}, J., {Schiminovich}, D., {et~al.} 2005, \apjl, 619,
  L1

\bibitem[{{Matthews} {et~al.}(2023){Matthews}, {Margutti}, {Metzger},
  {Milisavljevic}, {Migliori}, {Laskar}, {Brethauer}, {Berger}, {Chornock},
  {Drout}, \& {Ramirez-Ruiz}}]{2023RNAAS...7..126M}
{Matthews}, D., {Margutti}, R., {Metzger}, B.~D., {et~al.} 2023, Research Notes
  of the American Astronomical Society, 7, 126

\bibitem[{{Metzger}(2022)}]{2022ApJ...932...84M}
{Metzger}, B.~D. 2022, \apj, 932, 84

\bibitem[{{Michaely} {et~al.}(2016){Michaely}, {Ginzburg}, \&
  {Perets}}]{2016arXiv161000593M}
{Michaely}, E., {Ginzburg}, D., \& {Perets}, H.~B. 2016, arXiv e-prints,
  arXiv:1610.00593

\bibitem[{{Michaely} \& {Perets}(2016)}]{2016MNRAS.458.4188M}
{Michaely}, E. \& {Perets}, H.~B. 2016, \mnras, 458, 4188

\bibitem[{{Michaely} \& {Perets}(2020)}]{2020MNRAS.498.4924M}
{Michaely}, E. \& {Perets}, H.~B. 2020, \mnras, 498, 4924

\bibitem[{{Morscher} {et~al.}(2015){Morscher}, {Pattabiraman}, {Rodriguez},
  {Rasio}, \& {Umbreit}}]{2015ApJ...800....9M}
{Morscher}, M., {Pattabiraman}, B., {Rodriguez}, C., {Rasio}, F.~A., \&
  {Umbreit}, S. 2015, \apj, 800, 9

\bibitem[{{Ofek} {et~al.}(2021){Ofek}, {Adams}, {Waxman}, {Sharon}, {Kushnir},
  {Horesh}, {Ho}, {Kasliwal}, {Yaron}, {Gal-Yam}, {Kulkarni}, {Bellm}, {Masci},
  {Shupe}, {Dekany}, {Graham}, {Riddle}, {Duev}, {Andreoni}, {Mahabal}, \&
  {Drake}}]{2021ApJ...922..247O}
{Ofek}, E.~O., {Adams}, S.~M., {Waxman}, E., {et~al.} 2021, \apj, 922, 247

\bibitem[{{Ohlmann} {et~al.}(2016{\natexlab{a}}){Ohlmann}, {R{\"o}pke},
  {Pakmor}, \& {Springel}}]{2016ApJ...816L...9O}
{Ohlmann}, S.~T., {R{\"o}pke}, F.~K., {Pakmor}, R., \& {Springel}, V.
  2016{\natexlab{a}}, \apjl, 816, L9

\bibitem[{{Ohlmann} {et~al.}(2017){Ohlmann}, {R{\"o}pke}, {Pakmor}, \&
  {Springel}}]{2017A&A...599A...5O}
{Ohlmann}, S.~T., {R{\"o}pke}, F.~K., {Pakmor}, R., \& {Springel}, V. 2017,
  \aap, 599, A5

\bibitem[{{Ohlmann} {et~al.}(2016{\natexlab{b}}){Ohlmann}, {R{\"o}pke},
  {Pakmor}, {Springel}, \& {M{\"u}ller}}]{2016MNRAS.462L.121O}
{Ohlmann}, S.~T., {R{\"o}pke}, F.~K., {Pakmor}, R., {Springel}, V., \&
  {M{\"u}ller}, E. 2016{\natexlab{b}}, \mnras, 462, L121

\bibitem[{{Ondratschek} {et~al.}(2022){Ondratschek}, {R{\"o}pke}, {Schneider},
  {Fendt}, {Sand}, {Ohlmann}, {Pakmor}, \& {Springel}}]{2022A&A...660L...8O}
{Ondratschek}, P.~A., {R{\"o}pke}, F.~K., {Schneider}, F. R.~N., {et~al.} 2022,
  \aap, 660, L8

\bibitem[{{Pakmor} {et~al.}(2022){Pakmor}, {Callan}, {Collins}, {de Mink},
  {Holas}, {Kerzendorf}, {Kromer}, {Neunteufel}, {O'Brien}, {R{\"o}pke},
  {Ruiter}, {Seitenzahl}, {Shingles}, {Sim}, \&
  {Taubenberger}}]{2022MNRAS.517.5260P}
{Pakmor}, R., {Callan}, F.~P., {Collins}, C.~E., {et~al.} 2022, \mnras, 517,
  5260

\bibitem[{{Pakmor} {et~al.}(2013){Pakmor}, {Kromer}, {Taubenberger}, \&
  {Springel}}]{2013ApJ...770L...8P}
{Pakmor}, R., {Kromer}, M., {Taubenberger}, S., \& {Springel}, V. 2013, \apjl,
  770, L8

\bibitem[{{Pakmor} {et~al.}(2016){Pakmor}, {Springel}, {Bauer}, {Mocz},
  {Munoz}, {Ohlmann}, {Schaal}, \& {Zhu}}]{2016MNRAS.455.1134P}
{Pakmor}, R., {Springel}, V., {Bauer}, A., {et~al.} 2016, \mnras, 455, 1134

\bibitem[{{Pakmor} {et~al.}(2021){Pakmor}, {Zenati}, {Perets}, \&
  {Toonen}}]{2021MNRAS.503.4734P}
{Pakmor}, R., {Zenati}, Y., {Perets}, H.~B., \& {Toonen}, S. 2021, \mnras, 503,
  4734

\bibitem[{{Paxton} {et~al.}(2011){Paxton}, {Bildsten}, {Dotter}, {Herwig},
  {Lesaffre}, \& {Timmes}}]{2011ApJS..192....3P}
{Paxton}, B., {Bildsten}, L., {Dotter}, A., {et~al.} 2011, \apjs, 192, 3

\bibitem[{{Pechetti} {et~al.}(2022){Pechetti}, {Seth}, {Kamann}, {Caldwell},
  {Strader}, {den Brok}, {Luetzgendorf}, {Neumayer}, \&
  {Voggel}}]{2022ApJ...924...48P}
{Pechetti}, R., {Seth}, A., {Kamann}, S., {et~al.} 2022, \apj, 924, 48

\bibitem[{{Perets} {et~al.}(2011){Perets}, {Badenes}, {Arcavi}, {Simon}, \&
  {Gal-yam}}]{2011ApJ...730...89P}
{Perets}, H.~B., {Badenes}, C., {Arcavi}, I., {Simon}, J.~D., \& {Gal-yam}, A.
  2011, \apj, 730, 89

\bibitem[{{Perets} {et~al.}(2016){Perets}, {Li}, {Lombardi}, \&
  {Milcarek}}]{2016ApJ...823..113P}
{Perets}, H.~B., {Li}, Z., {Lombardi}, James~C., J., \& {Milcarek}, Stephen~R.,
  J. 2016, \apj, 823, 113

\bibitem[{{Perley} {et~al.}(2021){Perley}, {Ho}, {Yao}, {Fremling}, {Anderson},
  {Schulze}, {Kumar}, {Anupama}, {Barway}, {Bellm}, {Bhalerao}, {Chen}, {Duev},
  {Galbany}, {Graham}, {Gromadzki}, {Guti{\'e}rrez}, {Ihanec}, {Inserra},
  {Kasliwal}, {Kool}, {Kulkarni}, {Laher}, {Masci}, {Neill}, {Nicholl},
  {Pursiainen}, {van Roestel}, {Sharma}, {Sollerman}, {Walters}, \&
  {Wiseman}}]{2021MNRAS.508.5138P}
{Perley}, D.~A., {Ho}, A. Y.~Q., {Yao}, Y., {et~al.} 2021, \mnras, 508, 5138

\bibitem[{{Podsiadlowski} {et~al.}(2003){Podsiadlowski}, {Rappaport}, \&
  {Han}}]{2003MNRAS.341..385P}
{Podsiadlowski}, P., {Rappaport}, S., \& {Han}, Z. 2003, \mnras, 341, 385

\bibitem[{{Portegies Zwart} {et~al.}(1997){Portegies Zwart}, {Verbunt}, \&
  {Ergma}}]{1997A&A...321..207P}
{Portegies Zwart}, S.~F., {Verbunt}, F., \& {Ergma}, E. 1997, \aap, 321, 207

\bibitem[{{Prentice} {et~al.}(2018){Prentice}, {Maguire}, {Smartt}, {Magee},
  {Schady}, {Sim}, {Chen}, {Clark}, {Colin}, {Fulton}, {McBrien}, {O'Neill},
  {Smith}, {Ashall}, {Chambers}, {Denneau}, {Flewelling}, {Heinze}, {Holoien},
  {Huber}, {Kochanek}, {Mazzali}, {Prieto}, {Rest}, {Shappee}, {Stalder},
  {Stanek}, {Stritzinger}, {Thompson}, \& {Tonry}}]{2018ApJ...865L...3P}
{Prentice}, S.~J., {Maguire}, K., {Smartt}, S.~J., {et~al.} 2018, \apjl, 865,
  L3

\bibitem[{{Press} \& {Teukolsky}(1977)}]{1977ApJ...213..183P}
{Press}, W.~H. \& {Teukolsky}, S.~A. 1977, \apj, 213, 183

\bibitem[{{Pursiainen} {et~al.}(2018){Pursiainen}, {Childress}, {Smith},
  {Prajs}, {Sullivan}, {Davis}, {Foley}, {Asorey}, {Calcino}, {Carollo},
  {Curtin}, {D'Andrea}, {Glazebrook}, {Gutierrez}, {Hinton}, {Hoormann},
  {Inserra}, {Kessler}, {King}, {Kuehn}, {Lewis}, {Lidman}, {Macaulay},
  {M{\"o}ller}, {Nichol}, {Sako}, {Sommer}, {Swann}, {Tucker}, {Uddin},
  {Wiseman}, {Zhang}, {Abbott}, {Abdalla}, {Allam}, {Annis}, {Avila}, {Brooks},
  {Buckley-Geer}, {Burke}, {Carnero Rosell}, {Carrasco Kind}, {Carretero},
  {Castander}, {Cunha}, {Davis}, {De Vicente}, {Diehl}, {Doel}, {Eifler},
  {Flaugher}, {Fosalba}, {Frieman}, {Garc{\'\i}a-Bellido}, {Gruen}, {Gruendl},
  {Gutierrez}, {Hartley}, {Hollowood}, {Honscheid}, {James}, {Jeltema},
  {Kuropatkin}, {Li}, {Lima}, {Maia}, {Martini}, {Menanteau}, {Ogando},
  {Plazas}, {Roodman}, {Sanchez}, {Scarpine}, {Schindler}, {Smith},
  {Soares-Santos}, {Sobreira}, {Suchyta}, {Swanson}, {Tarle}, {Tucker},
  {Walker}, \& {DES Collaboration}}]{2018MNRAS.481..894P}
{Pursiainen}, M., {Childress}, M., {Smith}, M., {et~al.} 2018, \mnras, 481, 894

\bibitem[{{Reines}(2022)}]{2022NatAs...6...26R}
{Reines}, A.~E. 2022, Nature Astronomy, 6, 26

\bibitem[{{Rizzuto} {et~al.}(2023){Rizzuto}, {Naab}, {Rantala}, {Johansson},
  {Ostriker}, {Stone}, {Liao}, \& {Irodotou}}]{2023MNRAS.521.2930R}
{Rizzuto}, F.~P., {Naab}, T., {Rantala}, A., {et~al.} 2023, \mnras, 521, 2930

\bibitem[{{Ryu} {et~al.}(2024{\natexlab{a}}){Ryu}, {Amaro Seoane}, {Taylor}, \&
  {Ohlmann}}]{2024MNRAS.528.6193R}
{Ryu}, T., {Amaro Seoane}, P., {Taylor}, A.~M., \& {Ohlmann}, S.~T.
  2024{\natexlab{a}}, \mnras, 528, 6193

\bibitem[{{Ryu} {et~al.}(2024{\natexlab{b}}){Ryu}, {de Mink}, {Farmer},
  {Pakmor}, {Perna}, \& {Springel}}]{2024MNRAS.527.2734R}
{Ryu}, T., {de Mink}, S.~E., {Farmer}, R., {et~al.} 2024{\natexlab{b}}, \mnras,
  527, 2734

\bibitem[{{Ryu} {et~al.}(2020{\natexlab{a}}){Ryu}, {Krolik}, {Piran}, \&
  {Noble}}]{2020ApJ...904...98R}
{Ryu}, T., {Krolik}, J., {Piran}, T., \& {Noble}, S.~C. 2020{\natexlab{a}},
  \apj, 904, 98

\bibitem[{{Ryu} {et~al.}(2020{\natexlab{b}}){Ryu}, {Krolik}, {Piran}, \&
  {Noble}}]{2020ApJ...904...99R}
{Ryu}, T., {Krolik}, J., {Piran}, T., \& {Noble}, S.~C. 2020{\natexlab{b}},
  \apj, 904, 99

\bibitem[{{Ryu} {et~al.}(2020{\natexlab{c}}){Ryu}, {Krolik}, {Piran}, \&
  {Noble}}]{2020ApJ...904..100R}
{Ryu}, T., {Krolik}, J., {Piran}, T., \& {Noble}, S.~C. 2020{\natexlab{c}},
  \apj, 904, 100

\bibitem[{{Ryu} {et~al.}(2020{\natexlab{d}}){Ryu}, {Krolik}, {Piran}, \&
  {Noble}}]{2020ApJ...904..101R}
{Ryu}, T., {Krolik}, J., {Piran}, T., \& {Noble}, S.~C. 2020{\natexlab{d}},
  \apj, 904, 101

\bibitem[{{Ryu} {et~al.}(2024{\natexlab{c}}){Ryu}, {McKernan}, {Ford},
  {Cantiello}, {Graham}, {Stern}, \& {Leigh}}]{2024MNRAS.527.8103R}
{Ryu}, T., {McKernan}, B., {Ford}, K.~E.~S., {et~al.} 2024{\natexlab{c}},
  \mnras, 527, 8103

\bibitem[{{Ryu} {et~al.}(2023{\natexlab{a}}){Ryu}, {Perna}, {Pakmor}, {Ma},
  {Farmer}, \& {de Mink}}]{2023MNRAS.519.5787R}
{Ryu}, T., {Perna}, R., {Pakmor}, R., {et~al.} 2023{\natexlab{a}}, \mnras, 519,
  5787

\bibitem[{{Ryu} {et~al.}(2022){Ryu}, {Perna}, \& {Wang}}]{2022MNRAS.516.2204R}
{Ryu}, T., {Perna}, R., \& {Wang}, Y.-H. 2022, \mnras, 516, 2204

\bibitem[{{Ryu} {et~al.}(2023{\natexlab{b}}){Ryu}, {Valli}, {Pakmor}, {Perna},
  {de Mink}, \& {Springel}}]{2023MNRAS.525.5752R}
{Ryu}, T., {Valli}, R., {Pakmor}, R., {et~al.} 2023{\natexlab{b}}, \mnras, 525,
  5752

\bibitem[{{Salaris} \& {Weiss}(2002)}]{2002A&A...388..492S}
{Salaris}, M. \& {Weiss}, A. 2002, \aap, 388, 492

\bibitem[{{Sand} {et~al.}(2020){Sand}, {Ohlmann}, {Schneider}, {Pakmor}, \&
  {R{\"o}pke}}]{2020A&A...644A..60S}
{Sand}, C., {Ohlmann}, S.~T., {Schneider}, F. R.~N., {Pakmor}, R., \&
  {R{\"o}pke}, F.~K. 2020, \aap, 644, A60

\bibitem[{{Schneider} {et~al.}(2019){Schneider}, {Ohlmann}, {Podsiadlowski},
  {R{\"o}pke}, {Balbus}, {Pakmor}, \& {Springel}}]{2019Natur.574..211S}
{Schneider}, F. R.~N., {Ohlmann}, S.~T., {Podsiadlowski}, P., {et~al.} 2019,
  \nat, 574, 211

\bibitem[{{Sch{\"o}del} {et~al.}(2009){Sch{\"o}del}, {Merritt}, \&
  {Eckart}}]{2009A&A...502...91S}
{Sch{\"o}del}, R., {Merritt}, D., \& {Eckart}, A. 2009, \aap, 502, 91

\bibitem[{{S{\k{a}}dowski} {et~al.}(2014){S{\k{a}}dowski}, {Narayan},
  {McKinney}, \& {Tchekhovskoy}}]{2014MNRAS.439..503S}
{S{\k{a}}dowski}, A., {Narayan}, R., {McKinney}, J.~C., \& {Tchekhovskoy}, A.
  2014, \mnras, 439, 503

\bibitem[{{Smartt} {et~al.}(2018){Smartt}, {Clark}, {Smith}, {McBrien},
  {Maguire}, {O'Neil}, {Fulton}, {Magee}, {Prentice}, {Colin}, {Tonry},
  {Denneau}, {Stalder}, {Heinze}, {Weiland}, {Flewelling}, \&
  {Rest}}]{2018ATel11727....1S}
{Smartt}, S.~J., {Clark}, P., {Smith}, K.~W., {et~al.} 2018, The Astronomer's
  Telegram, 11727, 1

\bibitem[{{Springel}(2010)}]{2010MNRAS.401..791S}
{Springel}, V. 2010, \mnras, 401, 791

\bibitem[{{Stone} {et~al.}(2019){Stone}, {Kesden}, {Cheng}, \& {van
  Velzen}}]{2019GReGr..51...30S}
{Stone}, N.~C., {Kesden}, M., {Cheng}, R.~M., \& {van Velzen}, S. 2019, General
  Relativity and Gravitation, 51, 30

\bibitem[{{Stone} {et~al.}(2017){Stone}, {K{\"u}pper}, \&
  {Ostriker}}]{2017MNRAS.467.4180S}
{Stone}, N.~C., {K{\"u}pper}, A. H.~W., \& {Ostriker}, J.~P. 2017, \mnras, 467,
  4180

\bibitem[{{Timmes} \& {Swesty}(2000)}]{2000ApJS..126..501T}
{Timmes}, F.~X. \& {Swesty}, F.~D. 2000, \apjs, 126, 501

\bibitem[{{Tonry} {et~al.}(2018){Tonry}, {Denneau}, {Heinze}, {Stalder},
  {Smith}, {Smartt}, {Stubbs}, {Weiland}, \& {Rest}}]{2018PASP..130f4505T}
{Tonry}, J.~L., {Denneau}, L., {Heinze}, A.~N., {et~al.} 2018, \pasp, 130,
  064505

\bibitem[{{Udalski} {et~al.}(2015){Udalski}, {Szyma{\'n}ski}, \&
  {Szyma{\'n}ski}}]{2015AcA....65....1U}
{Udalski}, A., {Szyma{\'n}ski}, M.~K., \& {Szyma{\'n}ski}, G. 2015, \actaa, 65,
  1

\bibitem[{{Voges} {et~al.}(1999){Voges}, {Aschenbach}, {Boller},
  {Br{\"a}uninger}, {Briel}, {Burkert}, {Dennerl}, {Englhauser}, {Gruber},
  {Haberl}, {Hartner}, {Hasinger}, {K{\"u}rster}, {Pfeffermann}, {Pietsch},
  {Predehl}, {Rosso}, {Schmitt}, {Tr{\"u}mper}, \&
  {Zimmermann}}]{1999A&A...349..389V}
{Voges}, W., {Aschenbach}, B., {Boller}, T., {et~al.} 1999, \aap, 349, 389

\bibitem[{{Volonteri} {et~al.}(2021){Volonteri}, {Habouzit}, \&
  {Colpi}}]{2021NatRP...3..732V}
{Volonteri}, M., {Habouzit}, M., \& {Colpi}, M. 2021, Nature Reviews Physics,
  3, 732

\bibitem[{{Voss} \& {Gilfanov}(2007)}]{2007MNRAS.380.1685V}
{Voss}, R. \& {Gilfanov}, M. 2007, \mnras, 380, 1685

\bibitem[{{Wang} {et~al.}(2021){Wang}, {Perna}, \&
  {Armitage}}]{2021MNRAS.503.6005W}
{Wang}, Y.-H., {Perna}, R., \& {Armitage}, P.~J. 2021, \mnras, 503, 6005

\bibitem[{{Weinberger} {et~al.}(2020){Weinberger}, {Springel}, \&
  {Pakmor}}]{2020ApJS..248...32W}
{Weinberger}, R., {Springel}, V., \& {Pakmor}, R. 2020, \apjs, 248, 32

\bibitem[{{Weisskopf} {et~al.}(2000){Weisskopf}, {Tananbaum}, {Van Speybroeck},
  \& {O'Dell}}]{2000SPIE.4012....2W}
{Weisskopf}, M.~C., {Tananbaum}, H.~D., {Van Speybroeck}, L.~P., \& {O'Dell},
  S.~L. 2000, in Society of Photo-Optical Instrumentation Engineers (SPIE)
  Conference Series, Vol. 4012, X-Ray Optics, Instruments, and Missions III,
  ed. J.~E. {Truemper} \& B.~{Aschenbach}, 2--16

\bibitem[{{Xin} {et~al.}(2024){Xin}, {Haiman}, {Perna}, {Wang}, \&
  {Ryu}}]{2024ApJ...961..149X}
{Xin}, C., {Haiman}, Z., {Perna}, R., {Wang}, Y., \& {Ryu}, T. 2024, \apj, 961,
  149

\bibitem[{{Yao} {et~al.}(2022){Yao}, {Ho}, {Medvedev}, {Nayana}, {Perley},
  {Kulkarni}, {Chandra}, {Sazonov}, {Gilfanov}, {Khorunzhev}, {Khatami}, \&
  {Sunyaev}}]{2022ApJ...934..104Y}
{Yao}, Y., {Ho}, A. Y.~Q., {Medvedev}, P., {et~al.} 2022, \apj, 934, 104

\bibitem[{{Zauderer} {et~al.}(2011){Zauderer}, {Berger}, {Soderberg}, {Loeb},
  {Narayan}, {Frail}, {Petitpas}, {Brunthaler}, {Chornock}, {Carpenter},
  {Pooley}, {Mooley}, {Kulkarni}, {Margutti}, {Fox}, {Nakar}, {Patel},
  {Volgenau}, {Culverhouse}, {Bietenholz}, {Rupen}, {Max-Moerbeck}, {Readhead},
  {Richards}, {Shepherd}, {Storm}, \& {Hull}}]{2011Natur.476..425Z}
{Zauderer}, B.~A., {Berger}, E., {Soderberg}, A.~M., {et~al.} 2011, \nat, 476,
  425

\end{thebibliography}
%

\begin{appendix}

\onecolumn
\section{Summary of simulation results}

\begin{longtable}{llllllllllll}
    \caption{Post-disruption parameter values for our 58 simulations. }
    \label{tab:param_values}\\
    \hline\hline
     & $\ms$ & $\mbh$ & $b$ & $\Hc$ & $\Ls$ / $10^{46}$ & $\msr$ & $m_{\mathrm{\star \rightarrow BH}}$ & $m_{\mathrm{unbound}}$ & $\Lsr$ / $10^{48}$ & $a$ & $e$ \Tstrut \\
     & $[\Msun]$ & $[\Msun]$ &  &  & $[\gcmsqsinv]$ & $[\Msun]$ & $[\Msun]$ & $[\Msun]$ & $[\gcmsqsinv]$ & $[\au]$ &  \Bstrut \\
    \hline\hline
    1 & 0.5 & 10.0 & 0.70 & 0.25 & 2 & 0.00 & 0.29 & 0.21 & N/A & N/A & N/A \Tstrut \\
    2 & 0.5 & 10.0 & 0.70 & 0.33 & 2 & 0.00 & 0.22 & 0.28 & N/A & N/A & N/A \\
    3 & 0.5 & 10.0 & 0.70 & 0.50 & 2 & 0.00 & 0.23 & 0.17 & N/A & N/A & N/A \\
    4 & 0.5 & 10.0 & 0.70 & 0.75 & 2 & 0.15 & 0.21 & 0.14 & 87 & -0.07 & 1.06 \\
    5 & 0.5 & 10.0 & 0.70 & 1.00 & 2 & 0.25 & 0.14 & 0.11 & 97 & -0.27 & 1.02 \\
    6 & 0.5 & 10.0 & 0.70 & 1.50 & 2 & 0.46 & 0.03 & 0.01 & 117 & 1.91 & 0.995 \\
    7 & 0.5 & 10.0 & 0.70 & 2.00 & 2 & 0.50 & 0.00 & 0.00 & 17 & 2.61 & 0.996 \\
    8 & 0.5 & 10.0 & 0.70 & 2.50 & 2 & 0.50 & 0.00 & 0.00 & 1 & 42.3 & 1.00 \Bstrut \\
    \hline\hline
    9 & 0.5 & 40.0 & 0.70 & 0.25 & 2 & 0.00 & 0.27 & 0.23 & N/A & N/A & N/A \Tstrut \\
    10 & 0.5 & 40.0 & 0.70 & 0.33 & 2 & 0.00 & 0.30 & 0.20 & N/A & N/A & N/A \\
    11 & 0.5 & 40.0 & 0.70 & 0.50 & 2 & 0.00 & 0.30 & 0.20 & N/A & N/A & N/A \\
    12 & 0.5 & 40.0 & 0.70 & 0.75 & 2 & 0.00 & 0.31 & 0.19 & N/A & N/A & N/A \\
    13 & 0.5 & 40.0 & 0.70 & 1.00 & 2 & 0.21 & 0.18 & 0.11 & 76 & -0.35 & 1.03 \\
    14 & 0.5 & 40.0 & 0.70 & 1.50 & 2 & 0.46 & 0.03 & 0.01 & 110 & -0.11 & 1.00 \\
    15 & 0.5 & 40.0 & 0.70 & 2.00 & 2 & 0.50 & 0.00 & 0.00 & 15 & 11.4 & 0.998 \\
    16 & 0.5 & 40.0 & 0.70 & 2.50 & 2 & 0.50 & 0.00 & 0.00 & 1 & 161 & 1.00 \Bstrut \\
    \hline\hline
    17 & 1.0 & 10.0 & 0.00 & 0.25 & 12 & 0.04 & 0.42 & 0.54 & 2 & 0.02 & 0.916 \Tstrut \\
    18 & 1.0 & 10.0 & 0.00 & 0.33 & 12 & 0.33 & 0.27 & 0.40 & 101 & 0.04 & 0.923 \\
    19 & 1.0 & 10.0 & 0.00 & 0.50 & 12 & 0.68 & 0.11 & 0.21 & 410 & 0.10 & 0.945 \\
    20 & 1.0 & 10.0 & 0.00 & 0.75 & 12 & 0.89 & 0.05 & 0.06 & 354 & 0.32 & 0.972 \\
    21 & 1.0 & 10.0 & 0.00 & 1.00 & 12 & 0.97 & 0.02 & 0.01 & 199 & 1.15 & 0.989 \\
    22 & 1.0 & 10.0 & 0.00 & 1.50 & 12 & 1.00 & 0.00 & 0.00 & 49 & 13.2 & 0.999 \\
    23 & 1.0 & 10.0 & 0.00 & 2.00 & 12 & 1.00 & 0.00 & 0.00 & 10 & 54.6 & 1.00 \Bstrut \\
    \hline
    24 & 1.0 & 10.0 & 0.34 & 0.25 & 10 & 0.00 & 0.47 & 0.53 & N/A & N/A & N/A \Tstrut \\
    25 & 1.0 & 10.0 & 0.34 & 0.33 & 10 & 0.09 & 0.47 & 0.44 & 99 & -0.06 & 1.12 \\
    26 & 1.0 & 10.0 & 0.34 & 0.50 & 10 & 0.49 & 0.20 & 0.31 & 245 & 0.08 & 0.950 \\
    27 & 1.0 & 10.0 & 0.34 & 0.75 & 10 & 0.83 & 0.09 & 0.08 & 440 & 0.21 & 0.965 \\
    28 & 1.0 & 10.0 & 0.34 & 1.00 & 10 & 0.95 & 0.03 & 0.02 & 250 & 0.64 & 0.984 \\
    29 & 1.0 & 10.0 & 0.34 & 1.50 & 10 & 1.00 & 0.00 & 0.00 & 49 & 7.77 & 0.998 \\
    30 & 1.0 & 10.0 & 0.34 & 2.00 & 10 & 1.00 & 0.00 & 0.00 & 10 & 87.3 & 1.00 \Bstrut \\
    \hline
    31 & 1.0 & 10.0 & 0.70 & 0.25 & 16 & 0.00 & 0.37 & 0.63 & N/A & N/A & N/A \Tstrut \\
    32 & 1.0 & 10.0 & 0.70 & 0.33 & 16 & 0.00 & 0.49 & 0.51 & N/A & N/A & N/A \\
    33 & 1.0 & 10.0 & 0.70 & 0.50 & 16 & 0.34 & 0.29 & 0.37 & 163 & 0.28 & 0.986 \\
    34 & 1.0 & 10.0 & 0.70 & 0.75 & 16 & 0.68 & 0.16 & 0.16 & 440 & 0.2.2 & 0.968 \\
    35 & 1.0 & 10.0 & 0.70 & 1.00 & 16 & 0.90 & 0.06 & 0.04 & 387 & 0.35 & 0.974 \\
    36 & 1.0 & 10.0 & 0.70 & 1.50 & 16 & 1.00 & 0.00 & 0.00 & 67 & 3.43 & 0.996 \\
    37 & 1.0 & 10.0 & 0.70 & 2.00 & 16 & 1.00 & 0.00 & 0.00 & 8 & 25.5 & 0.999 \Bstrut \\
    \hline\hline
    38 & 1.0 & 40.0 & 0.00 & 0.25 & 12 & 0.07 & 0.45 & 0.48 & 56 & -0.05 & 1.14 \Tstrut \\
    39 & 1.0 & 40.0 & 0.00 & 0.33 & 12 & 0.28 & 0.27 & 0.45 & 66 & 0.17 & 0.969 \\
    40 & 1.0 & 40.0 & 0.00 & 0.50 & 12 & 0.63 & 0.16 & 0.21 & 340 & 0.58 & 0.984 \\
    41 & 1.0 & 40.0 & 0.00 & 0.75 & 12 & 0.88 & 0.06 & 0.06 & 332 & 2.24 & 0.994 \\
    42 & 1.0 & 40.0 & 0.00 & 1.00 & 12 & 0.96 & 0.02 & 0.02 & 192 & 9.63 & 0.998 \\
    43 & 1.0 & 40.0 & 0.00 & 1.50 & 12 & 1.00 & 0.00 & 0.00 & 50 & 102 & 1.00 \\
    44 & 1.0 & 40.0 & 0.00 & 2.00 & 12 & 1.00 & 0.00 & 0.00 & 8 & 153 & 1.00 \Bstrut \\
    \hline
    45 & 1.0 & 40.0 & 0.34 & 0.25 & 10 & 0.00 & 0.42 & 0.58 & N/A & N/A & N/A \Tstrut \\
    46 & 1.0 & 40.0 & 0.34 & 0.33 & 10 & 0.00 & 0.39 & 0.61 & N/A & N/A & N/A \\
    47 & 1.0 & 40.0 & 0.34 & 0.50 & 10 & 0.42 & 0.22 & 0.36 & 169 & 0.95 & 0.992 \\
    48 & 1.0 & 40.0 & 0.34 & 0.75 & 10 & 0.81 & 0.09 & 0.10 & 413 & 1.84 & 0.993 \\
    49 & 1.0 & 40.0 & 0.34 & 1.00 & 10 & 0.95 & 0.03 & 0.02 & 245 & 5.09 & 0.997 \\
    50 & 1.0 & 40.0 & 0.34 & 1.50 & 10 & 1.00 & 0.00 & 0.00 & 50 & 62.6 & 1.00 \\
    51 & 1.0 & 40.0 & 0.34 & 2.00 & 10 & 1.00 & 0.00 & 0.00 & 10 & 180 & 1.00 \Bstrut \\
    \hline
    52 & 1.0 & 40.0 & 0.70 & 0.25 & 16 & 0.00 & 0.55 & 0.45 & N/A & N/A & N/A \Tstrut \\
    53 & 1.0 & 40.0 & 0.70 & 0.33 & 16 & 0.00 & 0.50 & 0.50 & N/A & N/A & N/A \\
    54 & 1.0 & 40.0 & 0.70 & 0.50 & 16 & 0.21 & 0.39 & 0.40 & 156 & -0.12 & 1.06 \\
    55 & 1.0 & 40.0 & 0.70 & 0.75 & 16 & 0.63 & 0.18 & 0.19 & 365 & -4.17 & 1.00 \\
    56 & 1.0 & 40.0 & 0.70 & 1.00 & 16 & 0.89 & 0.06 & 0.05 & 368 & 2.91 & 0.995 \\
    57 & 1.0 & 40.0 & 0.70 & 1.50 & 16 & 1.00 & 0.00 & 0.00 & 66 & 29.4 & 0.999 \\
    58 & 1.0 & 40.0 & 0.70 & 2.00 & 16 & 1.00 & 0.00 & 0.00 & 8 & 119 & 1.00 \Bstrut \\
    \hline
\end{longtable}
\tablefoot{The columns, respectively, denote star mass $\ms$, BH mass $\mbh$, impact parameter $b$, central hydrogen abundance $\Hc$, initial (almost non-rotating) star spin angular momentum $\Ls$, remnant mass $\msr$, gas mass bound to BH $m_{\mathrm{\star \rightarrow BH}}$, gas mass unbound from the BH-star system $m_{\mathrm{unbound}}$, remnant spin angular momentum $\Lsr$, orbital semimajor axis $a$ and orbital eccentricity $e$.}

\end{appendix}

\end{document}